\documentclass[aps,prd]{revtex4-2}
\usepackage{graphicx}
\usepackage{amssymb}
\usepackage{enumitem}
\usepackage{float}

\begin{document}

\title{Gravitational lensing in the charged NUT--de Sitter spacetime}

\author{Torben C. Frost}

\affiliation{ZARM, University of Bremen, 28359 Bremen, Germany and Institute for Theoretical Physics, Leibniz University Hannover, 30167 Hannover, Germany\\
e-mail: torben.frost@zarm.uni-bremen.de}

\date{December 31, 2023}

\begin{abstract}
It is a long-standing open question if a gravitomagnetic charge, the gravitational
analogon to a hypothetical magnetic charge in electrodynamics, exists in nature.
It naturally occurs in certain exact solutions to Einstein's electrovacuum-field equations with cosmological constant.
The charged NUT--de Sitter metric is such a solution. It describes a black hole with
electric and gravitomagnetic charges and a cosmological constant. In this paper we will
address the question how we can observe the gravitomagnetic charge using gravitational
lensing. For this purpose we first solve the equations of motion for lightlike geodesics using
Legendre's canonical forms of the elliptic integrals and Jacobi's elliptic functions.
We fix a stationary observer in the domain of outer communication and introduce an orthonormal
tetrad. The orthonormal tetrad relates the direction under which the observer detects
a light ray to its latitude-longitude coordinates on the observer's celestial sphere.
In this parametrization we rederive the angular radius of the shadow, formulate
a lens map, discuss the redshift, and the travel time. We also discuss relevant
differences with respect to spherically symmetric and static spacetimes and how
we can use them to determine if an astrophysical black hole has a gravitomagnetic
charge.
\end{abstract}

\maketitle
\section{Introduction}
The charged NUT--de Sitter metric belongs to the more exotic solutions of Einstein's
electrovacuum-field equations with cosmological constant. It is axisymmetric and
stationary and belongs to the Pleba{\'n}ski-Demia{\'n}ski family of spacetimes of Petrov
type D \cite{Plebanski1976}. In addition to the mass parameter $m$, the electric
charge $e$ and the cosmological constant $\Lambda$ it contains two parameters $n$
and $C$. In analogy to a hypothetical magnetic monopole with magnetic charge $b$
the parameter $n$ is usually referred to as "gravitomagnetic charge." The parameter
$C$ is called the Manko-Ruiz parameter \cite{Manko2005}. The spacetime is usually interpreted
to describe a black hole; however, unlike the Reissner-Nordstr\"{o}m--de Sitter metric
it does not contain a curvature singularity at $r=0$.
The original Taub-NUT spacetime was discovered in two steps. First the time-dependent
part of the spacetime was discovered by Taub in 1951 \cite{Taub1951}. In 1963
Newman \emph{et al.} \cite{Newman1963} used the Newman-Penrose formalism
to derive three different metrics characterized by geodesic rays which do not diverge
or shear but curl. One of these metrics they identified as a generalization of
the Schwarzschild metric, the so-called NUT metric. Newman \emph{et al.}, and about one month
later Misner \cite{Misner1963}, also pointed out that Taub's solution can be interpreted as
an extension of their spacetime. Misner \cite{Misner1963} was also the first who referred to the
spacetime as "NUT space." In its original form the spacetime is asymptotically flat
in the sense that for $r\rightarrow \infty$ the Riemann tensor vanishes. However, the spacetime
does not become asymptotically Minkowskian \cite{Misner1963}. Misner also
noted that either the metric or the time coordinate $t$ has a singularity at $\vartheta=\pi$ (for this historical
reason the axial singularities are called Misner strings). Bonnor \cite{Bonnor1969}
investigated the nature of this singularity and came to the conclusion that it can
be interpreted as a semi-infinite massless rotating rod that serves as a source of angular momentum (see also the work of
Sackfield \cite{Sackfield1971}). He also pointed out that the strength of the Misner
string is directly related to the gravitomagnetic charge $n$. The parameter $C$
is also closely tied to the axial singularities. It was introduced by Manko and
Ruiz \cite{Manko2005} and can be used to control the number (one or two) and location
of the axial singularities. Analogously to the Schwarzschild metric the NUT metric
can also be generalized. According to Griffiths and Podolsk{\' y} \cite{Griffiths2009}
the NUT metric with electric charge was first discovered by Brill \cite{Brill1964}
(note that with a sufficiently large electric charge $e$ the charged NUT metric can also be interpreted
as a wormhole, see, e.g., Cl{\' e}ment \emph{et al.} \cite{Clement2016}) and the charged
NUT--de Sitter metric was discovered in 1972 by Ruban \cite{Ruban1972}. The charged
NUT--de Sitter metrics (whenever we use the plural we will refer to the whole family
of metrics with gravitomagnetic charge in the following) are interesting from the
physical perspective because they represent exact solutions to Einstein's electrovacuum-field
equations with cosmological constant which in addition to the mass parameter $m$ also incorporate a gravitomagnetic
mass $n$ (to maintain consistency throughout the paper hereafter we will continue to refer to
it as gravitomagnetic charge).
However, the presence of the Misner strings leads to two undesirable aspects. First,
although the Misner strings are massless it is unclear if geodesics can be continued
through the axes. While many authors advocate that the spacetime is geodesically incomplete,
see, e.g., the work in \cite{Misner1969,Miller1971,Kagramanova2010}, Cl{\' e}ment \emph{et al.} \cite{Clement2015}
investigated this aspect for the NUT metric and came to the conclusion that geodesics can be smoothly continued
through the Misner strings (we will see that for the spatial coordinates this argument
can also be transferred to all charged NUT--de Sitter metrics). The second problematic
aspect of the NUT metric is that close to the Misner strings it contains regions
with closed timelike curves. Misner \cite{Misner1963} demonstrated that the axial
singularities can be removed by introducing a periodic time coordinate; however,
this step does not alleviate the problem but actually makes it worse.
Using the periodic time coordinate the spacetime contains closed timelike curves
everywhere, which is even less desirable. The presence of closed timelike curves
makes the spacetime on the first view appear unphysical; however, the presence of
closed timelike curves is limited to a narrow region around the Misner strings.
Thus the NUT metric may still serve as an approximate model for a spacetime
with gravitomagnetism as long as these regions are excluded.\\
In astrophysical settings the gravitomagnetic charge is expected to be very small
\cite{LyndenBell1998,Rahvar2003}. Therefore, if we ever want to have a chance to
detect visible effects caused by the gravitomagnetic charge we need gravitationally
heavy objects. Supermassive black holes (SMBHs) at the center of galaxies are ideal
candidates for such objects. Because we are currently not able to send any probes to
SMBHs we have to rely on information carried to us by electromagnetic or gravitational
radiation.
Present-day gravitational wave detectors such as Laser Interferometer Gravitational Wave Observatory (LIGO) \cite{LIGOCollaboration2015},
Virgo \cite{Acernese2015}, and KAGRA \cite{Akutsu2018} so far only detected gravitational
waves from stellar mass binary black hole and neutron star mergers and thus even with very high accuracy
gravitational wave templates it is very likely that imprints of the gravitomagnetic
charge on the detected gravitational wave signals are impossible to resolve.
On the other hand recent technological advances in Very Large Baseline Interferometry (VLBI)
lead to the observation of the shadow of the supermassive black hole in the
galaxy M87 by the Event Horizon Telescope (EHT) \cite{EHTCollaboration2019a}. The
EHT has an angular resolution of about $25~\mu \mathrm{as}$ at a wavelength of 1.3~mm \cite{EHTCollaboration2019b}. This resolution
is high enough to demonstrate that M87 contains an object that casts a shadow; however,
the shape of the shadow is strongly blurred by the surrounding accretion disk and
thus without further information its exact shape is difficult to reconstruct from
observations alone. Because the resolution of ground-based VLBI is limited by the distribution
of radio telescopes on the surface of Earth we can only enhance it by extending VLBI to space.
Space VLBI reaches back to the late 1970s. The most recent space VLBI program
used the Spektr-R satellite \cite{Kardashev2013,Kardashev2017} as space-borne station
and was terminated in 2019. The antenna of Spektr-R was able to observe at four
wavelengths between 1 and 100 cm and thus did not operate in the millimeter/submillimeter
range required for VLBI observations of supermassive black holes. Satellite missions
attempting to achieve observations at these wavelengths are currently in their planning
stage and will allow enhanced observations of the shadow in M87 and, potentially,
also the observation of the centers of more distant galaxies. Therefore, from today's
perspective observing light gravitationally lensed by SMBHs promises the best chance
to detect effects caused by the presence of the gravitomagnetic charge $n$. \\
Gravitational lensing in the weak- and strong-field regimes of the NUT metric has
already been investigated by several authors. Gravitational lensing in the NUT metric
was first investigated by Zimmerman and Shahir in 1989 \cite{Zimmerman1989}. They
first showed that in the NUT metric all geodesics lie on spatial cones and then
calculated the bending angle up to the first nonvanishing order in $n$ for light rays on these cones. Up to first order in $m$ their
result was independently reproduced by Lynden-Bell and Nouri-Zonoz \cite{LyndenBell1998}.
In addition Lynden-Bell and Nouri-Zonoz defined a simple lens map. They determined area
magnification and the axial ratio of the image of a small circular source. In \cite{NouriZonoz1997}
Nouri-Zonoz and Lynden-Bell present a more thorough analytical approach to gravitational
lensing in the NUT metric. After first rederiving the light-bending formula on
a cone the authors proceed to define a different version of the lens equation and
the magnification factor. In addition, they derive the geometric time delay and
the Shapiro time delay between two images of the same source. Both works showed
that the presence of a gravitomagnetic charge is associated with a twist in the
observed lensing pattern. In \cite{Rahvar2003} Rahvar and Nouri-Zonoz used these
results to investigate gravitational microlensing in the NUT metric. While in all
previous works the deflection angle was calculated using a simple expansion, Halla
and Perlick \cite{Halla2020} used a different approach. Following the work of Werner
\cite{Werner2012} they used the Gauss-Bonnet theorem to derive the deflection angle.
The strong-field deflection limit was first investigated by Wei \emph{et al.} \cite{Wei2012}
for Kerr--NUT spacetimes. Using numerical and analytical methods the authors constructed
a lens equation for light rays in and close to the equatorial plane. In addition they
derived the critical curves and the caustic structure, and the magnification of the
images near the caustic points. Sharif and Iftikhar \cite{Sharif2016} investigated
strong gravitational lensing in the equatorial plane of accelerating Kerr-NUT black
holes. Finally, Grenzebach \emph{et al.} \cite{Grenzebach2014,Grenzebach2016} investigated
the photon region and the shadow of Kerr-Newman-NUT black holes with cosmological
constant. While all these works investigated gravitational lensing in different
NUT metrics, to the best of my knowledge in the charged NUT--de Sitter metrics an exact
analytic lens map has not been constructed so far. Therefore, the main aim of this
paper is to use exact analytical methods to investigate gravitational lensing for
arbitrary light rays in the charged NUT--de Sitter metrics. Geodesic motion in the
NUT metric was first investigated by Zimmerman and Shahir \cite{Zimmerman1989}.
After a thorough potential analysis Zimmerman and Shahir derived the time integral
for radial timelike geodesics and investigated timelike circular and elliptic bound orbits. In addition they derived
the deflection angle of light rays on spatial cones. The most thorough investigation of geodesic
motion was carried out by Kagramanova \emph{et al.} \cite{Kagramanova2010} using Weierstrass'
elliptic function and Weierstrass' $\zeta$ and $\sigma$ functions. However, for investigating gravitational lensing in the charged
NUT--de Sitter metrics these functions are rather impractical because in the equations for the time coordinate derived in \cite{Kagramanova2010}
during the integration procedure the branches of the logarithm have to be manually
adjusted for each light ray individually. This problem can be circumvented by using the canonical
forms of Legendre's elliptic integrals and Jacobi's elliptic functions. In general relativity
using Legendre's canonical forms of the elliptic integrals and Jacobi's elliptic functions
for solving the equations of motion has already a long tradition since the early
1920s. Forsyth \cite{Forsyth1920}, Morton \cite{Morton1921}, and Darwin \cite{Darwin1959}
used Jacobi's elliptic functions to solve and discuss lightlike and timelike geodesics
in the Schwarzschild metric. More recently Yang and Wang \cite{Yang2013} and Gralla
and Lupsasca \cite{Gralla2020} extended these investigations to lightlike geodesics
in the Kerr metric. In particular, the approach of Gralla and Lupsasca \cite{Gralla2020}
can be easily transferred to lightlike geodesics in the charged NUT--de Sitter metrics.
Therefore in the first part of this paper we will derive the solutions to the equations
of motion in terms of Legendre's elliptic integrals and Jacobi's elliptic functions following the approach
of Gralla and Lupsasca \cite{Gralla2020}.
In the second part of the paper we will investigate gravitational lensing in the
charged NUT--de Sitter metrics. We will construct an exact lens map following Frost
and Perlick \cite{Frost2021a} using the tetrad approach of Grenzebach \emph{et al.}
\cite{Grenzebach2015a}. We will use the tetrad approach to calculate the shadow
of the black hole, set up a lens equation, and discuss the redshift and the travel
time. \\
The remainder of this paper is structured as follows. In Sec.~\ref{Sec:NUT}
we will summarize the main properties of the charged NUT--de Sitter metrics. In
Sec.~\ref{Sec:EoM} we will discuss and solve the equations of motion. In Sec.~\ref{Sec:Lensing}
we will set up the lens map and discuss lensing features in the charged NUT--de Sitter
metrics, namely, the angular radius of the shadow, the lens equation, the redshift,
and the travel time. We will also comment on how the observed lensing features can
be used to measure the gravitomagnetic charge. In Sec.~\ref{Sec:Summary} we will
summarize our results and conclusions. Throughout the paper we will use geometric units such that
$c=G=1$. The metric signature is $\left(-,+,+,+\right)$.\\

\section{THE CHARGED NUT--$\mathrm{\textbf{de}}$ SITTER SPACETIME} \label{Sec:NUT}
The charged NUT--de Sitter metric belongs to the Pleba{\'n}ski-Demia{\'n}ski family
of electrovacuum spacetimes of Petrov type D \cite{Plebanski1976} and is an exact
solution of Einstein's electrovacuum-field equations with cosmological constant.
In Boyer-Lindquist-like coordinates its line element reads \cite{Griffiths2009}
\begin{eqnarray}\label{eq:line}
g_{\mu\nu}\mathrm{d}x^{\mu}\mathrm{d}x^{\nu}=-\frac{Q(r)}{\rho(r)}(\mathrm{d}t+2n(\cos\vartheta+C)\mathrm{d}\varphi)^2+\frac{\rho(r)}{Q(r)}\mathrm{d}r^2+\rho(r)(\mathrm{d}\vartheta^2+\sin^2\vartheta\mathrm{d}\varphi^2),
\end{eqnarray}
where
\begin{eqnarray}\label{eq:coeffP}
Q(r)=-\frac{\Lambda}{3}r^4+r^2(1-2\Lambda n^2)-2mr+e^2-n^2(1-\Lambda n^2),
\end{eqnarray}
and
\begin{eqnarray}\label{eq:coeffrho}
\rho(r)=r^2+n^2.
\end{eqnarray}
The metric is axisymmetric and stationary and for $\Lambda=0$ asymptotically flat
(note that here asymptotically flat means that the Riemann tensor vanishes but the
spacetime is not asymptotically Minkowskian \cite{Misner1963}).
It contains five parameters: the mass parameter $m$, the cosmological constant $\Lambda$,
the electric charge $e$, the gravitomagnetic charge $n$ and the so-called Manko-Ruiz
parameter $C$ (for more information regarding the Manko-Ruiz parameter see \cite{Manko2005}).
When we set $n=0$ the metric reduces to the Reissner-Nordstr\"{o}m--de Sitter family
of spacetimes which includes the Schwarzschild metric ($\Lambda=0$ and $e=0$), the
Schwarzschild--de Sitter metric ($e=0$) and the Reissner-Nordstr\"{o}m metric ($\Lambda=0$).
For $e=0$ and $\Lambda=0$ the metric reduces to the standard NUT metric. For $e=0$
it reduces to the NUT--de Sitter metric and for $\Lambda=0$ it reduces to the charged
NUT metric. \\
In this article we choose $\vartheta$ and $\varphi$ such that they represent coordinates
on the two-sphere $S^{2}$ and cover the range $\vartheta \in [0,\pi]$ and $\varphi\in [0,2\pi)$.
Although the spacetime is axisymmetric it retains some degree of "spherical symmetry."
As discussed in Newman \emph{et al.} \cite{Newman1963} for $C=-1$ and in
Halla and Perlick \cite{Halla2020} for arbitrary $C$ the spacetime admits four linearly
independent Killing vector fields. Three of these Killing vector fields generate
isometries isomorphic to the rotation group $SO(3,{\Bbb R})$ and thus the metric
is rotationally symmetric with respect to any radial direction (for more details
see Halla and Perlick \cite{Halla2020}). The Manko-Ruiz parameter $C$ can be removed
from Eq.~(\ref{eq:line}) using the coordinate transformation $\tilde{t}=t+2nC\varphi$.
Note that this transformation is not valid globally because the $\varphi$ coordinate
is periodic and the time coordinate $t$ is not. Therefore, charged NUT--de Sitter
spacetimes with arbitrary $C$ are locally isometric \cite{Grenzebach2014}. \\
From the theoretical perspective the five parameters $m$, $\Lambda$, $e$, $n$, and
$C$ can take any arbitrary real value. Luckily the symmetries of the spacetime and
observational experience allow us to reduce their range for our investigation of
gravitational lensing. First, astronomical observations show that all objects in
nature have a positive mass and thus we choose $m>0$. Second, cosmological observations
indicate that we live in an expanding Universe with positive cosmological constant
allowing us to choose $0\leq\Lambda<\Lambda_{\mathrm{C}}$. Third, in Eq. (\ref{eq:coeffP})
the electric charge $e$ only enters as square and since we only deal with light rays
we can choose $0\leq e\leq e_{\mathrm{C}}$ without loss of generality. Fourth and
last, the gravitomagnetic charge $n$ can be restricted considering the symmetries
of the spacetime. When we set $n \rightarrow -n$, $C\rightarrow -C$
and perform the coordinate transformation $\vartheta \rightarrow \pi-\vartheta$
the line element remains invariant. Consequently we can limit the gravitomagnetic
charge to $0\leq n<n_{\mathrm{C}}$ while the Manko-Ruiz parameter $C$ can take any
real number. Here, the three constants $\Lambda_{\mathrm{C}}$, $e_{\mathrm{C}}$,
and $n_{\mathrm{C}}$ are limiting values that are determined by the nature of the
desired spacetime. We will come back to these parameters when we discuss the singularities
of the spacetime below. The charged NUT--de Sitter metric offers several mathematical
peculiarities that may not be familiar to every reader. Thus in the following we
will provide a short summary of its physical properties before we move on to discuss
and solve the equations of motion. \\
The charged NUT--de Sitter metric admits several singularities. The line element
Eq. (\ref{eq:line}) maintains its structure independent of how we choose $\Lambda$
and $e$. Therefore, we will restrict our discussion to the NUT metric whenever
possible. We start by discussing the singularities of the metric associated with
the roots of $Q(r)=0$. In the charged NUT--de Sitter metric the equation $Q(r)=0$
can lead to up to four singularities. In this paper we want the metric to represent
black hole spacetimes and thus we have to choose $\Lambda_{\mathrm{C}}$, $e_{\mathrm{C}}$
and $n_{\mathrm{C}}$ such that all roots of the equation $Q(r)=0$ are real. In this
case all roots are coordinate singularities that can be removed using appropriate
coordinate transformations. Figure~1 shows the horizon structures of the NUT metric
[panel (a)], of the charged NUT metric [panels (b) and (c)], of the NUT--de Sitter metric
[panel (d)] and of the charged NUT--de Sitter metric [panels (e) and (f)]. In the NUT metric
Eq. (\ref{eq:coeffP}) reduces to $Q(r)=r^2-2mr-n^2$. We can immediately read that it
has two roots at
\begin{eqnarray}
r_{\pm}=m\pm\sqrt{m^2+n^2}.
\end{eqnarray}
For consistency with Fig.~1 from now on we will label them $r_{-}=r_{\mathrm{H},\mathrm{i}}$
and $r_{+}=r_{\mathrm{H},\mathrm{o}}$. For $r<r_{\mathrm{H},\mathrm{i}}$ and $r_{\mathrm{H},\mathrm{o}}<r$
the vector field $K_{t}=\partial_{t}$ is timelike and the vector field $K_{r}=\partial_{r}$
is spacelike. In these two domains the spacetime is stationary (except for a narrow
region close to the Misner string as we will discuss below). The domain $r_{\mathrm{H},\mathrm{o}}<r$
is usually referred to as domain of outer communication and will be of importance
in Secs. \ref{Sec:EoM} and \ref{Sec:Lensing}. Between the horizons $\partial_{t}$
is spacelike and $\partial_{r}$ is timelike. In this domain the spacetime is nonstationary.
When we add the electric charge $e$ the horizon $r_{\mathrm{H},\mathrm{i}}$ shifts
to larger $r$ and the horizon $r_{\mathrm{H},\mathrm{o}}$ shifts to smaller $r$.
When $e=e_{\mathrm{C}}$ both horizons coincide at $r_{\mathrm{H}}=m$. Adding the
cosmological constant $\Lambda$ gives rise to two additional, cosmological horizons
$r_{\mathrm{C}-}$ and $r_{\mathrm{C}+}$. Both cosmological horizons limit the stationary domains found
in the NUT metric to $r_{\mathrm{C}-}<r<r_{\mathrm{H},\mathrm{i}}$ and $r_{\mathrm{H},\mathrm{o}}<r<r_{\mathrm{C}+}$.
The two domains $r<r_{\mathrm{C}-}$ and $r_{\mathrm{C}+}<r$ are nonstationary. The function $\rho(r)$
is always positive. Consequently the charged NUT--de Sitter metric does not possess
a curvature singularity at $r=0$. This has an important implication for the whole
spacetime. Lightlike and timelike geodesics are not blocked at $r=0$ and thus the
$r$ coordinate covers the whole real axis ($r \in [-\infty,\infty]$). \\
In addition to the horizons the NUT metric has one or two conical singularities
on the axes. The exact number depends on the Manko-Ruiz parameter $C$. For $C=1$
the singularity is located on the axis $\vartheta=0$. For $C=-1$ the spacetime has
a singularity on the axis $\vartheta=\pi$. For all other choices of $C$ the spacetime
has singularities on both axes. For a more detailed discussion of the conical singularities
see Jefremov and Perlick \cite{Jefremov2016} and Halla and Perlick \cite{Halla2020}.
The allowed range of the time coordinate $t$ depends on how the NUT metric is
interpreted. As discussed before depending on the choice of the Manko-Ruiz parameter
$C$ the NUT metric has conical singularities (Misner strings) on one or both
axes \cite{Manko2005}. As discussed by Bonnor \cite{Bonnor1969} the Misner strings
can be interpreted as semi-infinite massless line sources of angular momentum and give
rise to the gravitomagnetic charge. Following the approach of Misner \cite{Misner1963}
we can remove them by introducing a periodic time coordinate. But, this comes at
a high price. In Misner's interpretation the periodic time coordinate leads to
closed timelike curves in the whole spacetime. Closed timelike curves violate
causality and are thus physically not desirable. Therefore in this paper we choose
to retain the Misner strings and have $t\in {\Bbb R}$. In this case the spacetime
also contains closed timelike curves whenever $g_{\varphi\varphi}\leq 0$ \cite{Bonnor1969};
however, these are confined to very narrow regions close to the Misner strings.
\begin{center}
    \begin{figure*}\label{fig:NUTHorizon}
	      \includegraphics[width=\textwidth]{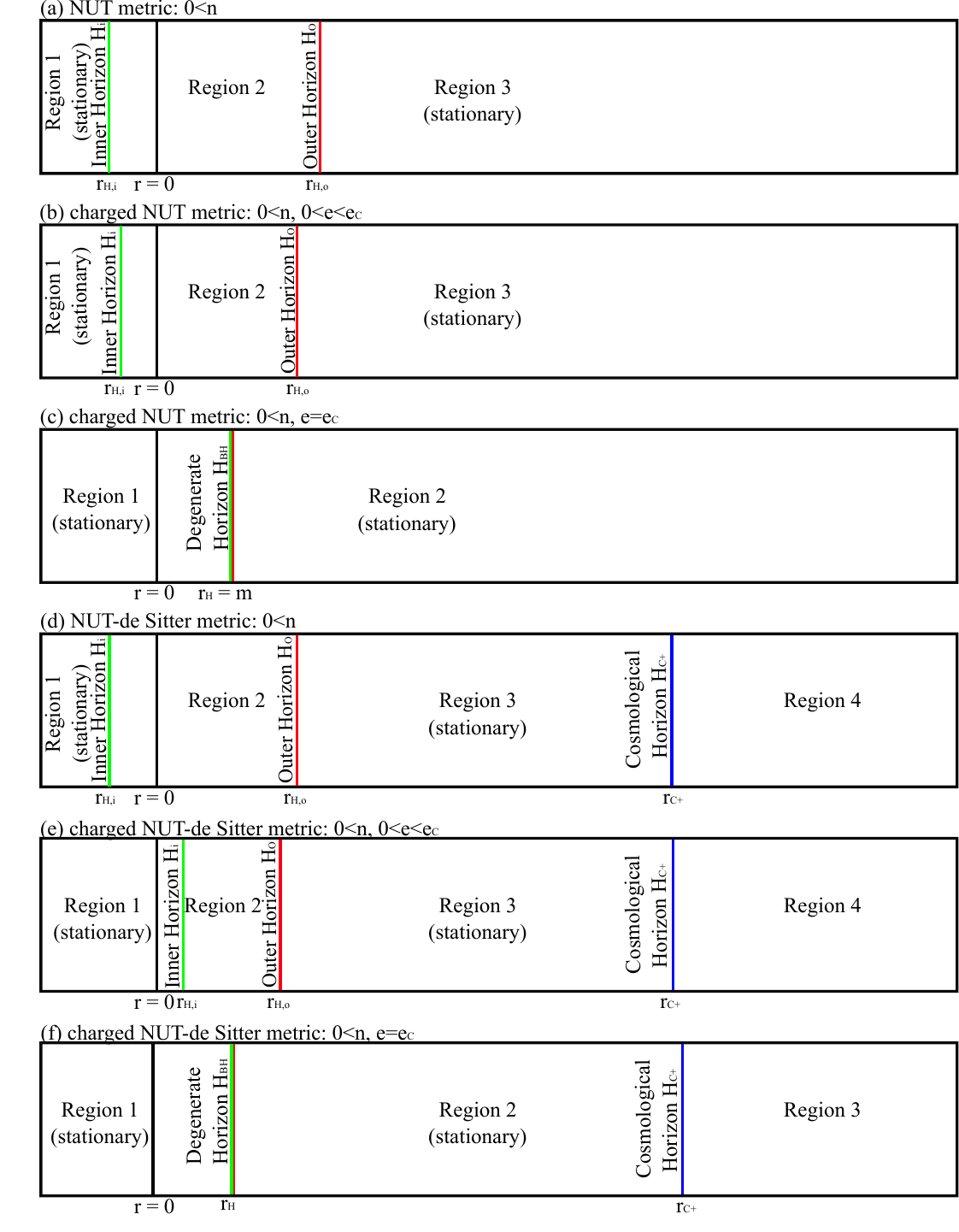}
	      \caption{Positions of the coordinate singularities in (a) the NUT metric,
        the charged NUT metric with (b) $0<e<e_{\mathrm{C}}=\sqrt{m^2+n^2}$, and (c)
        $e=e_{\mathrm{C}}=\sqrt{m^2+n^2}$, (d) the NUT--de Sitter metric and the charged
        NUT--de Sitter metric with (e) $0<e<e_{\mathrm{C}}$ and (f) $e=e_{\mathrm{C}}$.
        Note that the angular coordinates are suppressed and the cosmological horizon
        $H_{\mathrm{C}-}$ at $r_{\mathrm{C}-}$, the region $r<r_{\mathrm{C}-}$ and other singularities are
        not shown.}
    \end{figure*}
\end{center}

\section{Equations of Motion}\label{Sec:EoM}
For lightlike geodesics the charged NUT--de Sitter metric admits four constants of
motion. These are the Lagrangian $\mathcal{L}=0$, the energy of the light ray $E$,
the angular momentum about the $z$ axis $L_{z}$ and the Carter constant $K$. The
equations of motion are fully separable and read
\begin{eqnarray}\label{eq:EoMt}
\frac{\mathrm{d}t}{\mathrm{d}\lambda}=&\frac{\rho(r)^2}{Q(r)}E-2n(\cos\vartheta+C)\frac{L_{z}+2n\left(\cos\vartheta+C\right)E}{\sin^2\vartheta},
\end{eqnarray}
\begin{eqnarray}\label{eq:EoMr}
\left(\frac{\mathrm{d}r}{\mathrm{d}\lambda}\right)^2=\rho(r)^2E^2-Q(r)K,
\end{eqnarray}
\begin{eqnarray}\label{eq:EoMtheta}
\left(\frac{\mathrm{d}\vartheta}{\mathrm{d}\lambda}\right)^2=K-\frac{\left(L_{z}+2n\left(\cos\vartheta+C\right)E\right)^2}{\sin^2\vartheta},
\end{eqnarray}
\begin{eqnarray}\label{eq:EoMphi}
\frac{\mathrm{d}\varphi}{\mathrm{d}\lambda}=\frac{L_{z}+2n\left(\cos\vartheta+C\right)E}{\sin^2\vartheta}.
\end{eqnarray}
The parameter $\lambda$ is the Mino parameter \cite{Mino2003}. It is defined up
to an affine transformation and is related to the affine parameter $s$ by
\begin{eqnarray}\label{eq:MPAP}
\frac{\mathrm{d}\lambda}{\mathrm{d}s}=\frac{1}{\rho(r)}.
\end{eqnarray}
Equations~(\ref{eq:EoMtheta}) and (\ref{eq:EoMphi}) are independent of $\Lambda$ and
$e$ and as a consequence the conclusion of Cl\'{e}ment \emph{et al.} \cite{Clement2015} that the $\varphi$ coordinate is continuous
for lightlike geodesics crossing the Misner strings is valid for all charged NUT--de Sitter metrics.
The charged NUT--de Sitter metric does not possess an ergoregion and thus we are free to choose the
sign of $E$; however, to maintain comparability to Frost and Perlick \cite{Frost2021a,Frost2021b}
we will choose $E>0$. This implies that for future-directed lightlike geodesics
the Mino parameter $\lambda$ is increasing and for past-directed lightlike geodesics
the Mino parameter is decreasing. In the following we will first briefly discuss the
equations of motion. We will derive the radius coordinate of the photon sphere and the angles
of the photon cones. We already have to note here that the latter will only be valid for individual light rays. We will discuss
the turning points and solve the equations of motion for arbitrary initial conditions $(x_{i}^{\mu})=(x^{\mu}(\lambda_{i}))=(t_{i},r_{i},\vartheta_{i},\varphi_{i})$
following the procedures described in Gralla and Lupsasca \cite{Gralla2020} and
Frost and Perlick \cite{Frost2021a,Frost2021b}. In Sec.~\ref{Sec:Lensing} we
will then use the obtained solutions to discuss gravitational lensing in the charged
NUT--de Sitter metrics. For this purpose we only need the solutions to the equations
of motion in the domain of outer communication. Therefore we will limit our discussion
to lightlike geodesics with $r_{\mathrm{H},\mathrm{o}}<r(<r_{\mathrm{C}+})$.

\subsection{The $r$ motion}\label{Sec:EoMr}
\subsubsection{Potential and photon sphere}\label{Sec:PotandPhotr}
\begin{figure*}\label{fig:Vrpot}
  \centering
		\includegraphics[width=\textwidth]{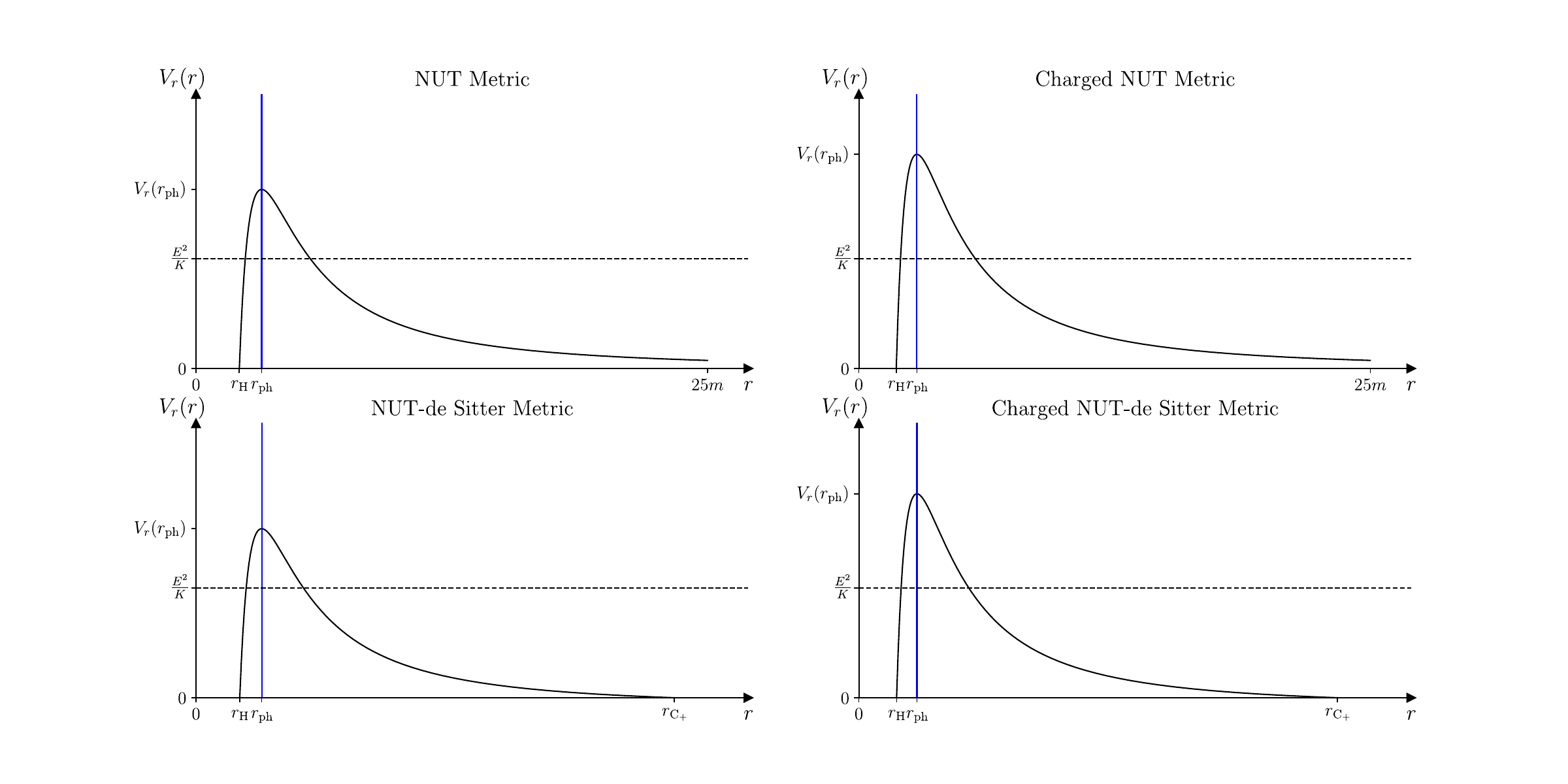}
	\caption{Potential $V_{r}(r)$ of the $r$ motion in the NUT metric (top left),
  the charged NUT metric (top right), the NUT--de Sitter metric (bottom left) and
  the charged NUT--de Sitter metric (bottom right) for $e=3m/4$, $\Lambda=1/(200m^2)$,
  and $n=m/2$. The axes have the same scale in all four plots. Note that due to
  spatial limitations we wrote $r_{\mathrm{H}}$ instead of $r_{\mathrm{H},\mathrm{o}}$.}
\end{figure*}
We begin with discussing the $r$ motion. Following \cite{Frost2021a} we
first rewrite Eq.~(\ref{eq:EoMr}) in terms of the potential $V_{r}(r)$:
\begin{equation}\label{eq:EoMrPot}
\frac{1}{\rho(r)^2K}\left(\frac{\mathrm{d}r}{\mathrm{d}\lambda}\right)^2+V_{r}(r)=\frac{E^2}{K},
\end{equation}
where
\begin{equation}
V_{r}(r)=\frac{Q(r)}{\rho(r)^2}.
\end{equation}
Figure~2 shows the potentials for the NUT metric (top left), the charged NUT metric
(top right), the NUT--de Sitter metric (bottom left) and the charged NUT--de Sitter
metric (bottom right) between the outer black hole horizon $r_{\mathrm{H},\mathrm{o}}$ and $r=25m$
($\Lambda=0$) or $r=r_{\mathrm{C+}}$ ($0<\Lambda<\Lambda_{\mathrm{C}}$). We see that in the NUT metric
(top left) the potential starts at $V_{r}(r_{\mathrm{H},\mathrm{o}})=0$, has a maximum
at $E^2/K=V_{r}(r_{\mathrm{ph}})$ and then it falls off to $V_{r}(r)=0$ for $r\rightarrow\infty$.
When we turn on the electric charge $e$ and the cosmological constant $\Lambda$
the basic structure of the potential remains the same and we only observe small
changes. When we turn on the electric charge $e$ the maximum of $V_{r}(r)$ increases
(top right). Turning on the cosmological constant on the other hand leads to a decrease
of the maximum of $V_{r}(r)$ (bottom). In addition, for $0<\Lambda<\Lambda_{\mathrm{C}}$ we have
$V_{r}(r_{\mathrm{C+}})=0$ at the cosmological horizon $r_{\mathrm{C+}}$. \\
At the maximum of $V_{r}(r)$ we have $\mathrm{d}r/\mathrm{d}\lambda=\mathrm{d}^2r/\mathrm{d}\lambda^2=0$.
When we now combine these two constraints we obtain the determining relation for
the radius coordinate of the photon sphere:
\begin{equation}\label{eq:rph}
r^3-\frac{3m}{1-\frac{4}{3}\Lambda n^2}r^2+\frac{2e^2-3n^2\left(1-\frac{4}{3}\Lambda n^2\right)}{1-\frac{4}{3}\Lambda n^2}r+\frac{mn^2}{1-\frac{4}{3}\Lambda n^2}=0.
\end{equation}
In our Universe we can safely assume that the cosmological constant $\Lambda$ and
the gravitomagnetic charge $n$ are very small. The consequence of this assumption is that the denominator of the
coefficients is always positive and we can read from the structure of Eq.~(\ref{eq:rph})
that one solution is always real and negative. In addition we can either have a
pair of complex conjugate roots or two real roots. In the following we agree to
choose $\Lambda$, $e$ and $n$ such that we always have two real positive roots.
We solve Eq.~(\ref{eq:rph}) using Cardano's method. We then label the three roots
such that $r_{\mathrm{ph}}>r_{\mathrm{ph}+}>r_{\mathrm{ph}-}$. The first root
$r_{\mathrm{ph}}$ lies in the domain of outer communication. In terms of the mass
parameter $m$ and the gravitomagnetic charge $n$ for the NUT metric it is explicitly
given in Jefremov and Perlick \cite{Jefremov2016} and for all NUT--de Sitter spacetimes it
is also contained as special case in the results of Grenzebach \emph{et al.} \cite{Grenzebach2014}.
Because $V_{r}(r_{\mathrm{ph}})$ has a maximum at $r_{\mathrm{ph}}$ this photon sphere
is unstable. An infinitesimal radial perturbation of these orbits has the consequence
that the light ray either falls into the black hole or escapes (across the cosmological
horizon) to infinity. The second photon sphere $r_{\mathrm{ph+}}$ also lies at positive
$r$ and corresponds to a minimum of $V_{r}(r)$. Thus it is stable. The third photon
sphere $r_{\mathrm{ph-}}$ lies in the region $r<0$ and is again unstable.
\begin{figure*}\label{fig:rph}
  \centering
		\includegraphics[width=0.4\textwidth]{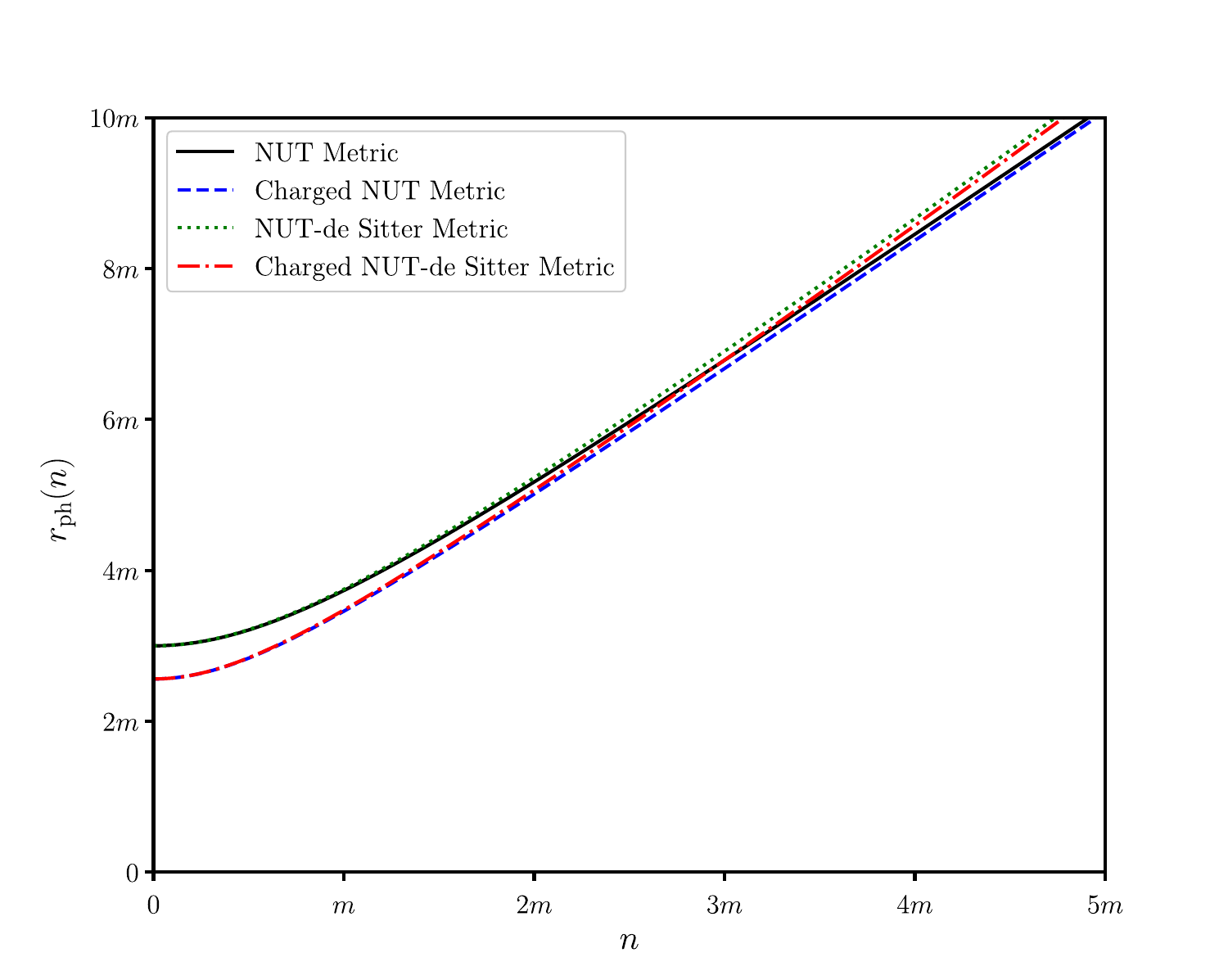}
	\caption{Radius coordinate of the photon sphere $r_{\mathrm{ph}}(n)$ as function of the gravitomagnetic charge $n$ for the NUT metric (black solid),
  the charged NUT metric (blue dashed), the NUT--de Sitter metric (green dotted) and the
  charged NUT--de Sitter metric (red dashed-dotted). The electric charge and the cosmological
  constant are $e=3m/4$ and $\Lambda=1/(200m^2)$, respectively.}
\end{figure*}
Figure~3 shows the radius coordinate of the photon sphere $r_{\mathrm{ph}}$ as function of
the gravitomagnetic charge $n$ for the NUT metric (black solid), the charged NUT
metric (blue dashed), the NUT--de Sitter metric (green dotted) and the charged NUT--de
Sitter metric (red dashed-dotted). For $e=0$ and $n=0$ (Schwarzschild--de Sitter limit)
the photon sphere is located at the radius coordinate $r_{\mathrm{ph},\mathrm{S}}=3m$.
For $e>0$ and $n=0$ (Reissner-Nordst\"{o}m--de Sitter limit) the photon sphere is located
at the radius coordinate
\begin{eqnarray}
r_{\mathrm{ph},\mathrm{RN}}=\frac{3m+\sqrt{9m^2-8e^2}}{2}.
\end{eqnarray}
When we now turn on the gravitomagnetic charge $n$ the photon sphere expands with
increasing $n$. While this observation applies to all four spacetimes there are
distinct differences when we turn on the electric charge $e$ and the cosmological
constant $\Lambda$. When we turn on the electric charge $e$ (but still keep $\Lambda=0$)
for $n=0$ the photon sphere is located at $r_{\mathrm{ph},\mathrm{RN}}<r_{\mathrm{ph},\mathrm{S}}$.
With increasing gravitomagnetic charge the photon sphere expands and approaches
the radius coordinate $r_{\mathrm{ph}}$ of the photon sphere in the NUT metric. When we
turn on the cosmological constant we observe something similar. For $n=0$ the photon
spheres are located at $r_{\mathrm{ph},\mathrm{RN}}<r_{\mathrm{ph},\mathrm{S}}$.
When we turn on the gravitomagnetic charge $n$ both photon spheres expand and the
radius coordinate $r_{\mathrm{ph}}$ of the photon sphere in the charged NUT--de Sitter
metric approaches the radius coordinate $r_{\mathrm{ph}}$ of the photon sphere $r_{\mathrm{ph}}$
in the NUT--de Sitter metric. However, compared to the NUT metric and the charged
NUT metric the photon spheres expand more rapidly with increasing gravitomagnetic
charge $n$.

\subsubsection{Types of motion}\label{sec:ToMr}
The potentials in Fig.~2 allow us to distinguish between the six following different
types of motion in the domain of outer communication:
\begin{enumerate}[label={(\arabic*)}]
    \item $E^2/K>V_{r}(r_{\mathrm{ph}-})$ and $K=0$: These geodesics do not have
          turning points in the domain of outer communication. We have one pair
          of complex conjugate purely imaginary double roots and label them such
          that $r_{1}=r_{3}=\bar{r}_{2}=\bar{r}_{4}=in$. These geodesics are the
          principal null geodesics of the charged NUT--de Sitter metrics.
    \item $E^2/K>V_{r}(r_{\mathrm{ph}-})$ and $K>0$: These geodesics do not have
          turning points in the domain of outer communication. We have two pairs
          of complex conjugate roots. We label them such that $r_{1}=\bar{r}_{2}=R_{1}+iR_{2}$
          and $r_{3}=\bar{r}_{4}=R_{3}+iR_{4}$. We always choose $R_{1}<R_{3}$ and
          $R_{2},R_{4}>0$.
    \item $E^2/K=V_{r}(r_{\mathrm{ph}-})$: These geodesics do not have turning points
          in the domain of outer communication. Two roots are real and equal.
          The other two roots are complex conjugate. We label the roots such
          that $r_{1}=r_{2}=r_{\mathrm{ph}-}$ and $r_{3}=\bar{r}_{4}=R_{3}+iR_{4}$.
          We always choose $R_{4}>0$.
    \item $V_{r}(r_{\mathrm{ph}-})>E^2/K>V_{r}(r_{\mathrm{ph}})$: These geodesics
          do not have turning points in the domain of outer communication. Two roots
          are real and two roots are complex conjugate. We label the roots such
          that $r_{1}>r_{2}$ and $r_{3}=\bar{r}_{4}=R_{3}+iR_{4}$. We always choose
          $R_{4}>0$.
    \item $E^2/K=V_{r}(r_{\mathrm{ph}})$: These geodesics do not have turning points
          in the domain of outer communication but four real roots. We label the
          roots such that $r_{1}=r_{2}=r_{\mathrm{ph}}>r_{3}>r_{4}$. These geodesics
          asymptotically come from or go to the photon sphere.
    \item $V_{r}(r_{\mathrm{ph}})>E^2/K$: These geodesics have turning points in
          the domain of outer communication. All four roots are real. We label the
          roots such that $r_{1}>r_{2}>r_{3}>r_{4}$. For $r_{\mathrm{H},\mathrm{o}}<r<r_{\mathrm{ph}}$
          these geodesics have a maximum at $r_{\mathrm{max}}=r_{2}$. For $r_{\mathrm{ph}}<r(<r_{\mathrm{C}+})$
          these geodesics have a minimum at $r_{\mathrm{min}}=r_{1}$.
\end{enumerate}
\subsubsection{Calculating $r(\lambda)$}\label{Sec:Solr}
Case 1: We have $E^2/K>V_{r}(r_{\mathrm{ph}-})$ and $K=0$. We will see in Sec.~\ref{Sec:Theta}
that these geodesics are the principal null geodesics of the charged NUT--de Sitter
metrics. In Eq.~(\ref{eq:EoMr}) we first set $K=0$ and get
\begin{eqnarray}\label{eq:EoMr1}
\left(\frac{\mathrm{d}r}{\mathrm{d}\lambda}\right)^2=\rho(r)^2E^2.
\end{eqnarray}
Equation~(\ref{eq:EoMr1}) can be solved in terms of elementary functions. We first separate variables and integrate.
Then we solve for $r$. With $i_{r_{i}}=\mathrm{sgn}(\left.\text{d}r/\text{d}\lambda\right|_{r=r_{i}})$
the solution reads
\begin{eqnarray}\label{eq:Solr1}
r(\lambda)=n\tan\left(\arctan\left(\frac{r_{i}}{n}\right)+i_{r_{i}}nE\left(\lambda-\lambda_{i}\right)\right).
\end{eqnarray}
Case 2: Lightlike geodesics with $E^2/K>V_{r}(r_{\mathrm{ph}-})$ and $K>0$ have
no turning points in the domain of outer communication. Here we first define two
new constants of motion \cite{Gralla2020,Byrd1954}:
\begin{eqnarray}\label{eq:S}
S=\sqrt{(R_{2}-R_{4})^2+(R_{1}-R_{3})^2},
\end{eqnarray}
and
\begin{eqnarray}\label{eq:Sbar}
\bar{S}=\sqrt{(R_{2}+R_{4})^2+(R_{1}-R_{3})^2},
\end{eqnarray}
and substitute
\begin{eqnarray}\label{eq:Subr1}
r=R_{1}-R_{2}\frac{g_{0}-\tan\chi}{1+g_{0}\tan\chi},
\end{eqnarray}
where
\begin{eqnarray}\label{eq:g0}
g_{0}=\sqrt{\frac{4R_{2}^2-(S-\bar{S})^2}{(S+\bar{S})^2-4R_{2}^2}}
\end{eqnarray}
to put Eq.~(\ref{eq:EoMr}) in the Legendre form Eq.~(\ref{eq:elldiff}). Now we
follow the steps described in Appendix~\ref{Sec:Ellfunc} and obtain the solution
$r(\lambda)$ in terms of Jacobi's elliptic $\mathrm{sc}$ function
\begin{eqnarray}\label{eq:Solr2}
r(\lambda)=R_{1}-R_{2}\frac{g_{0}-\mathrm{sc}\left(i_{r_{i}}\sqrt{E^2+\frac{\Lambda}{3}K}\frac{S+\bar{S}}{2}\left(\lambda-\lambda_{i}\right)+\lambda_{r_{i},k_{1}},k_{1}\right)}{1+g_{0}\mathrm{sc}\left(i_{r_{i}}\sqrt{E^2+\frac{\Lambda}{3}K}\frac{S+\bar{S}}{2}\left(\lambda-\lambda_{i}\right)+\lambda_{r_{i},k_{1}},k_{1}\right)},
\end{eqnarray}
where $\lambda_{r_{i},k_{1}}$, the initial condition $\chi_{i}$ and the square of
the elliptic modulus $k_{1}$ are given by
\begin{eqnarray}
\lambda_{r_{i},k_{1}}=F_{L}(\chi_{i},k_{1}),
\end{eqnarray}
\begin{eqnarray}\label{eq:chi1}
\chi_{i}=\mathrm{arctan}\left(\frac{r_{i}-R_{1}}{R_{2}}\right)+\mathrm{arctan}\left(g_{0}\right),
\end{eqnarray}
and
\begin{eqnarray}\label{eq:EM1}
k_{1}=\frac{4S\bar{S}}{(S+\bar{S})^2}.
\end{eqnarray}
Case 3: Lightlike geodesics with $E^2/K=V_{r}(r_{\mathrm{ph}-})$ have two equal
roots at $r_{1}=r_{2}=r_{\mathrm{ph}-}$. We first express the right-hand side of
Eq.~(\ref{eq:EoMr}) in terms of the roots. Then we separate variables and integrate
from $r(\lambda_{i})=r_{i}$ to $r(\lambda)$ and obtain
\begin{equation}\label{eq:Intr1}
\lambda-\lambda_{i}=\frac{i_{r_{i}}}{\sqrt{E^2+\frac{\Lambda}{3}K}}\int_{r_{i}}^{r(\lambda)}\frac{\mathrm{d}r'}{(r'-r_{1})\sqrt{(R_{3}-r')^2+R_{4}^2}}.
\end{equation}
The integral on the right-hand side of Eq.~(\ref{eq:Intr1}) has the form of Eq.~(\ref{eq:EMInt3}).
Now we follow the steps described in Appendix~\ref{Sec:EMInt1} and integrate.
Then we insert Eq.~(\ref{eq:EMInt3}), solve for $r$ and obtain
\begin{eqnarray}\label{eq:Solr3}
\hspace*{-0.2cm}r(\lambda)=r_{1}+\frac{(R_{3}-r_{1})^2+R_{4}^2}{R_{3}-r_{1}+R_{4}\mathrm{sinh}\left(\mathrm{arsinh}\left(\frac{r_{1}-R_{3}}{R_{4}}+\frac{(R_{3}-r_{1})^2+R_{4}^2}{R_{4}(r_{i}-r_{1})}\right)-i_{r_{i}}\sqrt{\left(E^2+\frac{\Lambda}{3}K\right)\left((R_{3}-r_{1})^2+R_{4}^2\right)}\left(\lambda-\lambda_{i}\right)\right)}.
\end{eqnarray}
Case 4: Lightlike geodesics with $V_{r}(r_{\mathrm{ph}-})>E^2/K>V_{r}(r_{\mathrm{ph}})$
have no turning points in the domain of outer communication. Two of the
roots are real. Using the two real roots $r_{1}$ and $r_{2}$ and the real and imaginary
parts $R_{3}$ and $R_{4}$ of the complex conjugate roots $r_{3}$ and $r_{4}$ we
first define two new constants of motion $R$ and $\bar{R}$:
\begin{eqnarray}\label{eq:NTC1}
R=\sqrt{(R_{3}-r_{1})^2+R_{4}^2},
\end{eqnarray}
\begin{eqnarray}\label{eq:NTC2}
\bar{R}=\sqrt{(R_{3}-r_{2})^2+R_{4}^2}.
\end{eqnarray}
Then we use the transformation \cite{Gralla2020,Hancock1917}
\begin{eqnarray}\label{eq:Sub2}
r=\frac{r_{1}\bar{R}-r_{2}R+(r_{1}\bar{R}+r_{2}R)\cos\chi}{\bar{R}-R+(\bar{R}+R)\cos\chi}
\end{eqnarray}
to put Eq.~(\ref{eq:EoMr}) into the Legendre form Eq.~(\ref{eq:elldiff}). Then we
follow the steps described in Appendix~\ref{Sec:Ellfunc} to obtain $r(\lambda)$
in terms of Jacobi's elliptic $\mathrm{cn}$ function:
\begin{eqnarray}\label{eq:Solr4}
r(\lambda)=\frac{r_{1}\bar{R}-r_{2}R+(r_{1}\bar{R}+r_{2}R)\mathrm{cn}\left(i_{r_{i}}\sqrt{\left(E^2+\frac{\Lambda}{3}K\right)R\bar{R}}\left(\lambda-\lambda_{i}\right)+\lambda_{r_{i},k_{2}},k_{2}\right)}{\bar{R}-R+(\bar{R}+R)\mathrm{cn}\left(i_{r_{i}}\sqrt{\left(E^2+\frac{\Lambda}{3}K\right)R\bar{R}}\left(\lambda-\lambda_{i}\right)+\lambda_{r_{i},k_{2}},k_{2}\right)},
\end{eqnarray}
where $\lambda_{r_{i},k_{2}}$, the initial condition $\chi_{i}$ and
the square of the elliptic modulus $k_{2}$ are given by
\begin{eqnarray}\label{eq:IC2}
\lambda_{r_{i},k_{2}}=F_{L}\left(\chi_{i},k_{2}\right),
\end{eqnarray}
\begin{eqnarray}\label{eq:Ichi1}
\chi_{i}=\arccos\left(\frac{(r_{i}-r_{2})R-(r_{i}-r_{1})\bar{R}}{(r_{i}-r_{2})R+(r_{i}-r_{1})\bar{R}}\right),
\end{eqnarray}
and
\begin{eqnarray}\label{eq:MOD2}
k_{2}=\frac{(R+\bar{R})^2-(r_{1}-r_{2})^2}{4R\bar{R}}.
\end{eqnarray}
Case 5: Lightlike geodesics with $E^2/K=V_{r}(r_{\mathrm{ph}})$ have two equal roots
at $r_{1}=r_{2}=r_{\mathrm{ph}}>r_{3}>r_{4}$. These are either lightlike geodesics
trapped on the photon sphere $r=r_{\mathrm{ph}}$ or lightlike geodesics asymptotically
coming from or going to the photon sphere. In the former case the solution to Eq.~(\ref{eq:EoMr})
is $r(\lambda)=r_{\mathrm{ph}}$. In the latter case we first rewrite the right-hand side
of Eq.~(\ref{eq:EoMr}) in terms of the roots
\begin{eqnarray}\label{eq:EoMr5}
\hspace*{-0.5cm}\left(\frac{\mathrm{d}r}{\mathrm{d}\lambda}\right)^2=\left(E^2+\frac{\Lambda}{3}K\right)(r-r_{\mathrm{ph}})^2(r-r_{3})(r-r_{4}).
\end{eqnarray}
Now we substitute
\begin{eqnarray}\label{eq:Sub3}
r=r_{3}+\frac{3a_{3,r}}{12y-a_{2,r}},
\end{eqnarray}
where
\begin{eqnarray}\label{eq:NTC3}
a_{2,r}=6\left(E^2+\frac{\Lambda}{3}K\right)r_{3}^2+2n^2E^2-(1-2\Lambda n^2)K,
\end{eqnarray}
\begin{eqnarray}\label{eq:NTC4}
a_{3,r}=4\left(E^2+\frac{\Lambda}{3}K\right)r_{3}^3+2\left(2n^2E^2-(1-2\Lambda n^2)K\right)r_{3}+2mK,
\end{eqnarray}
and obtain
\begin{eqnarray}\label{eq:EoMy}
\left(\frac{\mathrm{d}y}{\mathrm{d}\lambda}\right)^2=4(y-y_{\mathrm{ph}})^2(y-y_{1}).
\end{eqnarray}
$y_{\mathrm{ph}}$ and $y_{1}$ are related to the radius coordinate of the photon sphere  $r_{\mathrm{ph}}$
and the root $r_{4}$ by Eq.~(\ref{eq:Sub3}), respectively. It is easy to show that $y_{1}<y_{\mathrm{ph}}$
and $y_{1}<y$. Now we have to distinguish between lightlike geodesics between outer black
hole horizon $r_{\mathrm{H},\mathrm{o}}$ and photon sphere $r_{\mathrm{ph}}$ and
lightlike geodesics between photon sphere $r_{\mathrm{ph}}$ and infinity ($\Lambda=0$)
or cosmological horizon $r_{\mathrm{C}+}$ ($0<\Lambda<\Lambda_{\mathrm{C}}$). In the former case we have $y_{\mathrm{ph}}<y$
and in the latter case we have $y<y_{\mathrm{ph}}$. Now we separate variables and integrate
from $y(\lambda_{i})=y_{i}$ to $y(\lambda)$ and obtain
\begin{equation}\label{eq:Inty}
\lambda-\lambda_{i}=-\frac{i_{r_{i}}}{2}\int_{y_{i}}^{y(\lambda)}\frac{\mathrm{d}y'}{\sqrt{(y'-y_{\mathrm{ph}})^2(y'-y_{1})}}.
\end{equation}
For $r_{\mathrm{H},\mathrm{o}}<r<r_{\mathrm{ph}}$ we rewrite the right-hand side
of Eq.~(\ref{eq:Inty}) in terms of the integral $I_{6}$ given by Eq.~(\ref{eq:EMInty1})
in Appendix~\ref{Sec:EMInt2}. Now we follow the steps described in Appendix~\ref{Sec:EMInt2}
and obtain the right-hand side of Eq.~(\ref{eq:EMInty1}). After inserting $I_{6}$
in Eq.~(\ref{eq:Inty}) we solve for $r$ and obtain
\begin{eqnarray}\label{eq:Solr5}
r(\lambda)=r_{3}-\frac{(r_{\mathrm{ph}}-r_{3})(r_{3}-r_{4})}{r_{\mathrm{ph}}-r_{3}-(r_{\mathrm{ph}}-r_{4})\mathrm{coth}^2\left(\mathrm{arcoth}\left(\sqrt{\frac{(r_{i}-r_{4})(r_{\mathrm{ph}}-r_{3})}{(r_{i}-r_{3})(r_{\mathrm{ph}}-r_{4})}}\right)+i_{r_{i}}\sqrt{a_{r}}\left(\lambda-\lambda_{i}\right)\right)},
\end{eqnarray}
where
\begin{eqnarray}\label{eq:NTC5}
a_{r}=\frac{\left(2\left(E^2+\frac{\Lambda}{3}K\right)r_{3}^3+(2n^2E^2-(1-2\Lambda n^2)K)r_{3}+mK\right)(r_{\mathrm{ph}}-r_{4})}{2(r_{\mathrm{ph}}-r_{3})(r_{3}-r_{4})}.
\end{eqnarray}
Analogously for $r_{\mathrm{ph}}<r(<r_{\mathrm{C}+})$ we rewrite the right-hand side of Eq.~(\ref{eq:Inty})
in terms of the integral $I_{8}$ given by Eq.~(\ref{eq:EMInty3}) in Appendix~\ref{Sec:EMInt2}.
Again we integrate following the steps described in Appendix~\ref{Sec:EMInt2} and obtain the right-hand
side of Eq.~(\ref{eq:EMInty3}). After inserting $I_{8}$ in Eq.~(\ref{eq:Inty})
we solve for $r$ and obtain
\begin{eqnarray}\label{eq:Solr6}
r(\lambda)=r_{3}-\frac{(r_{\mathrm{ph}}-r_{3})(r_{3}-r_{4})}{r_{\mathrm{ph}}-r_{3}-(r_{\mathrm{ph}}-r_{4})\mathrm{tanh}^2\left(\mathrm{artanh}\left(\sqrt{\frac{(r_{i}-r_{4})(r_{\mathrm{ph}}-r_{3})}{(r_{i}-r_{3})(r_{\mathrm{ph}}-r_{4})}}\right)-i_{r_{i}}
\sqrt{a_{r}}\left(\lambda-\lambda_{i}\right)\right)}.
\end{eqnarray}
Case 6: Lightlike geodesics with $V_{r}(r_{\mathrm{ph}})>E^2/K$ have turning points
in the domain of outer communication. We have to distinguish between lightlike
geodesics between outer black hole horizon $r_{\mathrm{H},\mathrm{o}}$ and photon sphere
$r_{\mathrm{ph}}$ and lightlike geodesics between photon sphere $r_{\mathrm{ph}}$
and infinity ($\Lambda=0$) or the cosmological horizon $r_{\mathrm{C}+}$ ($0<\Lambda<\Lambda_{\mathrm{C}}$). We start with solving
Eq.~(\ref{eq:EoMr}) for lightlike geodesics in the domain $r_{\mathrm{H},\mathrm{o}}<r<r_{\mathrm{ph}}$.
Here we first substitute \cite{Gralla2020,Hancock1917}
\begin{eqnarray}\label{eq:Sub4}
r=r_{1}-\frac{(r_{1}-r_{2})(r_{1}-r_{3})}{r_{1}-r_{3}-(r_{2}-r_{3})\sin^2\chi}
\end{eqnarray}
to put Eq.~(\ref{eq:EoMr}) into the Legendre form Eq.~(\ref{eq:elldiff}).
Then we follow the steps described in Appendix~\ref{Sec:Ellfunc} to obtain $r(\lambda)$
in terms of Jacobi's elliptic $\mathrm{sn}$ function:
\begin{eqnarray}\label{eq:Solr7}
r(\lambda)=r_{1}-\frac{(r_{1}-r_{2})(r_{1}-r_{3})}{r_{1}-r_{3}-(r_{2}-r_{3})\mathrm{sn}^2\left(\frac{i_{r_{i}}}{2}\sqrt{\left(E^2+\frac{\Lambda}{3}K\right)(r_{1}-r_{3})(r_{2}-r_{4})}\left(\lambda_{i}-\lambda\right)+\lambda_{r_{i},k_{3}},k_{3}\right)},
\end{eqnarray}
where $\lambda_{r_{i},k_{3}}$, the initial condition $\chi_{i}$ and the square of
the elliptic modulus $k_{3}$ are given by
\begin{eqnarray}\label{eq:IC3}
\lambda_{r_{i},k_{3}}=F_{L}\left(\chi_{i},k_{3}\right),
\end{eqnarray}
\begin{eqnarray}\label{eq:Ichi3}
\chi_{i}=\arcsin\left(\sqrt{\frac{(r_{2}-r_{i})(r_{1}-r_{3})}{(r_{1}-r_{i})(r_{2}-r_{3})}}\right),
\end{eqnarray}
and
\begin{eqnarray}\label{eq:MOD3}
k_{3}=\frac{(r_{2}-r_{3})(r_{1}-r_{4})}{(r_{1}-r_{3})(r_{2}-r_{4})}.
\end{eqnarray}
Analogously for $r_{\mathrm{ph}}<r(<r_{\mathrm{C+}})$ we first substitute \cite{Gralla2020,Hancock1917}
\begin{eqnarray}\label{eq:Sub5}
r=r_{2}+\frac{(r_{1}-r_{2})(r_{2}-r_{4})}{r_{2}-r_{4}-(r_{1}-r_{4})\sin^2\chi}
\end{eqnarray}
to put Eq.~(\ref{eq:EoMr}) into the Legendre form Eq.~(\ref{eq:elldiff}). Then we
again follow the steps described in Appendix~\ref{Sec:Ellfunc} and obtain $r(\lambda)$
in terms of Jacobi's elliptic $\mathrm{sn}$ function:
\begin{eqnarray}\label{eq:Solr8}
r(\lambda)=r_{2}+\frac{(r_{1}-r_{2})(r_{2}-r_{4})}{r_{2}-r_{4}-(r_{1}-r_{4})\mathrm{sn}^2\left(\frac{i_{r_{i}}}{2}\sqrt{\left(E^2+\frac{\Lambda}{3}K\right)(r_{1}-r_{3})(r_{2}-r_{4})}\left(\lambda-\lambda_{i}\right)+\lambda_{r_{i},k_{3}},k_{3}\right)}.
\end{eqnarray}
Here $\lambda_{r_{i},k_{3}}$ and $k_{3}$ are given by Eq.~(\ref{eq:IC3}) and Eq.~(\ref{eq:MOD3}),
respectively, and the initial condition $\chi_{i}$ reads
\begin{eqnarray}\label{eq:Ichi4}
\chi_{i}=\arcsin\left(\sqrt{\frac{(r_{i}-r_{1})(r_{2}-r_{4})}{(r_{i}-r_{2})(r_{1}-r_{4})}}\right).
\end{eqnarray}

\subsection{The $\vartheta$ motion}\label{Sec:Theta}
For discussing the $\vartheta$ motion we first rewrite Eq.~(\ref{eq:EoMtheta}) in
terms of $x=\cos\vartheta$:
\begin{eqnarray}\label{eq:EoMx}
\left(\frac{\mathrm{d}x}{\mathrm{d}\lambda}\right)^2=(1-x^2)K-(L_{z}+2n(x+C)E)^2.
\end{eqnarray}
From the structure of Eq.~(\ref{eq:EoMx}) we can immediately read that for $K=0$
the right-hand side has to vanish. This simultaneously implies that we have $\mathrm{d}\varphi/\mathrm{d}\lambda=0$
and thus these are the principal null geodesics of the charged NUT--de Sitter metrics.
Similarly it is very easy to show that for very specific combinations of the constants
of motion the right-hand side of Eq.~(\ref{eq:EoMx}) vanishes. In both cases the
lightlike geodesics lie on cones of constant $\vartheta$ that have to fulfill the
constraints $\mathrm{d}x/\mathrm{d}\lambda=\mathrm{d}^2x/\mathrm{d}\lambda^2=0$.
From the second constraint we now immediately obtain the angle of the cones in terms
of the constants of motion. It reads
\begin{eqnarray}
\vartheta_{\mathrm{ph}}=\arccos\left(-\frac{2nE(2nEC+L_{z})}{4n^2E^2+K}\right).
\end{eqnarray}
Under the premise that we have $K\neq0$ we can use both constraints to rewrite the
Carter constant $K$ in terms of $E$ and $L_{z}$:
\begin{eqnarray}
K=(2nEC+L_{z})^2-4n^2E^2.
\end{eqnarray}
In analogy to the charged C--de Sitter metrics discussed in Frost and Perlick \cite{Frost2021a}
and Frost \cite{Frost2021b} we will call these cones \emph{individual photon cones}.
However, we have to emphasize that contrary to the charged C--de Sitter metrics in
which all geodesics tangential to the photon cone remain on the photon cone, in
the charged NUT--de Sitter metrics this is only the case for very specific lightlike geodesics.
In both cases, the principal null geodesics and the geodesics on the photon cones,
the solution to Eq.~(\ref{eq:EoMtheta}) is easy to obtain. It reads $\vartheta(\lambda)=\vartheta_{i}$.
All other geodesics oscillate between the two turning points,
\begin{eqnarray}\label{eq:xmin}
x_{\mathrm{min}}=\cos\vartheta_{\mathrm{min}}=\frac{\sqrt{K(K+4n^2E^2-(2nEC+L_{z})^2)}-2nE(2nEC+L_{z})}{K+4n^2E^2},
\end{eqnarray}
\begin{eqnarray}\label{eq:xmax}
x_{\mathrm{max}}=\cos\vartheta_{\mathrm{max}}=-\frac{\sqrt{K(K+4n^2E^2-(2nEC+L_{z})^2)}+2nE(2nEC+L_{z})}{K+4n^2E^2}.
\end{eqnarray}
As we can see $x_{\mathrm{min}}\neq -x_{\mathrm{max}}$ and thus the $\vartheta$
motion is not symmetric with respect to the plane $\vartheta=\pi/2$. For these geodesics
we can rewrite Eq.~(\ref{eq:EoMx}) in terms of an elementary integral that can be
easily calculated. After the integration we solve for $\vartheta$ and obtain as
solution to Eq.~(\ref{eq:EoMtheta})
\begin{eqnarray}\label{eq:Soltheta}
\hspace*{-0.7cm}\vartheta(\lambda)=\arccos\left(\frac{\sqrt{K(K+4n^2E^2-(2nEC+L_{z})^2)}}{K+4n^2E^2}\sin\left(a_{\vartheta}-i_{\vartheta{i}}\sqrt{K+4n^2E^2}(\lambda-\lambda_{i})\right)-\frac{2nE(2nEC+L_{z})}{K+4n^2E^2}\right),
\end{eqnarray}
where
\begin{eqnarray}\label{eq:Coefftheta}
a_{\vartheta}=\arcsin\left(\frac{(K+4n^2E^2)\cos\vartheta_{i}+2nE(2nEC+L_{z})}{\sqrt{K(K+4n^2E^2-(2nEC+L_{z})^2)}}\right),
\end{eqnarray}
and $i_{\vartheta_{i}}=\mathrm{sgn}\left(\left.\mathrm{d}\vartheta/\mathrm{d}\lambda\right|_{\vartheta=\vartheta_{i}}\right)$.
Structurally this solution is the same as Eq.~(32) in Kagramanova \emph{et al.} \cite{Kagramanova2010}
and it can be easily rewritten in the form of Eq.~(3.12) in Cl\'{e}ment \emph{et al.} \cite{Clement2015}.

\subsection{The $\varphi$ motion}\label{Sec:Phi}
For properly discussing the $\varphi$ motion in the charged NUT--de Sitter metrics
we have to consider several peculiarities. As stated in Zimmerman and Shahir \cite{Zimmerman1989} and in Halla and Perlick \cite{Halla2020},
all lightlike geodesics are contained in cones. These cones can point in arbitrary
directions and therefore lightlike geodesics can orbit any axis in space. This has
the consequence that not all geodesics perform a full $2\pi$ orbit about the $z$
axis. When the cones point away from the $z$ axis and the axis is not enclosed
by the cone the $\varphi$ motion reverses and the geodesic changes direction. In
addition it has long been an open question if the Misner strings are transparent
or opaque. When they are opaque all lightlike geodesics terminate at the Misner strings
and cannot be continued. In this case the Misner strings cast a shadow. However,
Cl\'{e}ment \emph{et al.} \cite{Clement2015} demonstrated that for lightlike
geodesics crossing the Misner strings the $\varphi$ motion is continuous. This strongly
advocates that it is transparent. Therefore, in this paper we will assume that the
Misner strings are transparent and do not cast a shadow. \\
When we want to integrate Eq.~(\ref{eq:EoMphi}) we have to distinguish the same three types of motion as in Sec.~\ref{Sec:Theta}
for $\vartheta$. We start with the principal null geodesics. Principal null geodesics
have $K=0$ and the right-hand side of Eq.~(\ref{eq:EoMphi}) vanishes. Therefore
the solution to Eq.~(\ref{eq:EoMphi}) simply reads $\varphi(\lambda)=\varphi_{i}$.
In the second case we have $K=(2nEC+L_{z})^2-4n^2E^2$. These are geodesics moving
on \emph{individual photon cones}. Here, the right-hand side of Eq.~(\ref{eq:EoMphi})
is constant and after a simple integration the solution reads
\begin{eqnarray}\label{eq:phiint}
\varphi(\lambda)=\varphi_{i}+\frac{\left(L_{z}+2n\left(\cos\vartheta_{\mathrm{ph}}+C\right)E\right)\left(\lambda-\lambda_{i}\right)}{\sin^2\vartheta_{\mathrm{ph}}}.
\end{eqnarray}
All other geodesics oscillate between the turning points $\vartheta_{\mathrm{min}}$
and $\vartheta_{\mathrm{max}}$ of the $\vartheta$ motion. Here, we proceed as follows. We
first replace $x=\cos\vartheta$ on the right-hand side of Eq.~(\ref{eq:EoMphi}):
\begin{eqnarray}\label{eq:EoMphix}
\frac{\mathrm{d}\varphi}{\mathrm{d}\lambda}=\frac{L_{z}+2n(x+C)E}{1-x^2}.
\end{eqnarray}
Now we perform a partial fraction decomposition,
\begin{eqnarray}\label{eq:PFDx}
\frac{1}{1-x^2}=\frac{1}{2}\left(\frac{1}{1-x}+\frac{1}{1+x}\right),
\end{eqnarray}
and rewrite Eq.~(\ref{eq:EoMphix}) as
\begin{eqnarray}\label{eq:phix}
\frac{\mathrm{d}\varphi}{\mathrm{d}\lambda}=\frac{L_{z}+2nE(1+C)}{2(1-x)}+\frac{L_{z}-2nE(1-C)}{2(1+x)}.
\end{eqnarray}
Now we resubstitute $x=\cos\vartheta$ and insert Eq.~(\ref{eq:Soltheta}) in Eq.~(\ref{eq:phix}). Then we integrate over $\lambda$.
The solution to Eq.~(\ref{eq:EoMphi}) now reads [see also Eq.~(43) in Kagramanova \emph{et al.} \cite{Kagramanova2010} and Eq.~(3.16) in Cl\'{e}ment \emph{et al.} \cite{Clement2015} for alternative formulations]
\begin{eqnarray}\label{eq:solphi}
\varphi(\lambda)=&\varphi_{i}+i_{\vartheta_{i}}\left(\arctan\left(c_{\vartheta,1}\left(\tan\left(\frac{\tilde{\lambda}(\lambda_{i})}{2}\right)-c_{\vartheta,2}\right)\right)-\arctan\left(c_{\vartheta,1}\left(\tan\left(\frac{\tilde{\lambda}(\lambda)}{2}\right)-c_{\vartheta,2}\right)\right)\right.\\ \nonumber
&\left.+\arctan\left(c_{\vartheta,3}\left(\tan\left(\frac{\tilde{\lambda}(\lambda)}{2}\right)+c_{\vartheta,4}\right)\right)-\arctan\left(c_{\vartheta,3}\left(\tan\left(\frac{\tilde{\lambda}(\lambda_{i})}{2}\right)+c_{\vartheta,4}\right)\right)\right),
\end{eqnarray}
where $c_{\vartheta,1}$, $c_{\vartheta,2}$, $c_{\vartheta,3}$, $c_{\vartheta,4}$ and
$\tilde{\lambda}(\lambda)$ are given by
\begin{eqnarray}\label{eq:coefftheta12}
c_{\vartheta,1}=\frac{K+4n^2E^2+2nE(2nEC+L_{z})}{\sqrt{K+4n^2E^2}(2nE(1+C)+L_{z})},~~~c_{\vartheta,2}=\frac{\sqrt{K(K+4n^2E^2-(2nEC+L_{z})^2)}}{K+4n^2E^2+2nE(2nEC+L_{z})},
\end{eqnarray}
\begin{eqnarray}\label{eq:coefftheta34}
c_{\vartheta,3}=\frac{K+4n^2E^2-2nE(2nEC+L_{z})}{\sqrt{K+4n^2E^2}(2nE(1-C)-L_{z})},~~~c_{\vartheta,4}=\frac{\sqrt{K(K+4n^2E^2-(2nEC+L_{z})^2)}}{K+4n^2E^2-2nE(2nEC+L_{z})},
\end{eqnarray}
\begin{eqnarray}\label{eq:Minotilde}
\tilde{\lambda}(\lambda)=\arcsin\left(\frac{(K+4n^2E^2)\cos\vartheta_{i}+2nE(2nEC+L_{z})}{\sqrt{K(K+4n^2E^2-(2nEC+L_{z})^2)}}\right)-i_{\vartheta_{i}}\sqrt{K+4n^2E^2}\left(\lambda-\lambda_{i}\right).
\end{eqnarray}
Note that for the explicit calculation of $\varphi(\lambda)$ the multivaluedness
of the arctan has to be appropriately considered.

\subsection{The time coordinate $t$}\label{Sec:EoMt}
Equation~(\ref{eq:EoMt}) has two terms that separately depend on $r$ and $\vartheta$.
In the following we will demonstrate how to calculate both components. For this
purpose we first integrate Eq.~(\ref{eq:EoMt}) over $\lambda$ and rewrite it as follows:
\begin{eqnarray}\label{eq:EoMtpart}
t(\lambda)=t_{i}+t_{r}(\lambda)+t_{\vartheta}(\lambda),
\end{eqnarray}
where the $r$-dependent integral reads
\begin{eqnarray}\label{eq:EoMtr}
t_{r}(\lambda)=\int_{\lambda_{i}}^{\lambda}\frac{\rho(r(\lambda'))^2 E\mathrm{d}\lambda'}{Q(r(\lambda'))},
\end{eqnarray}
and the $\vartheta$-dependent integral reads
\begin{eqnarray}\label{eq:EoMttheta}
t_{\vartheta}(\lambda)=-2n\int_{\lambda_{i}}^{\lambda}(\cos\vartheta(\lambda')+C)\frac{(L_{z}+2n(\cos\vartheta(\lambda')+C)E)\mathrm{d}\lambda'}{1-\cos^2\vartheta(\lambda')}.
\end{eqnarray}
\subsubsection{Calculating $t_{\vartheta}(\lambda)$}\label{Sec:ttheta}
We start with evaluating the integral on the right-hand side of $t_{\vartheta}(\lambda)$ in Eq.~(\ref{eq:EoMttheta}).
We have to distinguish the same three different types of motion as in Sec.~\ref{Sec:Theta}
for the $\vartheta$ motion. For $K=0$ the right-hand side of Eq.~(\ref{eq:EoMttheta})
vanishes and we have $t_{\vartheta}(\lambda)=0$. For lightlike geodesics on \emph{individual photon cones}
we have $K=(2nEC+L_{z})^2-4n^2E^2$ and thus the right-hand side of Eq.~(\ref{eq:EoMttheta})
is constant. We integrate over $\lambda$ and get
\begin{eqnarray}\label{eq:EoMtthetaturn}
t_{\vartheta}(\lambda)=-2n(\cos\vartheta_{\mathrm{ph}}+C)\frac{(L_{z}+2n(\cos\vartheta_{\mathrm{ph}}+C)E)(\lambda-\lambda_{i})}{\sin^2\vartheta_{\mathrm{ph}}}.
\end{eqnarray}
In all remaining cases the lightlike geodesics oscillate between the turning points
$\vartheta_{\mathrm{min}}$ and $\vartheta_{\mathrm{max}}$. Here we first substitute $x=\cos\vartheta$
and perform a partial fraction decomposition using Eq.~(\ref{eq:PFDx}). We restructure
and integrate the constant term. Now $t_{\vartheta}(\lambda)$ reads
\begin{eqnarray}
&t_{\vartheta}(\lambda)=4n^2E(\lambda-\lambda_{i})\\ \nonumber
&+n\left((1-C)(L_{z}-2nE(1-C))\int_{\lambda_{i}}^{\lambda}\frac{\mathrm{d}\lambda'}{1+x(\lambda')}-(1+C)(L_{z}+2nE(1+C))\int_{\lambda_{i}}^{\lambda}\frac{\mathrm{d}\lambda'}{1-x(\lambda')}\right).
\end{eqnarray}
Now we insert $x(\lambda)=\cos\vartheta(\lambda)$ and calculate the remaining two
integrals. After integration $t_{\vartheta}(\lambda)$ reads
\begin{eqnarray}\label{eq:solttheta}
  &t_{\vartheta}(\lambda)=4n^2E(\lambda-\lambda_{i})\\ \nonumber
  &+i_{\vartheta_{i}}2n\left((1-C)\left(\arctan\left(c_{\vartheta,3}\left(\tan\left(\frac{\tilde{\lambda}(\lambda)}{2}\right)+c_{\vartheta,4}\right)\right)-\arctan\left(c_{\vartheta,3}\left(\tan\left(\frac{\tilde{\lambda}(\lambda_{i})}{2}\right)+c_{\vartheta,4}\right)\right)\right)\right.\\ \nonumber
  &\left.+(1+C)\left(\arctan\left(c_{\vartheta,1}\left(\tan\left(\frac{\tilde{\lambda}(\lambda)}{2}\right)-c_{\vartheta,2}\right)\right)-\arctan\left(c_{\vartheta,1}\left(\tan\left(\frac{\tilde{\lambda}(\lambda_{i})}{2}\right)-c_{\vartheta,2}\right)\right)\right)\right),
\end{eqnarray}
where the coefficients $c_{\vartheta,1}$, $c_{\vartheta,2}$, $c_{\vartheta,3}$ and
$c_{\vartheta,4}$ are given by Eqs.~(\ref{eq:coefftheta12}) and (\ref{eq:coefftheta34})
and $\tilde{\lambda}(\lambda)$ is given by Eq.~(\ref{eq:Minotilde}). Note that for
the explicit calculation of $t_{\vartheta}(\lambda)$ the multivaluedness
of the arctan has to be appropriately considered. In addition we
note that structurally Eq.~(\ref{eq:solttheta}) is the same as Eq.~(45)
 in Kagramanova \emph{et al.} \cite{Kagramanova2010} and Eq.~(4.23) in
Cl\'{e}ment \emph{et al.} \cite{Clement2015}.

\subsubsection{Calculating $t_{r}(\lambda)$}\label{Sec:tr}
Now we turn to the $r$-dependent part of the time coordinate $t_{r}(\lambda)$. Here
we have to distinguish the same six types of motion as for the $r$ motion. We
start by separating variables in Eq.~(\ref{eq:EoMr}). Then we rewrite Eq.~(\ref{eq:EoMtr}) as integral over $r$. Now it reads
\begin{eqnarray}\label{eq:EoMtrcoord}
t_{r}(\lambda)=\int_{r_{i}...}^{...r(\lambda)}\frac{\rho(r')^2E\mathrm{d}r'}{Q(r')\sqrt{\rho(r')^2E^2-Q(r')K}}.
\end{eqnarray}
Here, the dots in the limits shall indicate that we have to split the integral at the
turning points and the sign of the root in the denominator has to be chosen according to the direction
of the $r$ motion. In addition for explicitly integrating Eq.~(\ref{eq:EoMtrcoord})
for each type of motion we have to distinguish four different cases. These are
$r_{\mathrm{H},\mathrm{i}}<r_{\mathrm{H},\mathrm{o}}$ for the NUT metric and the
charged NUT metric, $r_{\mathrm{H},\mathrm{i}}=r_{\mathrm{H},\mathrm{o}}=r_{\mathrm{H}}$
for the extremally charged NUT metric, $r_{\mathrm{C}-}<r_{\mathrm{H},\mathrm{i}}<r_{\mathrm{H},\mathrm{o}}<r_{\mathrm{C}+}$
for the NUT--de Sitter metric and the charged NUT--de Sitter metric and $r_{\mathrm{C}-}<r_{\mathrm{H},\mathrm{i}}=r_{\mathrm{H},\mathrm{o}}=r_{\mathrm{H}}<r_{\mathrm{C}+}$
for the extremally charged NUT--de Sitter metric. Due to the sheer number of integrals
we cannot explicitly demonstrate how to calculate each of them here.
We only provide the exact equations for the time coordinate in
all four cases for the principal null geodesics. For all other types of motion we
only briefly describe the steps of the integration procedure. We proceed in the same order as in Sec.~\ref{Sec:EoMr}.\\
Case 1: We start with the principal null geodesics with $E^2/K>V_{r}(r_{\mathrm{ph}-})$
and $K=0$. In this case Eq.~(\ref{eq:EoMtrcoord}) reduces to
\begin{eqnarray}\label{eq:EoMtrrad}
t_{r}(\lambda)=i_{r_{i}}\int_{r_{i}}^{r(\lambda)}\frac{\rho(r')\mathrm{d}r'}{Q(r')}.
\end{eqnarray}
Now we restructure $\rho(r')/Q(r')$ such that only terms with $r'$ in the nominator
or the denominator remain. Then we perform a partial fraction decomposition and integrate.
The resulting expressions for $t_{r}(\lambda)$ are given in terms of simple elementary
functions. In the case of the NUT metric and the charged NUT metric with
$r_{\mathrm{H},\mathrm{i}}<r_{\mathrm{H},\mathrm{o}}$ $t_{r}(\lambda)$ reads
\begin{eqnarray}\label{eq:CNUTSoltradial}
t_{r}(\lambda)=i_{r_{i}}\left(r(\lambda)-r_{i}+\frac{r_{\mathrm{H},\mathrm{o}}^2+n^2}{r_{\mathrm{H},\mathrm{o}}-r_{\mathrm{H},\mathrm{i}}}\ln\left(\frac{r(\lambda)-r_{\mathrm{H},\mathrm{o}}}{r_{i}-r_{\mathrm{H},\mathrm{o}}}\right)+\frac{r_{\mathrm{H},\mathrm{i}}^2+n^2}{r_{\mathrm{H},\mathrm{o}}-r_{\mathrm{H},\mathrm{i}}}\ln\left(\frac{r_{i}-r_{\mathrm{H},\mathrm{i}}}{r(\lambda)-r_{\mathrm{H},\mathrm{i}}}\right)\right),
\end{eqnarray}
while for the extremally charged NUT metric with $r_{\mathrm{H},\mathrm{i}}=r_{\mathrm{H},\mathrm{o}}=r_{\mathrm{H}}$ it reads
\begin{eqnarray}\label{eq:eCNUTSoltradial}
t_{r}(\lambda)=i_{r_{i}}\left(r(\lambda)-r_{i}+2r_{\mathrm{H}}\ln\left(\frac{r(\lambda)-r_{\mathrm{H}}}{r_{i}-r_{\mathrm{H}}}\right)+(r_{\mathrm{H}}^2+n^2)\left(\frac{1}{r_{i}-r_{\mathrm{H}}}-\frac{1}{r(\lambda)-r_{\mathrm{H}}}\right)\right).
\end{eqnarray}
Analogously in the case of the NUT--de Sitter and
the charged NUT--de Sitter metrics with $r_{\mathrm{C}-}<r_{\mathrm{H},\mathrm{i}}<r_{\mathrm{H},\mathrm{o}}<r_{\mathrm{C}+}$ we obtain for $t_{r}(\lambda)$
\begin{eqnarray}\label{eq:CdSNUTSoltradial}
&t_{r}(\lambda)=i_{r_{i}}\frac{3}{\Lambda}\left(
\frac{\left(r_{\mathrm{C}+}^2+n^2\right)\ln\left(\frac{r_{\mathrm{C}+}-r_{i}}{r_{\mathrm{C}+}-r(\lambda)}\right)}{(r_{\mathrm{C}+}-r_{\mathrm{H},\mathrm{o}})(r_{\mathrm{C}+}-r_{\mathrm{H},\mathrm{i}})(r_{\mathrm{C}+}-r_{\mathrm{C}-})}+\frac{\left(r_{\mathrm{H},\mathrm{o}}^2+n^2\right)\ln\left(\frac{r(\lambda)-r_{\mathrm{H},\mathrm{o}}}{r_{i}-r_{\mathrm{H},\mathrm{o}}}\right)}{(r_{\mathrm{C}+}-r_{\mathrm{H},\mathrm{o}})(r_{\mathrm{H},\mathrm{o}}-r_{\mathrm{H},\mathrm{i}})(r_{\mathrm{H},\mathrm{o}}-r_{\mathrm{C}-})}\right.\\ \nonumber
&\left.+\frac{\left(r_{\mathrm{H},\mathrm{i}}^2+n^2\right)\ln\left(\frac{r_{i}-r_{\mathrm{H},\mathrm{i}}}{r(\lambda)-r_{\mathrm{H},\mathrm{i}}}\right)}{(r_{\mathrm{C}+}-r_{\mathrm{H},\mathrm{i}})(r_{\mathrm{H},\mathrm{o}}-r_{\mathrm{H},\mathrm{i}})(r_{\mathrm{H},\mathrm{i}}-r_{\mathrm{C}-})}+\frac{\left(r_{\mathrm{C}-}^2+n^2\right)\ln\left(\frac{r(\lambda)-r_{\mathrm{C}-}}{r_{i}-r_{\mathrm{C}-}}\right)}{(r_{\mathrm{C}+}-r_{\mathrm{C}-})(r_{\mathrm{H},\mathrm{o}}-r_{\mathrm{C}-})(r_{\mathrm{H},\mathrm{i}}-r_{\mathrm{C}-})}\right).
\end{eqnarray}
Finally for the extremally charged NUT--de Sitter metric with $r_{\mathrm{C}-}<r_{\mathrm{H},\mathrm{i}}=r_{\mathrm{H},\mathrm{o}}=r_{\mathrm{H}}<r_{\mathrm{C}+}$ $t_{r}(\lambda)$ becomes
\begin{eqnarray}\label{eq:eCdSNUTSoltradial}
&t_{r}(\lambda)=i_{r_{i}}\frac{3}{\Lambda}\left(
\frac{\left(r_{\mathrm{C}+}^2+n^2\right)\ln\left(\frac{r_{\mathrm{C}+}-r_{i}}{r_{\mathrm{C}+}-r(\lambda)}\right)}{(r_{\mathrm{C}+}-r_{\mathrm{H}})^2(r_{\mathrm{C}+}-r_{\mathrm{C}-})}
+\frac{(r_{\mathrm{C}+}^2-r_{\mathrm{C}-}^2)(r_{\mathrm{H}}^2-n^2)+2r_{\mathrm{H}}(r_{\mathrm{C}+}(r_{\mathrm{C}-}^2+n^2)-r_{\mathrm{C}-}(r_{\mathrm{C}+}^2+n^2))}{(r_{\mathrm{C}+}-r_{\mathrm{C}-})(r_{\mathrm{C}+}-r_{\mathrm{H}})^2(r_{\mathrm{H}}-r_{\mathrm{C}-})^2}\ln\left(\frac{r(\lambda)-r_{\mathrm{H}}}{r_{i}-r_{\mathrm{H}}}\right)\right.\\ \nonumber
&\left.+\frac{r_{\mathrm{H}}^2+n^2}{(r_{\mathrm{C}+}-r_{\mathrm{H}})(r_{\mathrm{H}}-r_{\mathrm{C}-})}\left(\frac{1}{r_{i}-r_{\mathrm{H}}}-\frac{1}{r(\lambda)-r_{\mathrm{H}}}\right)+\frac{r_{\mathrm{C}-}^2+n^2}{(r_{\mathrm{C}+}-r_{\mathrm{C}-})(r_{\mathrm{H}}-r_{\mathrm{C}-})^2}\ln\left(\frac{r(\lambda)-r_{\mathrm{C}-}}{r_{i}-r_{\mathrm{C}-}}\right)\right).
\end{eqnarray}
Case 2: These are geodesics with $E^2/K>V_{r}(r_{\mathrm{ph-}})$ and $K>0$. They
do not have turning points in the domain of outer communication. We
perform a partial fraction decomposition of $\rho(r')^2/Q(r')$ and rewrite the
right-hand side of Eq.~(\ref{eq:EoMtr}) in terms of the elliptic integrals $t_{r,1}(r_{i},r)$
and $t_{r,2}(r_{i},r)$ given by Eqs.~(\ref{eq:T1}) and (\ref{eq:T2}) in Appendix~\ref{Sec:Ell1}.
Now we substitute using Eq.~(\ref{eq:Subr1}) to rewrite the integrals in terms
of Legendre's elliptic integral of the first kind and the two nonstandard elliptic integrals $G_{L}(\chi_{i},\chi,k_{1},n_{k})$
and $H_{L}(\chi_{i},\chi,k_{1},n_{k})$ given by Eqs.~(\ref{eq:GL}) and (\ref{eq:HL}).
We rewrite the latter in terms of elementary functions and Legendre's elliptic integrals
of the first, second and third kind using Eqs.~(\ref{eq:GLint})--(\ref{eq:GTL}).\\
Case 3: These are geodesics with $E^2/K=V_{r}(r_{\mathrm{ph-}})$. They have a double
root at $r_{\mathrm{ph-}}$ and do not have turning points in the domain of outer communication.
We first perform a partial fraction decomposition of $\rho(r')^2/Q(r')$. Then we perform
a second partial fraction decomposition and restructure the right-hand side of
Eq.~(\ref{eq:EoMtrcoord}) such that it only contains the elementary integrals $I_{1}-I_{5}$
given by Eqs.~(\ref{eq:EMInt1})--(\ref{eq:EMInt5}) in Appendix~\ref{Sec:EMInt1}.\\
Case 4: These are geodesics with $V_{r}(r_{\mathrm{ph}-})>E^2/K>V_{r}(r_{\mathrm{ph}})$.
These geodesics have two real roots but no turning points in the domain of outer
communication. Again we perform a partial fraction decomposition of $\rho(r')^2/Q(r')$.
Then we use Eq.~(\ref{eq:Sub2}) to rewrite the right-hand side of Eq.~(\ref{eq:EoMtrcoord})
in terms of Legendre's elliptic integral of the first kind and the two nonstandard elliptic
integrals $I_{L}(\chi_{i},\chi,k_{2},n_{k})$ and $J_{L}(\chi_{i},\chi,k_{2},n_{k})$. We rewrite $I_{L}(\chi_{i},\chi,k_{2},n_{k})$
and $J_{L}(\chi_{i},\chi,k_{2},n_{k})$ as Eqs.~(\ref{eq:ILint}), (\ref{eq:ILTint}), and (\ref{eq:JLTint}) as
described in Appendix~\ref{Sec:Ell2}.\\
Case 5: These are geodesics with $E^2/K=V_{r}(r_\mathrm{ph})$. They either asymptotically
come from or asymptotically go to the photon sphere at $r_{\mathrm{ph}}$. Here we
have to distinguish three cases. In the first case we have $r(\lambda)=r_{\mathrm{ph}}$.
These are lightlike geodesics trapped on the photon sphere. Here, the right-hand
side of Eq.~(\ref{eq:EoMtr}) is constant. After a simple integration with respect
to the Mino parameter $t_{r}(\lambda)$ now reads
\begin{eqnarray}\label{eq:Soltphot}
t_{r}(\lambda)=\frac{\rho(r_{\mathrm{ph}})^2E(\lambda-\lambda_{i})}{Q(r_{\mathrm{ph}})}.
\end{eqnarray}
The other two cases only differ with respect to one term and an overall sign.
In the first case we have $r_{\mathrm{H},\mathrm{o}}<r<r_{\mathrm{ph}}$ and in the second case we have $r_{\mathrm{ph}}<r(<r_{\mathrm{C}+})$.
Again we first perform a partial fraction decomposition of $\rho(r')^2/Q(r')$.
We substitute using Eq.~(\ref{eq:Sub3}) and subsequently perform a partial fraction
decomposition with respect to $y$. Now we sort all terms such
that only integrals given by $I_{6}$--$I_{9}$ [Eqs.~(\ref{eq:EMInty1})--(\ref{eq:EMInty4})
in Appendix~\ref{Sec:EMInt2}] remain. Here, the main difference between $r_{\mathrm{H},\mathrm{o}}<r<r_{\mathrm{ph}}$
and $r_{\mathrm{ph}}<r(<r_{\mathrm{C}+})$ is that the term
containing $1/(y-y_{\mathrm{ph}})$ is given by $I_{6}$ [Eq.~(\ref{eq:EMInty1})]
for the former and by $I_{8}$ [Eq.~(\ref{eq:EMInty3})] for the latter.\\
Case 6: These geodesics are characterized by $V_{r}(r_{\mathrm{ph}})>E^2/K$ and
can have a turning point. For lightlike geodesics with $r_{\mathrm{H},\mathrm{o}}<r<r_{\mathrm{ph}}$
this turning point is always a maximum at $r_{\mathrm{max}}=r_{2}$ and for $r_{\mathrm{ph}}<r(<r_{\mathrm{C}+})$
this turning point is always a minimum at $r_{\mathrm{min}}=r_{1}$. Again we perform a partial fraction decomposition
of $\rho(r')^2/Q(r')$. We substitute using Eq.~(\ref{eq:Sub4}) for $r_{\mathrm{H},\mathrm{o}}<r<r_{\mathrm{ph}}$
and Eq.~(\ref{eq:Sub5}) for $r_{\mathrm{ph}}<r(<r_{\mathrm{C}+})$. Now we sort all terms and rewrite them as
Legendre's elliptic integrals of the first and third kind as well as the nonstandard elliptic
integral $M_{L}(\chi_{i},\chi,k_{3},n_{k})$. For the latter we now evoke Eq.~(\ref{eq:MLint})
in Appendix~\ref{Sec:Ell3} to rewrite it in terms of elementary functions and Legendre's elliptic
integrals of the first, second and third kind.

\section{GRAVITATIONAL LENSING}\label{Sec:Lensing}
\subsection{Orthonormal tetrad and the angles on the observer's celestial sphere}\label{Sec:TandCS}
\begin{figure*}\label{fig:CSphere}
  \includegraphics{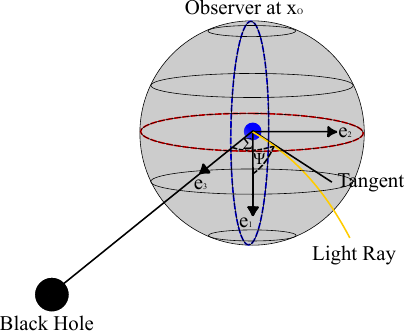}
\caption{Illustration of the lens-observer geometry and the orthonormal tetrad vectors
$e_{1}$, $e_{2}$ and $e_{3}$. The observer is located at $x_{O}=(x_{O}^{\mu})$. A light
ray is detected coming from the latitude $\Sigma$ and the longitude $\Psi$ on the observer's
celestial sphere.}
\end{figure*}
The ultimate goal of theoretical predictions is to be verified by observations. In
astronomy these observations are performed using telescopes on Earth's surface or
in orbits around Earth. For astronomical observations it is a common standard to
take the target of the observation as the center of the image and then divide the sky
using a coordinate grid whose angular coordinates are measured from the target.
Therefore, it will make our results much easier comparable to astronomical observations
when we adapt this approach to our theoretical predictions. \\
For achieving this goal we first introduce a stationary observer at coordinates $(x_{O}^{\mu})=(t_{O},r_{O},\vartheta_{O},\varphi_{O})$
in the domain of outer communication between photon sphere and infinity or cosmological
horizon for the (charged) NUT metric and the (charged) NUT--de Sitter metric, respectively.
Here, the symmetries of the spacetimes allow us to set $t_{O}=0$ and $\varphi_{O}=0$.
Now we choose the black hole as the target of our observation. In the next step we introduce
an orthonormal tetrad $e_{0}$, $e_{1}$, $e_{2}$, and $e_{3}$ as illustrated in Fig.~4
following the approach of Grenzebach \emph{et al.} \cite{Grenzebach2015a}:
\begin{eqnarray}\label{eq:e0}
e_{0}=\left.\sqrt{\frac{\rho(r)}{Q(r)}}\partial_{t}\right|_{(x_{O}^{\mu})},
\end{eqnarray}
\begin{eqnarray}\label{eq:e1}
e_{1}=\left.\frac{1}{\sqrt{\rho(r)}}\partial_{\vartheta}\right|_{(x_{O}^{\mu})},
\end{eqnarray}
\begin{eqnarray}\label{eq:e2}
e_{2}=\left.-\frac{\partial_{\varphi}-2n\left(\cos\vartheta+C\right)\partial_{t}}{\sqrt{\rho(r)}\sin\vartheta}\right|_{(x_{O}^{\mu})},
\end{eqnarray}
\begin{eqnarray}\label{eq:e3}
e_{3}=\left.-\sqrt{\frac{Q(r)}{\rho(r)}}\partial_{r}\right|_{(x_{O}^{\mu})},
\end{eqnarray}
where $e_{0}$ is the four-velocity vector of the observer. Now we introduce latitude
and longitude coordinates $\Sigma$ and $\Psi$ such that the latitude $\Sigma$ is
measured from $e_{3}$ and the longitude $\Psi$ is measured from $e_{1}$ in the direction
of $e_{2}$. In the next step we have to relate the three constants of motion $E$,
$L_{z}$ and $K$ to the angular coordinates on the observer's celestial sphere. For
this purpose let us consider the tangent vector of a light ray in Mino parametrization:
\begin{eqnarray}\label{eq:eta}
\frac{\mathrm{d}\eta}{\mathrm{d}\lambda}=\frac{\mathrm{d}t}{\mathrm{d}\lambda}\partial_{t}+\frac{\mathrm{d}r}{\mathrm{d}\lambda}\partial_{r}+\frac{\mathrm{d}\vartheta}{\mathrm{d}\lambda}\partial_{\vartheta}+\frac{\mathrm{d}\varphi}{\mathrm{d}\lambda}\partial_{\varphi}.
\end{eqnarray}
At the position of the observer we can also write the tangent vector of the light
ray in terms of the orthonormal tetrad and the angles $\Sigma$ and $\Psi$ on the
observer's celestial sphere as
\begin{eqnarray}\label{eq:etatetrad}
\frac{\mathrm{d}\eta}{\mathrm{d}\lambda}=\sigma\left(-e_{0}+\sin\Sigma\cos\Psi e_{1}+\sin\Sigma\sin\Psi e_{2}+\cos\Sigma e_{3}\right),
\end{eqnarray}
where $\sigma$ is a normalization constant. In Mino parametrization the normalization
constant $\sigma$ is given by
\begin{eqnarray}\label{eq:sigma}
\sigma=g\left(\frac{\mathrm{d}\eta}{\mathrm{d}\lambda},e_{0}\right).
\end{eqnarray}
The Mino parameter is defined up to an affine transformation and therefore we can
choose $\sigma=-\rho(r_{O})$ without loss of generality. We insert $\sigma$ and
Eqs.~(\ref{eq:e0})--(\ref{eq:e3}) in Eq.~(\ref{eq:etatetrad}) and compare coefficients with Eq.~(\ref{eq:eta})
evaluated at the position of the observer. Solving for $E$, $L_{z}$ and $K$ now leads to
the following relations between the constants of motion $E$, $L_{z}$ and $K$ and the angles $\Sigma$
and $\Psi$ on the observer's celestial sphere:
\begin{eqnarray}\label{eq:CoME}
E=\sqrt{\frac{Q(r_{O})}{\rho(r_{O})}},
\end{eqnarray}
\begin{eqnarray}\label{eq:CoMLz}
L_{z}=\sqrt{\rho(r_{O})}\sin\vartheta_{O}\sin\Sigma\sin\Psi-2n(\cos\vartheta_{O}+C)\sqrt{\frac{Q(r_{O})}{\rho(r_{O})}},
\end{eqnarray}
\begin{eqnarray}\label{eq:CoMK}
K=\rho(r_{O})\sin^2\Sigma.
\end{eqnarray}

\subsection{The shadow}
\begin{figure*}\label{fig:Shadow}
  \includegraphics{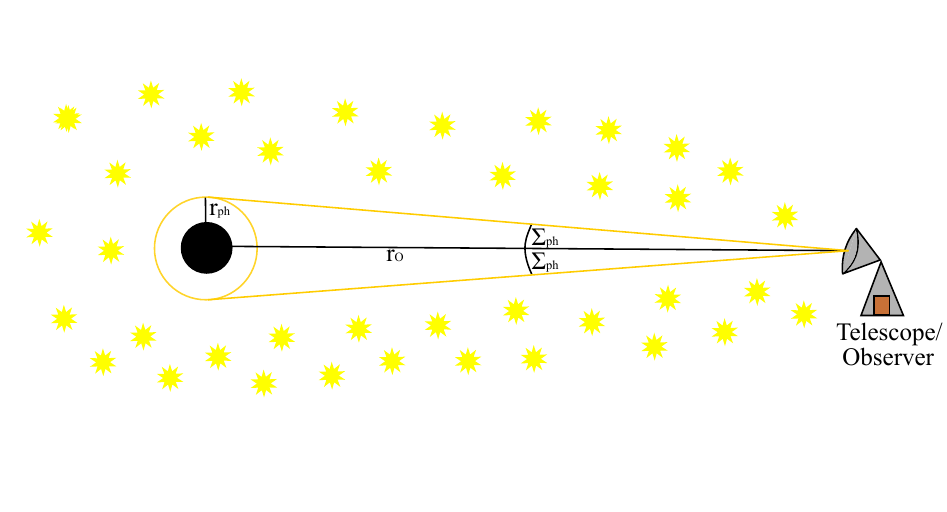}
\caption{Illustration of the construction of the shadow of a black hole. The black
circle marks the region behind the horizon of the black hole. The yellow circle marks
the photon sphere, the yellow stars are light sources and the yellow lines symbolize
lightlike geodesics asymptotically coming from the photon sphere.}
\end{figure*}
When we consider gravitational lensing in a black hole spacetime one of the most
easily accessible features is the shadow of the black hole. Although the shadow
is a very idealized concept it is very characteristic and therefore in this section
we calculate the angular radius of the shadow on the celestial sphere of an observer in the spacetime of a charged NUT--de Sitter black
hole. For this purpose let us consider the same observer as in Sec.~\ref{Sec:TandCS}
fixed at coordinates $(x_{O}^{\mu})$. As illustrated in Fig.~5 we distribute
light sources everywhere except between the black hole and the observer. The light sources
are now associated with brightness on the observer's celestial sphere while the
void is associated with darkness on the observer's celestial sphere. This dark area
is the shadow of the black hole. The boundary between brightness and darkness exactly
marks the direction of light rays asymptotically coming from the photon sphere.
These light rays have exactly the same constants of motion as light rays on the
photon sphere. In addition light rays asymptotically coming from the photon sphere
have $\left.\mathrm{d}r/\mathrm{d}\lambda\right|_{r=r_{\mathrm{ph}}}=0$. We now
use this fact and the relations Eqs.~(\ref{eq:CoME}) and (\ref{eq:CoMK}) between the constants of motion $E$ and $K$ and
the celestial latitude $\Sigma$ to evaluate Eq.~(\ref{eq:EoMr}) at $r=r_\mathrm{ph}$.
We solve for $\Sigma=\Sigma_{\mathrm{ph}}$ and obtain for the angular radius of
the shadow of a charged NUT--de Sitter black hole
\begin{eqnarray}\label{eq:ShadowCNdS}
\Sigma_{\mathrm{ph}}=\arcsin\left(\frac{\rho(r_{\mathrm{ph}})}{\rho(r_{O})}\sqrt{\frac{Q(r_{O})}{Q(r_{\mathrm{ph}})}}\right).
\end{eqnarray}
Note that this equation is structurally the same for all charged NUT--de Sitter metrics.
The obtained result is already contained as special case in the results of Grenzebach \emph{et al.} \cite{Grenzebach2014}; however, to our knowledge an explicit equation has not been
derived yet. For $n\rightarrow 0$ $\Sigma_{\mathrm{ph}}$ reduces to the angular radius
of the shadow of the Reissner-Nordstr\"{o}m--de Sitter family of spacetimes:
\begin{eqnarray}\label{eq:ShadowRNdS}
\Sigma_{\mathrm{ph},RNdS}=\arcsin\left(\frac{r_{\mathrm{ph}}}{r_{O}}\sqrt{\frac{\tilde{Q}(r_{O})}{\tilde{Q}(r_{\mathrm{ph}})}}\right),
\end{eqnarray}
where $\tilde{Q}(r)=Q(r)/r^2$. In particular it reduces to Synge's formula \cite{Synge1966}
for the Schwarzschild metric when $\Lambda \rightarrow0$, $e\rightarrow 0$ and $n\rightarrow 0$.
Although the charged NUT--de Sitter metric is only stationary and axisymmetric it is
not surprising that the shadow is circular because of the metric's $SO(3,\mathbb{R})$ symmetry.
Figure~6 shows plots of the angular radius of the shadow $\Sigma_{\mathrm{ph}}$
as function of the gravitomagnetic charge $n$ for the NUT metric (top left), the
charged NUT metric with $e=3m/4$ (top right), the NUT--de Sitter metric with $\Lambda=1/(200m^2)$
(bottom left) and the charged NUT--de Sitter metric with $\Lambda=1/(200m^2)$ and
$e=3m/4$ (bottom right) for $r_{O}=4m$ (black solid), $r_{O}=6m$ (blue dashed), $r_{O}=8m$
(green dotted) and $r_{O}=10m$ (red dashed-dotted). With increasing distance
of the observer from the black hole $\Sigma_{\mathrm{ph}}$ decreases. In addition with
increasing gravitomagnetic charge $n$ the photon sphere expands and the angular
radius of the shadow increases. For $r_{\mathrm{ph}}\rightarrow r_{O}$ we have $\Sigma_{\mathrm{ph}}\rightarrow\pi/2$
and the shadow covers half of the observer's sky. For $r_{\mathrm{H},\mathrm{o}}<r_{O}<r_{\mathrm{ph}}$ (not
shown) the complement of the shadow, usually also referred to as escape cone, shrinks while $r_{O}$ approaches the outer black
hole horizon $r_{\mathrm{H},\mathrm{o}}$. When we turn on the electric charge $e$
(top right) the angular radius of the shadow shrinks slightly because in the
presence of the electric charge $r_{\mathrm{ph}}$ is slightly smaller. As a consequence
$\Sigma_{\mathrm{ph}}$ approaches $\pi/2$ for larger $n$. Something similar happens
when we turn on the cosmological constant $\Lambda$. We can see in the bottom
panels that for $\Lambda>0$ the angular radius of the shadow $\Sigma_{\mathrm{ph}}$
also slightly decreases. With increasing $n$ the photon sphere expands
and $r_{\mathrm{ph}}$ approaches $r_{O}$ slightly faster than for the NUT metric and the charge NUT metric.
As a consequence $\Sigma_{\mathrm{ph}}$ approaches $\pi/2$ for slightly smaller gravitomagnetic charges $n$. However,
unlike for the Schwarzschild--de Sitter and the Reissner-Nordstr\"{o}m--de Sitter
metrics, for which the radius coordinate $r_{\mathrm{ph}}$ of the photon sphere is independent of the cosmological
constant, we cannot only attribute these effects to Eq.~(\ref{eq:ShadowCNdS}) but also have
to consider the effect of the cosmological constant on the radius coordinate of the photon
sphere $r_{\mathrm{ph}}$ itself [as determined from Eq.~(\ref{eq:rph})]. As discussed
in Sec.~\ref{Sec:PotandPhotr} when we turn on the cosmological constant $\Lambda$ the
photon sphere expands much faster for increasing $n$ compared to $\Lambda=0$ and
this effect gets stronger the larger the gravitomagnetic charge $n$. This leads to the observed fact that
$\Sigma_{\mathrm{ph}}$ approaches $\pi/2$ already for smaller $n$.\\
\begin{figure*}\label{fig:Sigmaph}
\begin{tabular}{cc}
  NUT Metric & Charged NUT Metric\\[6pt]
  \includegraphics[width=90mm]{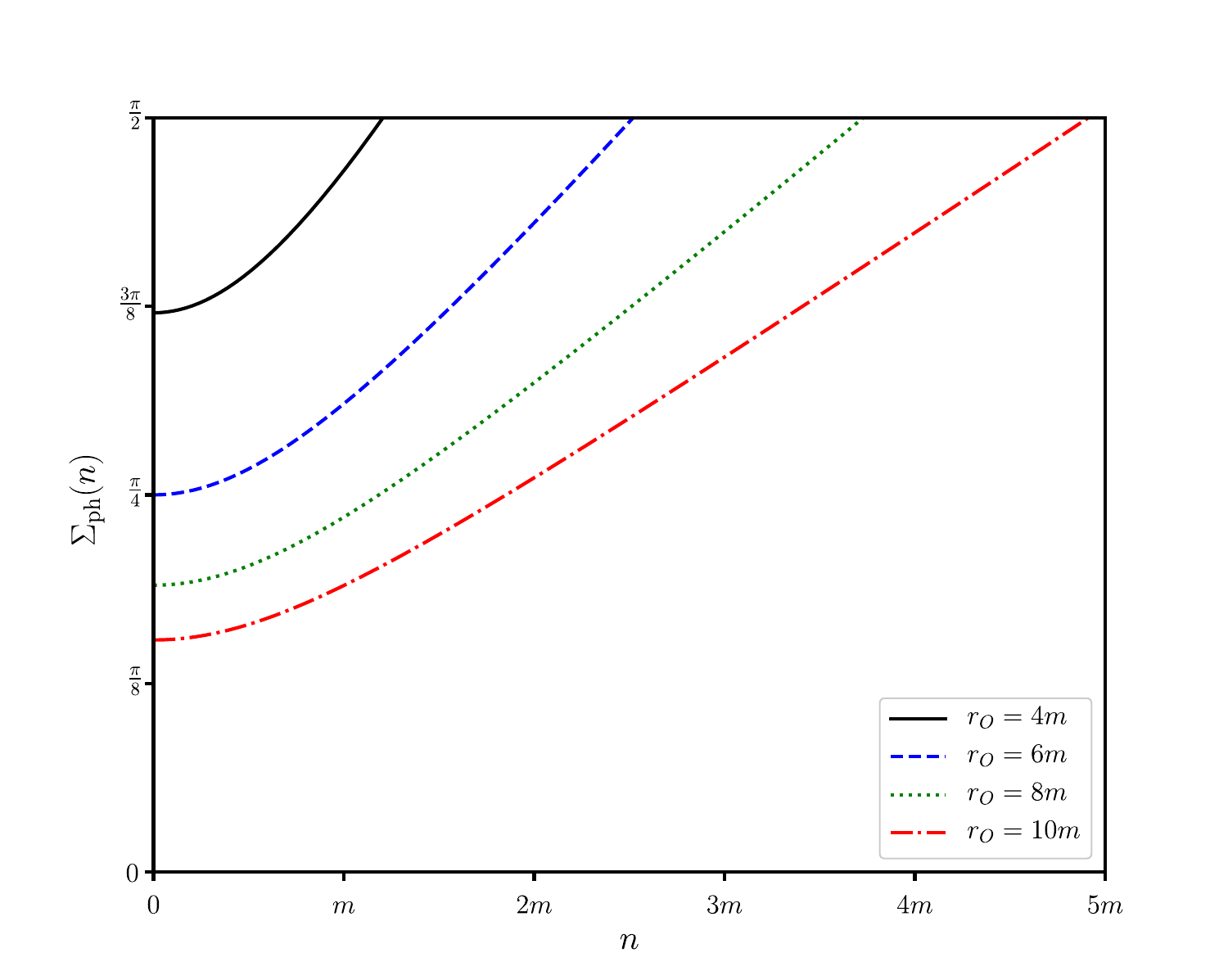} &   \includegraphics[width=90mm]{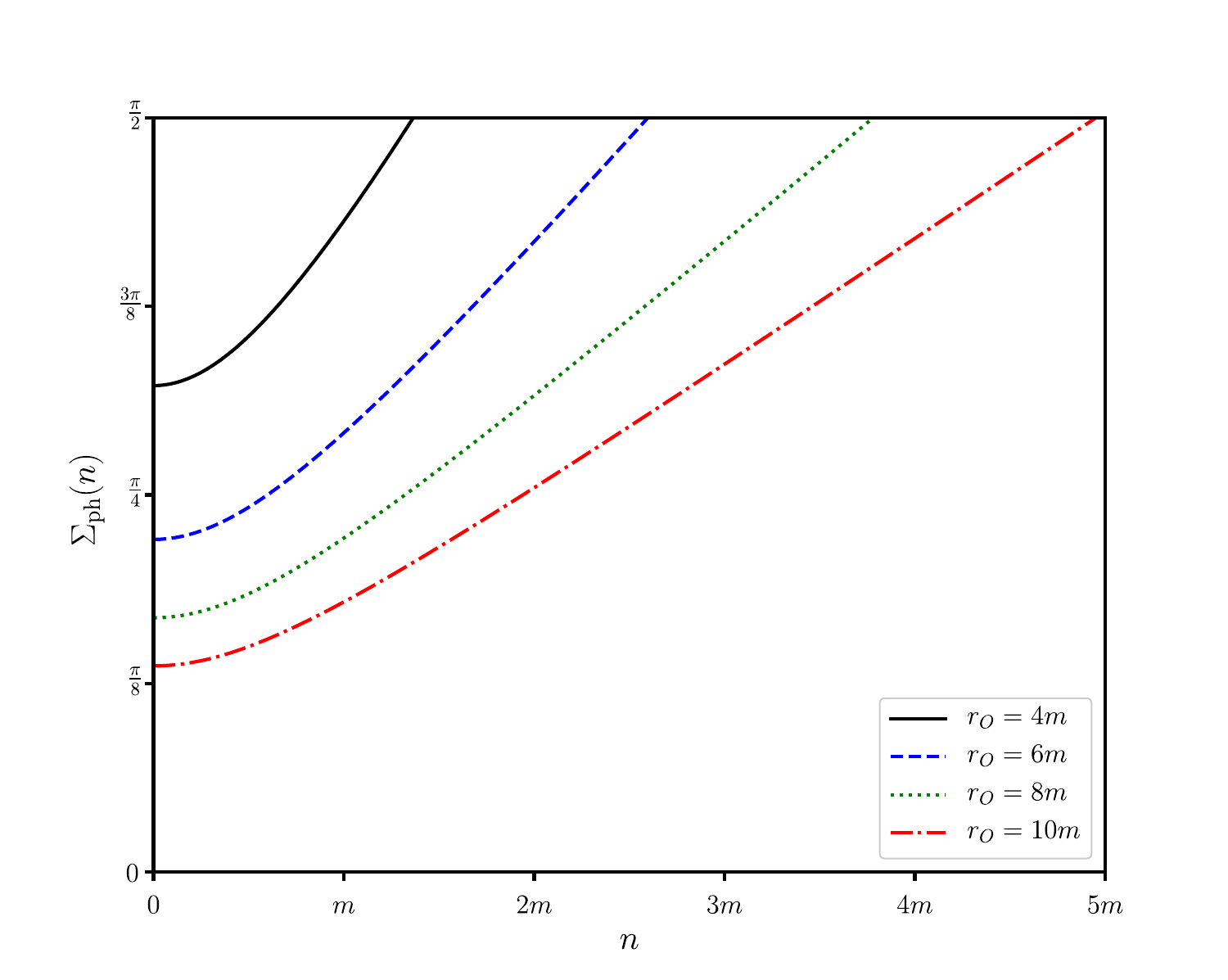} \\
  NUT-de Sitter Metric& Charged NUT-de Sitter Metric\\[6pt]
  \includegraphics[width=90mm]{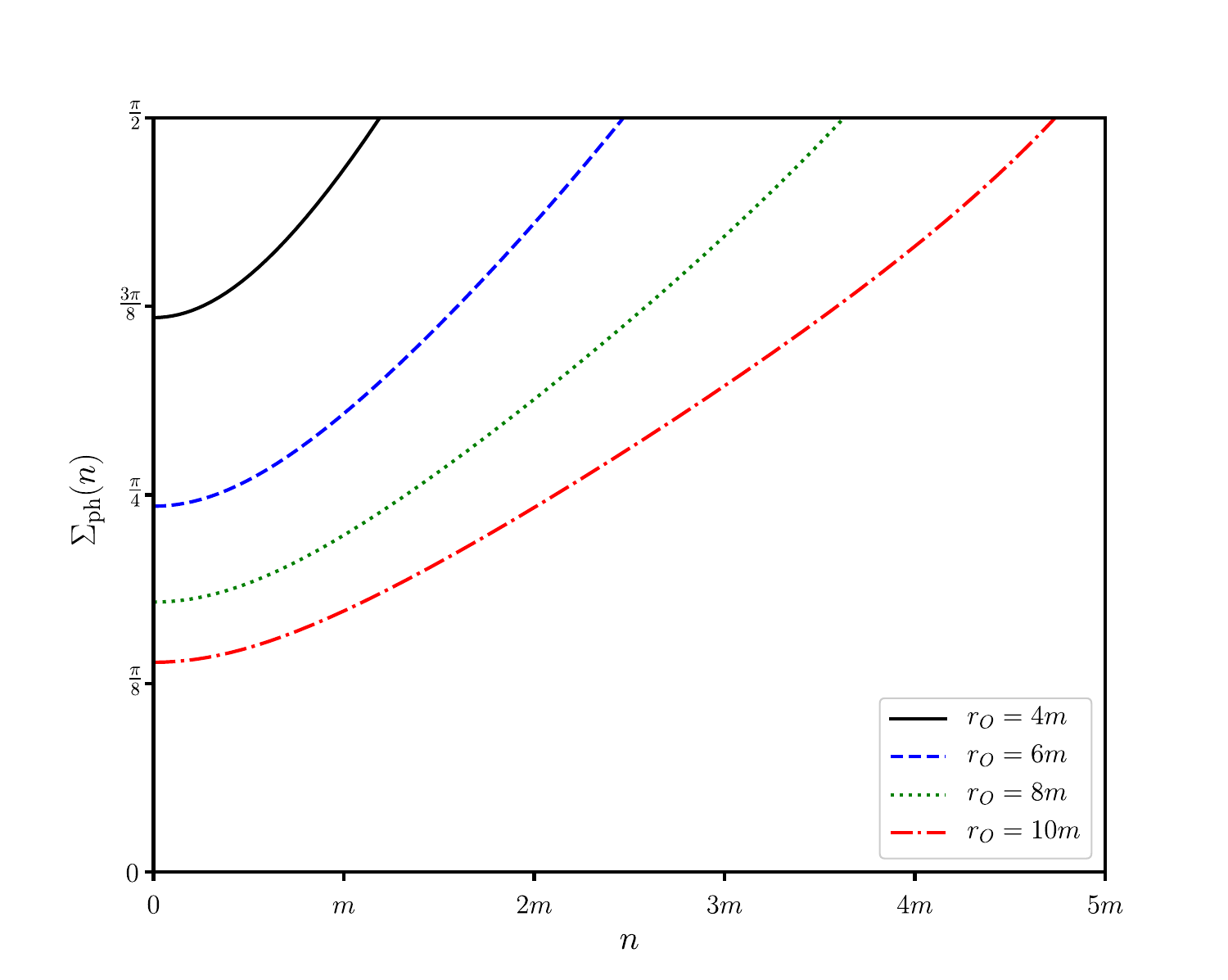} &   \includegraphics[width=90mm]{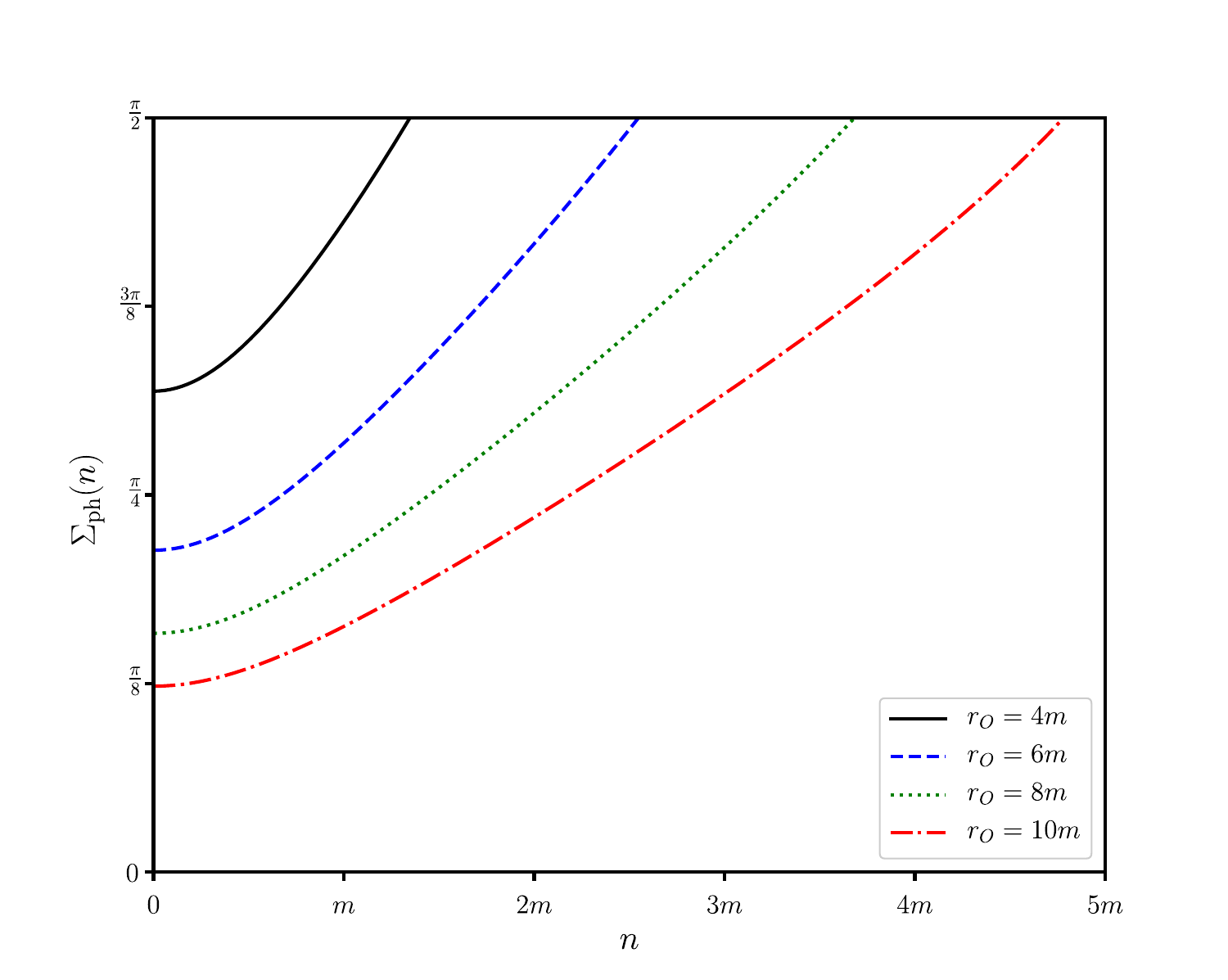} \\
\end{tabular}
\caption{Angular radius of the shadow $\Sigma_{\mathrm{ph}}$ for observers at radii
$r_{O}=4m$ (black), $r_{O}=6m$ (blue dashed), $r_{O}=8m$ (green dotted) and $r_{O}=10m$
(red dashed-dotted) for the NUT metric (top left), the charged NUT metric with $e=3m/4$
(top right), the NUT--de Sitter metric with $\Lambda=1/(200m^2)$ (bottom left), and
the charged NUT--de Sitter metric with $\Lambda=1/(200m^2)$, $e=3m/4$ (bottom right).}
\end{figure*}
In this paper we only considered stationary observers. So the immediate question arises
how the shadow would look like for an observer moving at a constant velocity $v$.
Here, we can draw insight by having a look at the reasoning of Penrose \cite{Penrose1959}
on the appearance of a moving sphere to a resting observer. For a resting observer
a moving sphere always appears to be circular independent of how it moves relative
to the observer. Now we can always find a coordinate system in which the observer
is moving relative to the resting sphere. In the new coordinate system the sphere
is at rest while the observer moves. Therefore, both a resting observer and a moving
observer see a sphere as a circle on their skies. We can now immediately transfer this
reasoning to the shadow. The photon sphere takes the role of the sphere and the
shadow is seen by a distant observer as a circle of darkness. The angular radius of
the shadow on the celestial sphere of the moving observer can then be calculated
from the angular radius of the shadow on the celestial sphere of the resting observer
by applying the aberration formula. Because the aberration formula maps circles
on circles the shadow is circular for both the resting and the moving observers.\\
How can we now use our insights to measure the gravitomagnetic charge from observations
of the shadow? As discussed above the gravitomagnetic charge $n$ affects the size
of the shadow. Even if the gravitomagnetic charge is only very small it will lead
to a larger angular radius $\Sigma_{\mathrm{ph}}$ of the shadow compared to a spacetime
without gravitomagnetic charge. However, as we can read from Eqs.~(\ref{eq:ShadowCNdS})
and (\ref{eq:ShadowRNdS}) as long as we have vanishing spin the shadow is always
circular independent of the presence of the gravitomagnetic charge. To make things
worse also observers around black holes potentially described by the charged C--de Sitter
metrics, which describe charged accelerating black holes with a cosmological constant,
see a circular shadow \cite{Frost2021a,Frost2021b}. While it is true that the angular
radius of the shadow in the C--de Sitter metrics decreases with increasing acceleration
parameter as the observer approaches the acceleration horizon, in reality we can
expect both, the acceleration parameter and the gravitomagnetic charge and also
their effects on the shadow to be very small. Therefore even if we only consider the Pleba\'{n}ski-Demia\'{n}ski
class we have a degeneracy between 12 spacetimes that can potentially describe
black holes with circular shadows in nature. Because we do not \emph{a priori} know the
distance between Earth and an observed astrophysical black hole we cannot lift this degeneracy using
observations of the shadow alone. Mars \emph{et al.} \cite{Mars2018} showed
that for Kerr-Newman black holes and observers that are not located on the axis of symmetry
this degeneracy is lifted. They also concluded that for Kerr-Newman-NUT
black holes the only parameter that cannot be determined from observations of the
shadow alone is the gravitomagnetic charge $n$. However, in this paper we do not
consider the spin and therefore we need additional observables that help us to distinguish
between the shadows in different spacetimes and to potentially measure the gravitomagnetic charge $n$.

\subsection{The lens equation}
We now move on to define the lens map. The most general version of a general relativistic
lens map was first introduced by Frittelli and Newman \cite{Frittelli1999} and later adapted
to spherically symmetric spacetimes by Perlick \cite{Perlick2004a}. Only recently
the approach of Perlick was adapted to axisymmetric spacetimes in Frost and Perlick
\cite{Frost2021a} and Frost \cite{Frost2021b}. We now apply their approach to the
charged NUT--de Sitter metrics. For this purpose we proceed as follows.\\
We first distribute light sources on a two-sphere $S_{L}^2$ at the radius coordinate $r_{L}$.
We place the stationary observer with coordinates $(x_{O}^{\mu})$ at a radius coordinate
$r_{\mathrm{ph}}<r_{O}<r_{L}(<r_{\mathrm{C}+})$ and construct its past light cone. We follow all lightlike
geodesics on this cone back into the past. Some of these geodesics will intersect
with the two-sphere $S_{L}^{2}$ while others will intersect with the outer black hole horizon
$r_{\mathrm{H},\mathrm{o}}$ and end up in the black hole. The geodesics that intersect
with the two-sphere $S_{L}^{2}$ now constitute a map from the celestial coordinates
$\Sigma$ and $\Psi$ on the celestial sphere of the observer to the angular coordinates
$\vartheta_{L}(\Sigma,\Psi)$ and $\varphi_{L}(\Sigma,\Psi)$ on the two-sphere of light
sources $S_{L}^2$:
\begin{eqnarray}\label{eq:LensE}
(\Sigma,\Psi)\rightarrow (\vartheta_{L}(\Sigma,\Psi),\varphi_{L}(\Sigma,\Psi)).
\end{eqnarray}
This is our lens equation. For the calculation of the lens map we now employ the
solutions for $\vartheta(\lambda)$ and $\varphi(\lambda)$ calculated in Secs.~\ref{Sec:Theta}
and \ref{Sec:Phi}. We express the constants of motion in Eqs.~(\ref{eq:Soltheta}), (\ref{eq:Coefftheta}),
and (\ref{eq:solphi})--(\ref{eq:Minotilde}) by Eqs.~(\ref{eq:CoME})--(\ref{eq:CoMK}). Now we choose
$\lambda_{O}=0$ and thus the only thing left to do is to eliminate the unknown
$\lambda_{L}<\lambda_{O}$. We can calculate it from the radius coordinate of the observer
$r_{O}$ and the radius coordinate $r_{L}$ at which the light ray intersects with
the two-sphere $S_{L}^{2}$. For this purpose we separate variables in Eq.~(\ref{eq:EoMr})
and integrate. Now we have to distinguish two different types of lightlike geodesics.
The first type of lightlike geodesics has a turning point at the radius coordinate
$r_{\mathrm{min}}=r_{1}$. In this case $\lambda_{L}$ becomes
\begin{eqnarray}\label{eq:Minomin}
\lambda_{L}=\int_{r_{O}}^{r_{\mathrm{min}}}-\int_{r_{\mathrm{min}}}^{r_{L}}\frac{\sqrt{\rho(r_{O})}\mathrm{d}r'}{\sqrt{\rho(r')^2Q(r_{O})-Q(r')\rho(r_{O})^2\sin^2\Sigma}}.
\end{eqnarray}
The second type of lightlike geodesics does not have a turning point and is propagating
in the radial direction outward. In this case $\lambda_{L}$ reads
\begin{eqnarray}\label{eq:MinoNT}
\hspace*{-0.5cm}\lambda_{L}=-\int_{r_{O}}^{r_{L}}\frac{\sqrt{\rho(r_{O})}\mathrm{d}r'}{\sqrt{\rho(r')^2Q(r_{O})-Q(r')\rho(r_{O})^2\sin^2\Sigma}}.
\end{eqnarray}
For the calculation of the lens map we now rewrite $\lambda_{L}$ in terms of Legendre's elliptic
integral of the first kind or when possible in terms of elementary functions. We calculate
$\vartheta_{L}(\Sigma,\Psi)$ and $\varphi_{L}(\Sigma,\Psi)$ as described in Secs.~\ref{Sec:Theta}
and \ref{Sec:Phi}. For a fast and efficient calculation of the
lens equation and the travel time in Sec.~\ref{Sec:TravTime} their evaluation was implemented
in the programming language JULIA \cite{Bezanson2017}. For the visual representation
we follow the color conventions of Bohn \emph{et al.} \cite{Bohn2015} illustrated in Fig.~7
with a small modification which will be described below.\\
Figure~8 shows the lens map for an observer located at $r_{O}=8m$ and $\vartheta_{O}=\pi/2$
and a sphere of light sources $S_{L}^2$ at the radius coordinate $r_{L}=9m$ for the Schwarzschild
metric (top left), the NUT metric with $n=m/100$ (top right), $n=m/10$ (bottom left)
and $n=m/2$ (bottom right). The Misner string is located at $\vartheta=0$ ($C=1$). The observer looks in the direction of the
black hole. The black circle in the center is the shadow of the black hole. In the
Schwarzschild metric the lens map is rotationally symmetric. The rings around the
center represent images of different orders. Here, we say that an image is of order
$n_{\mathrm{im}}$ when the absolute value of the covered angle $\Delta\varphi_{L}$
fulfills the relation $(n_{\mathrm{im}}-1)\pi<\left|\Delta\varphi_{L}\right|<n_{\mathrm{im}}\pi$.
The outer, strongly colored ring represents images of first order, while the second,
fainter colored ring represents images of second order. Closer to the shadow we
can also see images of third and, when we zoom in, images of fourth order. The borders
between the images of different orders are the critical curves. Patches with the
same color represent images from light sources on the same quadrant on the sphere
of light sources. In our representation we slightly deviate from the representation
of Bohn \emph{et al.} \cite{Bohn2015} as we represent images of odd order by stronger colors
than images of even order. When we now turn on the gravitomagnetic charge $n$ the
patches on the observer's sky start to become twisted and the formerly separated
areas in the rings with images of first and second order connect. This effect becomes
stronger the larger the gravitomagnetic charge $n$. The pattern of the lens map
is symmetric under rotations by $\pi$. The images of first and second order from the
same quadrant on the two-sphere $S_{L}^2$ are separated by sharp lines. The geodesics
exactly on these lines cross the axes at least once (here we have to note that these
geodesics can only cross one axis, either $\vartheta=0$ or $\vartheta=\pi$). In the lower two panels we also observe odd order images
close to the shadow at $\Psi=0$ (red) and $\Psi=\pi$ (blue). A closer investigation
reveals that formally these are images of first order. The associated lightlike
geodesics move on cones not enclosing the axes and thus along these geodesics the
direction of the $\varphi$ motion reverses. Considering the observed lensing pattern it is now an
interesting question how the critical curves of the NUT metric look. In Fig.~9
we show an enlarged view of the lens map between $\Psi=\pi$ and $\Psi=9\pi/8$ for the NUT metric with $n=m/2$ with 16
times higher $\Psi$ resolution than for Fig.~8. The sharp boundaries still remain and therefore we can exclude
with high certainty that they are artifacts of too-scarce point sampling. However,
although these lines separate images of first and second order it is rather unlikely
that they are part of the critical curves for three reasons. First of all, although
not clearly visible in the top right panel of Fig.~8 they form as soon as we turn
on the gravitomagnetic charge $n$. As discussed above for the Schwarzschild metric the
critical curves are circles and \emph{a priori} there seems to be no reason why
this should suddenly change. Second, the NUT metric maintains an $SO(3,\mathbb{R})$
symmetry which also strongly suggests that the critical curves are likely to be
circles. Third, if we have a closer look images of first and second order and images of third and fourth order seem to
be clearly separated by circles indicating that this boundary is a critical curve.
Settling this question would require a more detailed analysis of the geodesic
motion in the NUT metric or exactly deriving the determining relation for the critical
curves. Both are beyond the scope of this paper and will be part of future work. \\
Figure~10 shows the lens maps of the Reissner-Nordstr\"{o}m metric (top left), the
charged NUT metric (top right), the Schwarzschild--de Sitter metric (middle left),
the NUT--de Sitter metric (middle right), the Reissner-Nordstr\"{o}m--de Sitter metric
(bottom left) and the charged NUT--de Sitter metric (bottom right) with $\Lambda=1/(200m^2)$,
$e=3m/4$ and $n=m/2$ in the respective cases for an observer at $r_{O}=8m$ and $\vartheta_{O}=\pi/2$
and light sources distributed on the two-sphere $S_{L}^{2}$ with radius coordinate $r_{L}=9m$.
The Misner string is located at $\vartheta=0$ ($C=1$). As soon as we turn on the electric charge $e$ and the cosmological constant $\Lambda$
the shadow shrinks; however, the overall pattern of the lens map remains the same.\\
The twist observed in Figs.~8 and 10 has already been observed by
Lynden-Bell and Nouri-Zonoz \cite{NouriZonoz1997,LyndenBell1998} in the weak-field
limit. When we observe a circular shadow this is one of two recognizable characteristics
indicating the presence of the gravitomagnetic charge $n$. This twist
can potentially be observed when we observe multiple images from light sources at
approximately the same distance from the black hole, e.g., in a star cluster or
a galaxy cluster. Identifying enough images and their positions on the sky will allow
us to construct a partial lens map and potentially infer the magnitude of the twist.
From the determined magnitude of the twist we can then draw conclusions on the magnitude
of the gravitomagnetic charge $n$. Although this partial lens map may allow us to
draw conclusions on the presence and potentially the magnitude of the gravitomagnetic
charge $n$ it will not allow us to lift the degeneracy with respect to the cosmological
constant $\Lambda$, the electric charge $e$, $r_{O}$ and $r_{L}$.\\
\begin{figure*}\label{fig:SphereoLS}
    \includegraphics[width=0.6\textwidth]{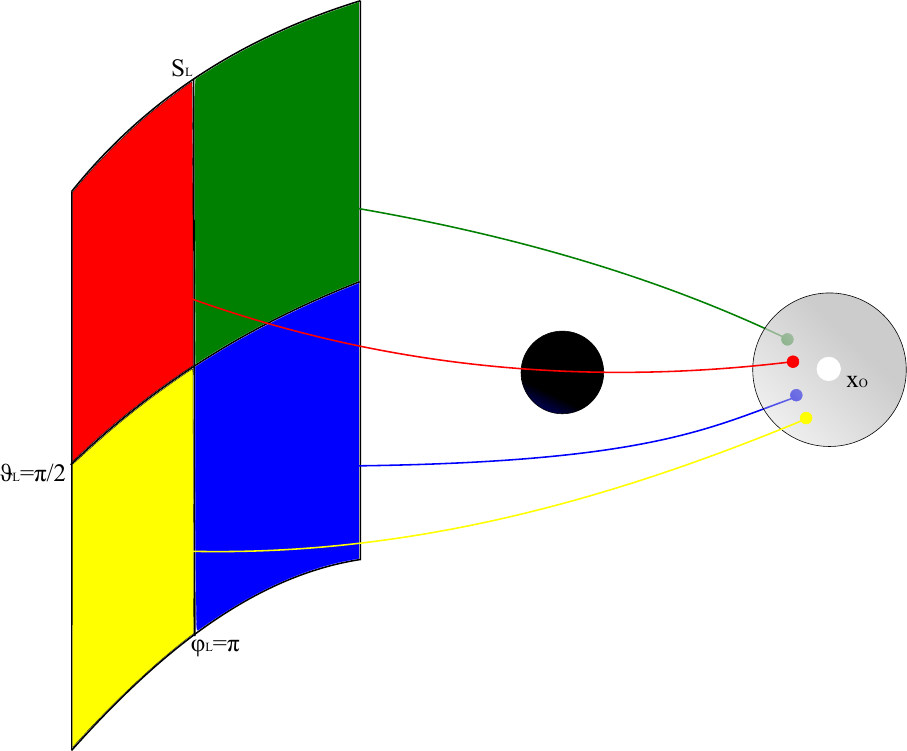}
    \caption{A simple illustration of the lens map. The black sphere in the center
    represents the black hole. The white dot and the gray area surrounding it are the
    observer and its celestial sphere at $x_{O}=(x_{O}^{\mu})$. The colored area
    represents a patch on the two-sphere of light sources $S_{L}^{2}$ with coordinate radius $r_{L}$.
    The two-sphere $S_{L}^{2}$ is colored using the convention in Bohn \emph{et al.}
    \cite{Bohn2015}. Extended to the whole sphere we color it as follows: $0\leq\vartheta_{L}\leq\pi/2$
    and $0\leq\varphi_{L}<\pi$: green, $\pi/2<\vartheta_{L}\leq\pi$ and
    $0\leq\varphi_{L}<\pi$: blue, $0\leq\vartheta_{L}\leq \pi/2$ and
    $\pi\leq\varphi_{L}<2\pi$: red, $\pi/2<\vartheta_{L}\leq\pi$ and
    $\pi\leq\varphi_{L}<2\pi$: yellow. The colored lines represent lightlike
    geodesics emitted by light sources on each patch of the two-sphere $S_{L}^{2}$.}
\end{figure*}
\begin{figure*}\label{fig:LensMap1}
\begin{tabular}{cc}
  Schwarzschild Metric& NUT Metric $n=m/100$ \\[6pt]
  \includegraphics[width=90mm]{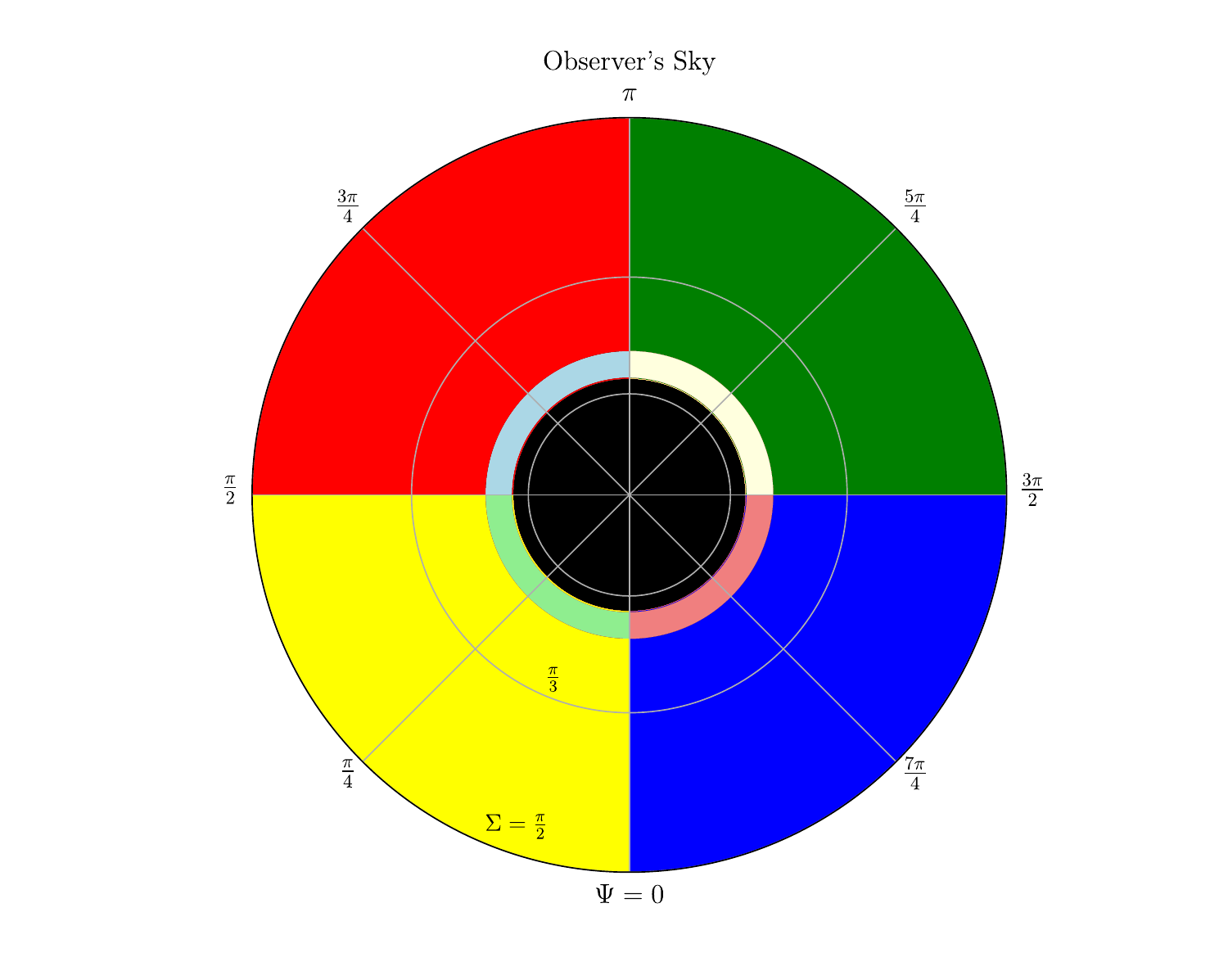} &   \includegraphics[width=90mm]{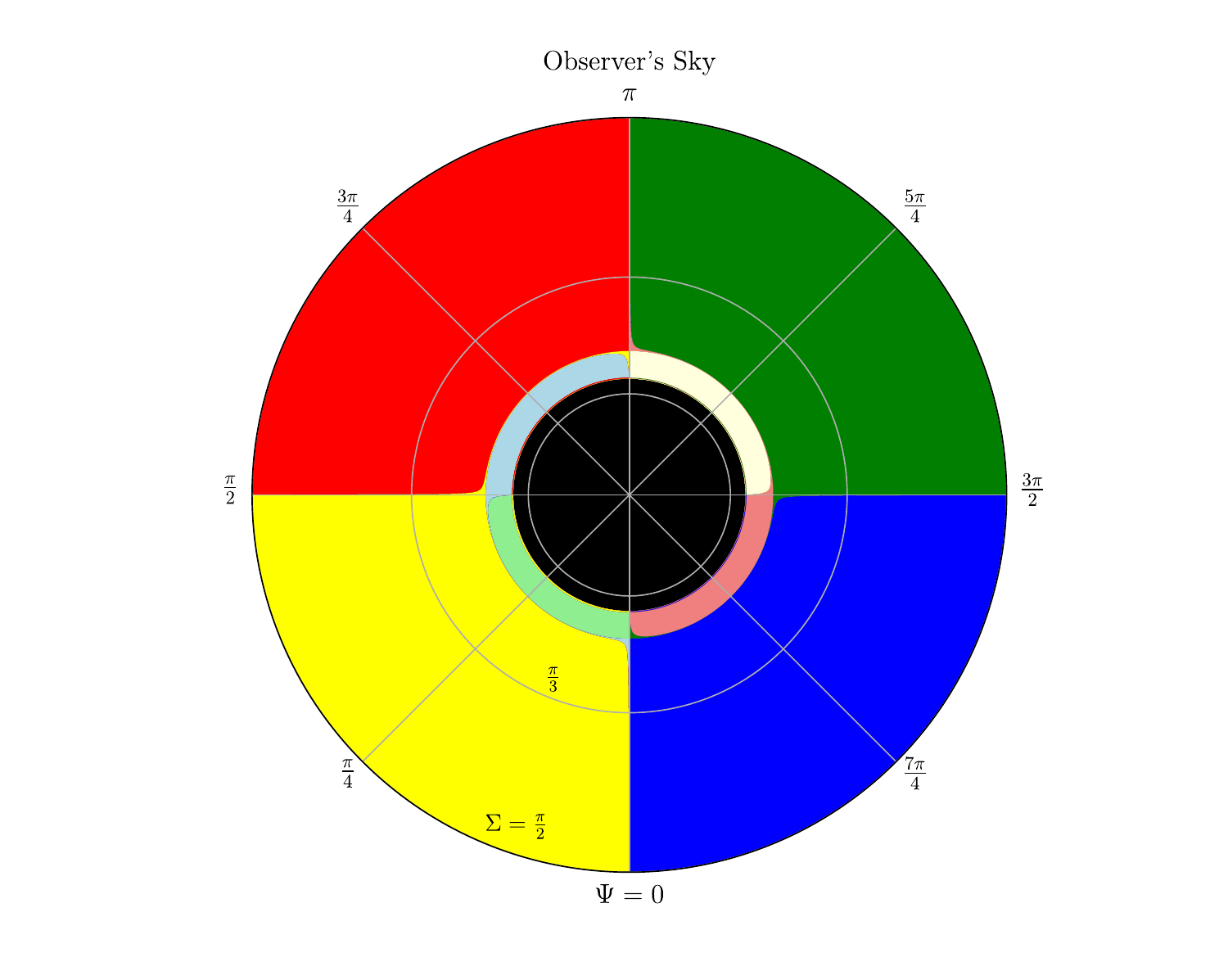} \\
  NUT Metric $n=m/10$ & NUT Metric $n=m/2$ \\[6pt]
  \includegraphics[width=90mm]{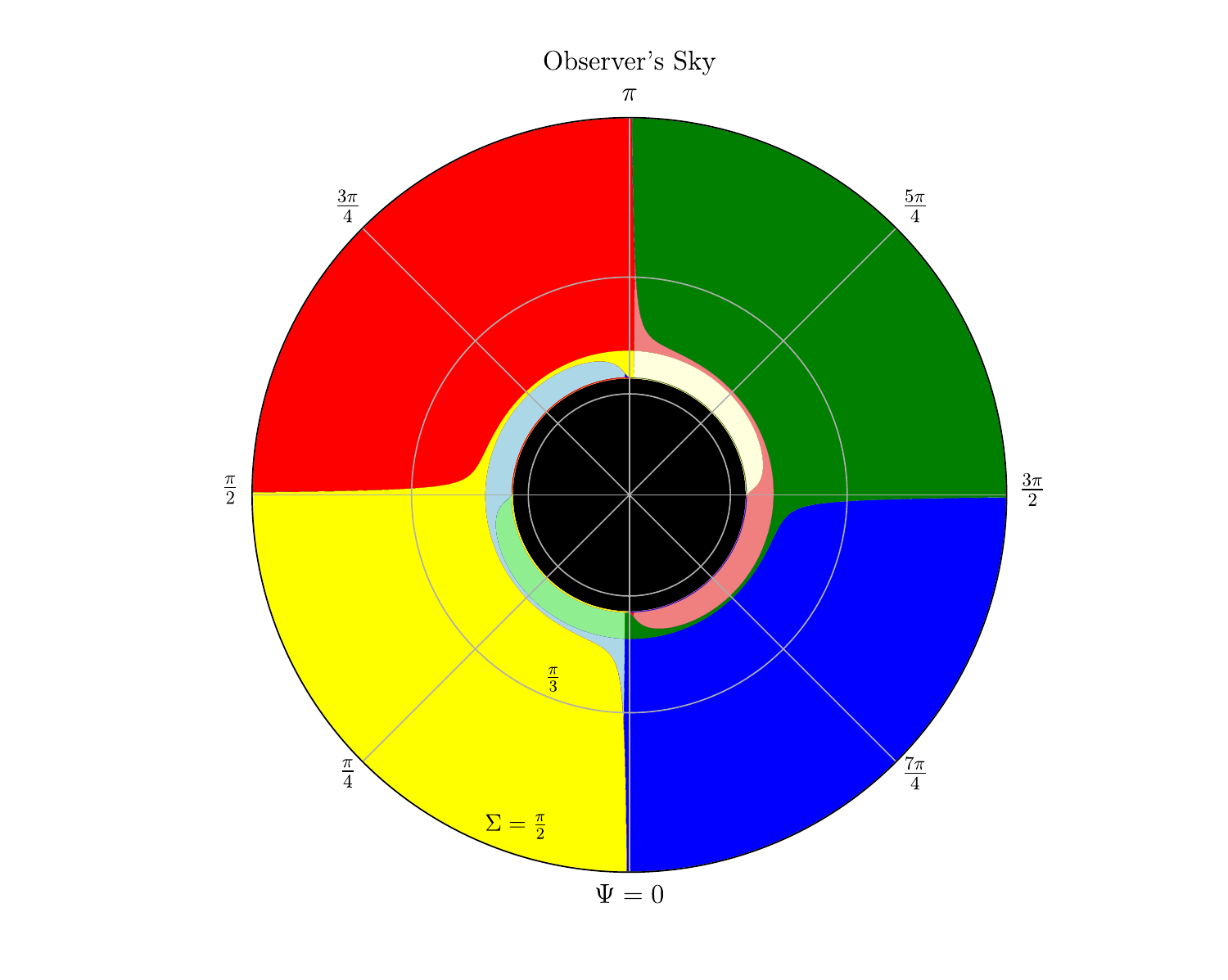} &   \includegraphics[width=90mm]{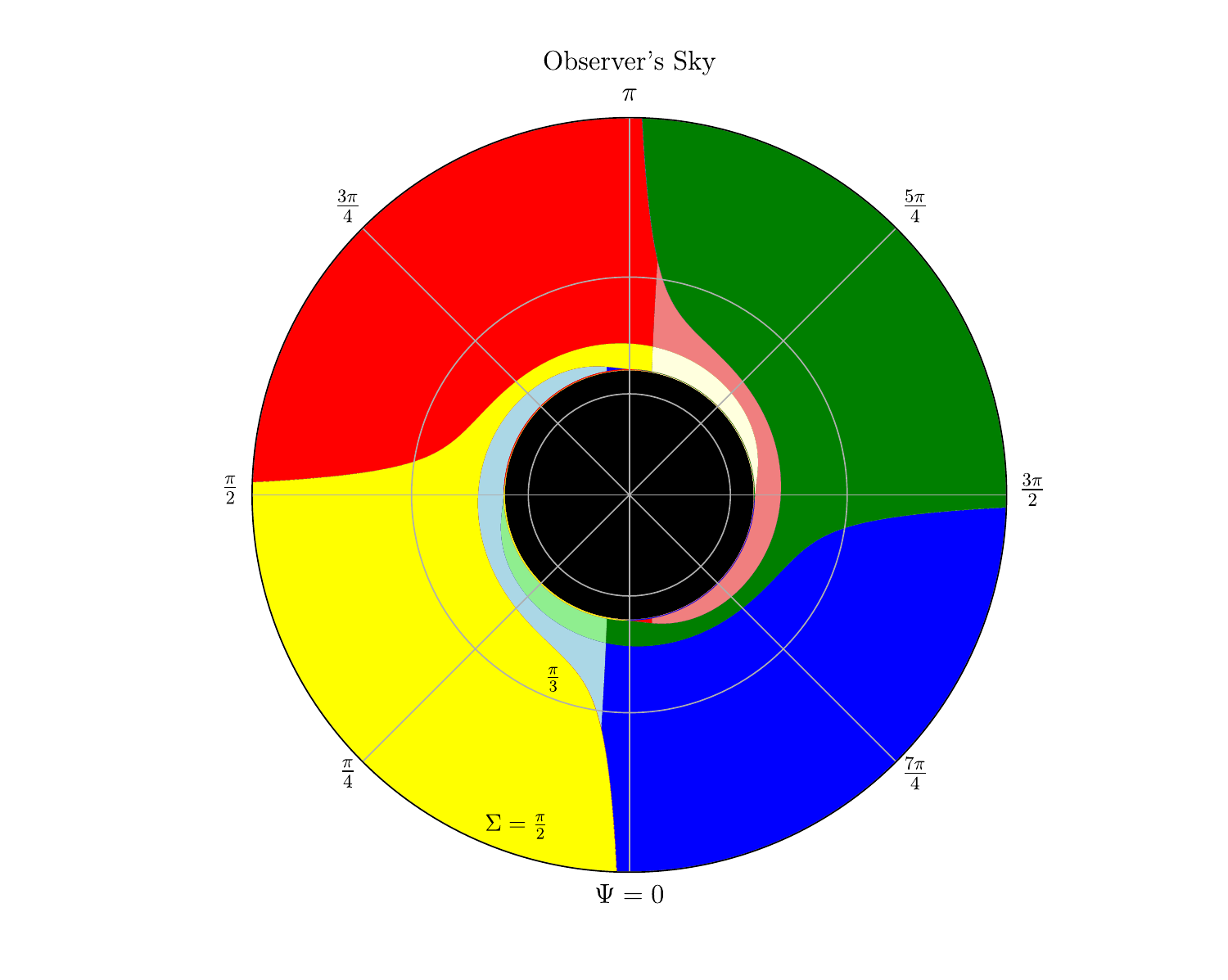} \\
\end{tabular}
\caption{Lens maps for light rays emitted by light sources located on the
two-sphere $S_{L}^2$ at the radius coordinate $r_{L}=9m$ and detected by an observer located at $r_{O}=8m$,
$\vartheta_{O}=\pi/2$ in the Schwarzschild metric (top left) and in the NUT metric
with $n=m/100$ (top right), $n=m/10$ (bottom left), and $n=m/2$ (bottom right).
The Misner string is located at $\vartheta=0$ ($C=1$).}
\end{figure*}
\begin{figure}[h]\label{fig:LensMap2}
\includegraphics[width=90mm]{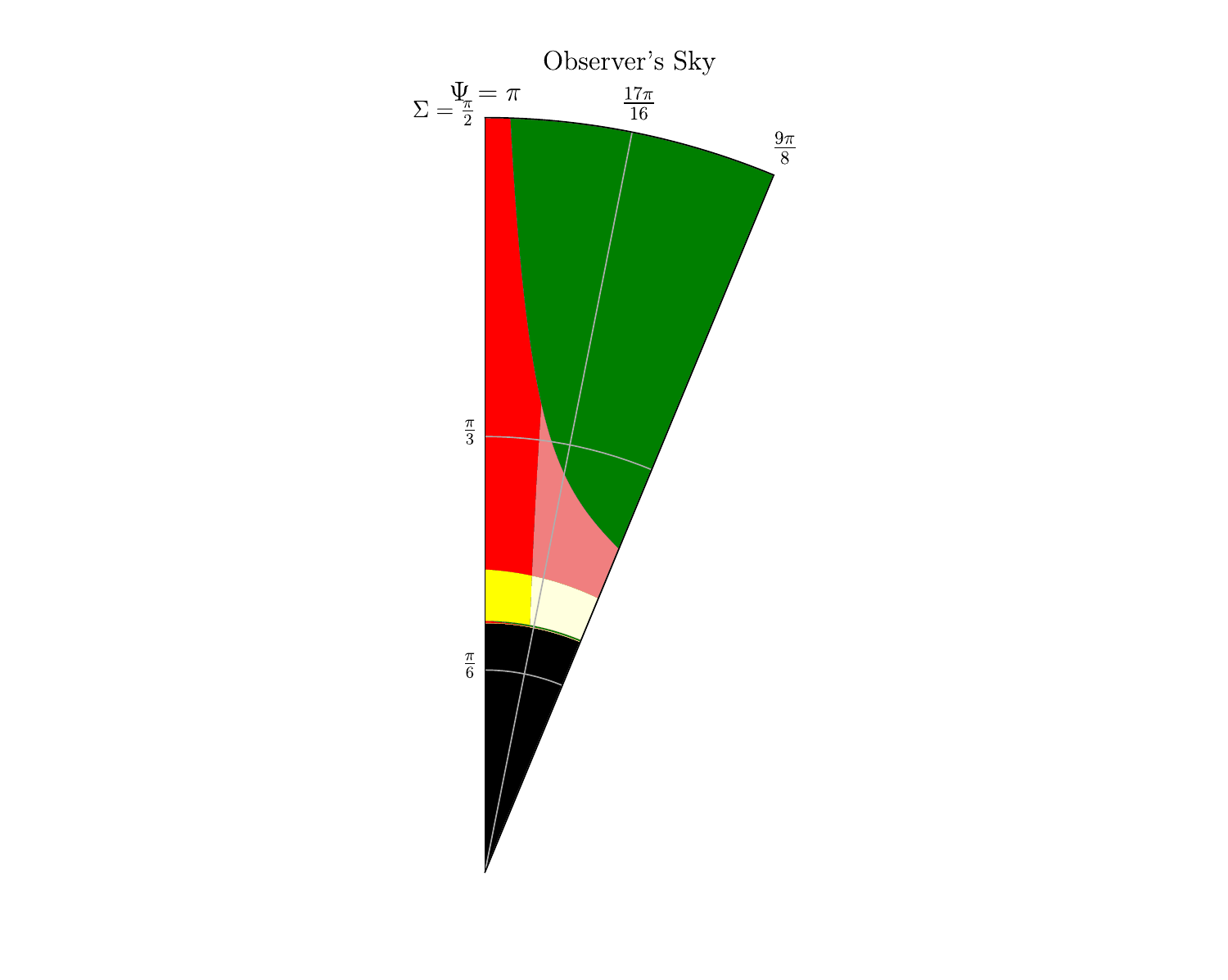}
\caption{Enlarged view of the lens map in Fig.~8 (bottom right) between $\Psi=\pi$
and $\Psi=9\pi/8$ for light rays emitted by light sources located on the two-sphere
$S_{L}^2$ at the radius coordinate $r_{L}=9m$ and detected by an observer located at $r_{O}=8m$,
$\vartheta_{O}=\pi/2$ in the NUT metric with $n=m/2$. The Misner string is located
at $\vartheta=0$ ($C=1$).}
\end{figure}
\begin{figure*}[h]\label{fig:LensMap3}
\begin{tabular}{cc}
  Reissner-Nordstr\"{o}m Metric& Charged NUT Metric\\[6pt]
  \includegraphics[width=89mm]{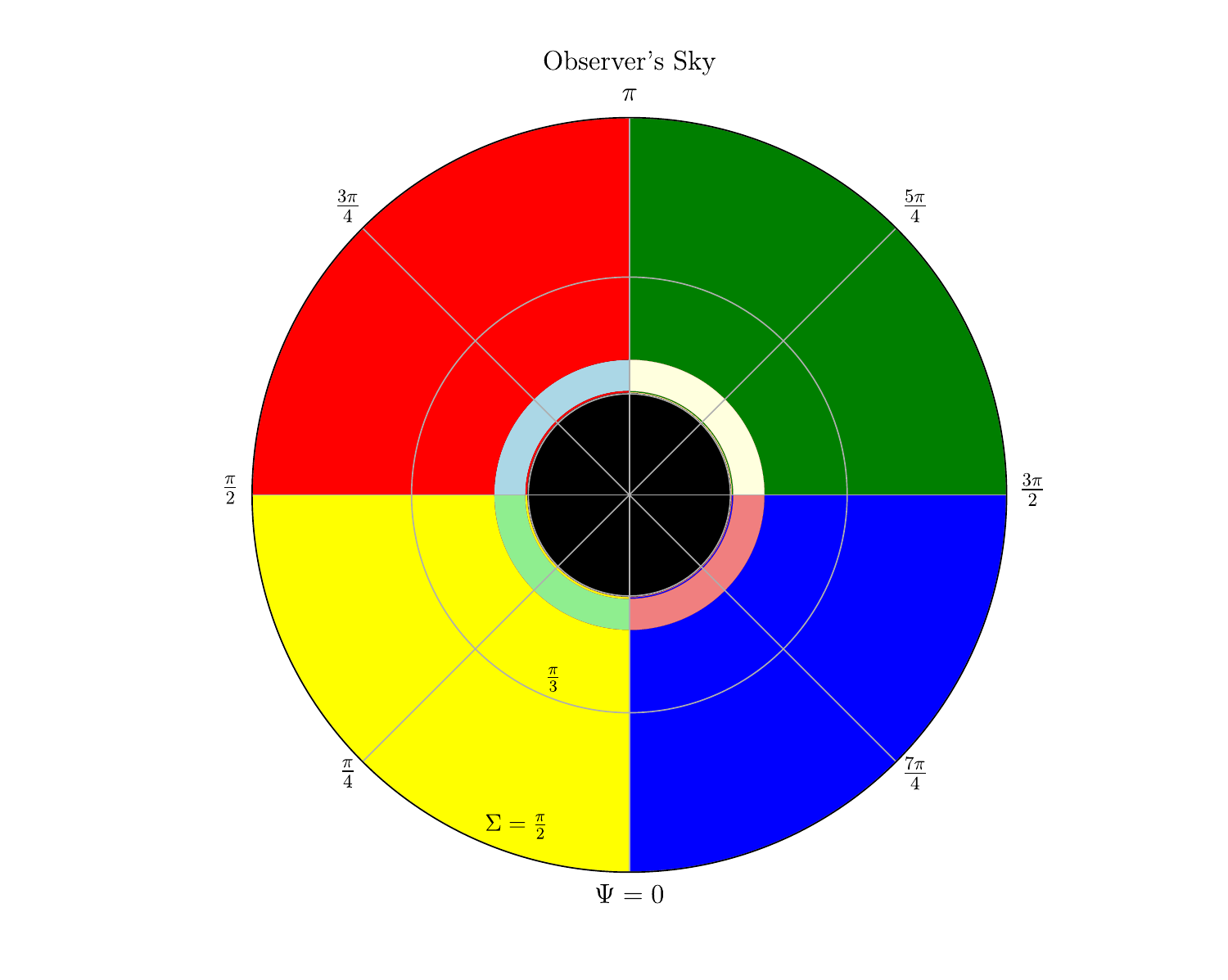} &   \includegraphics[width=89mm]{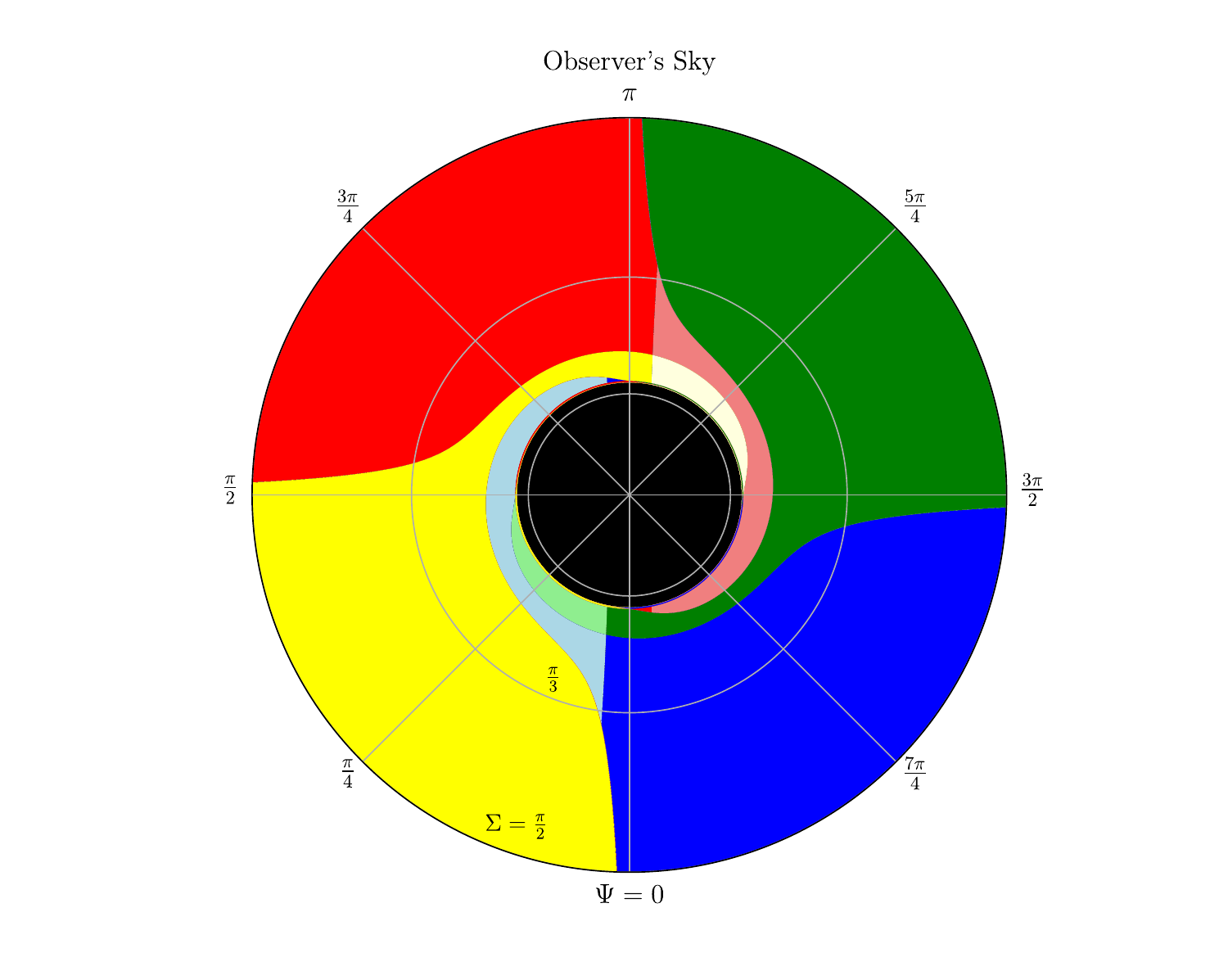} \\
  Schwarzschild-de Sitter Metric& NUT-de Sitter metric \\[6pt]
  \includegraphics[width=89mm]{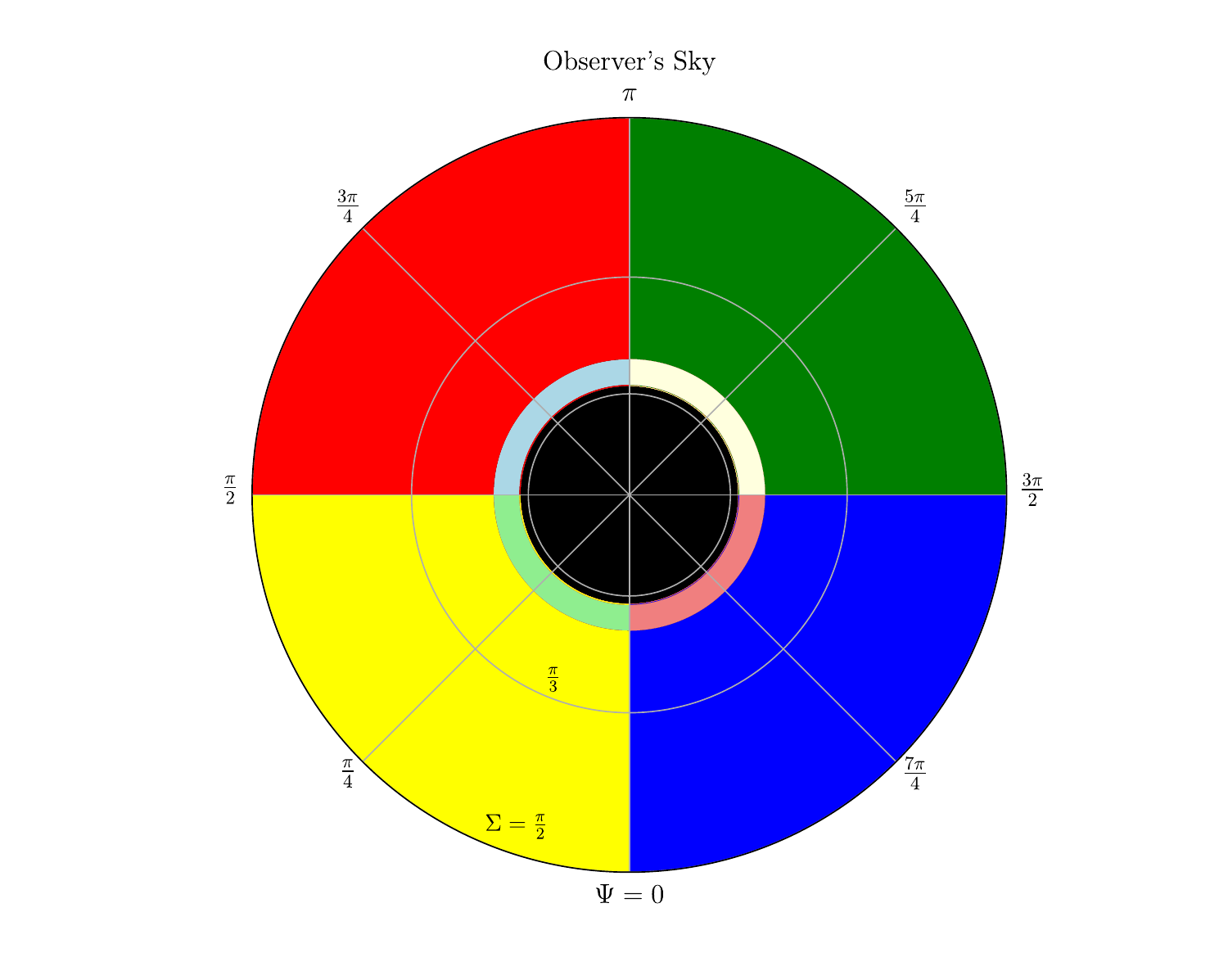} &   \includegraphics[width=89mm]{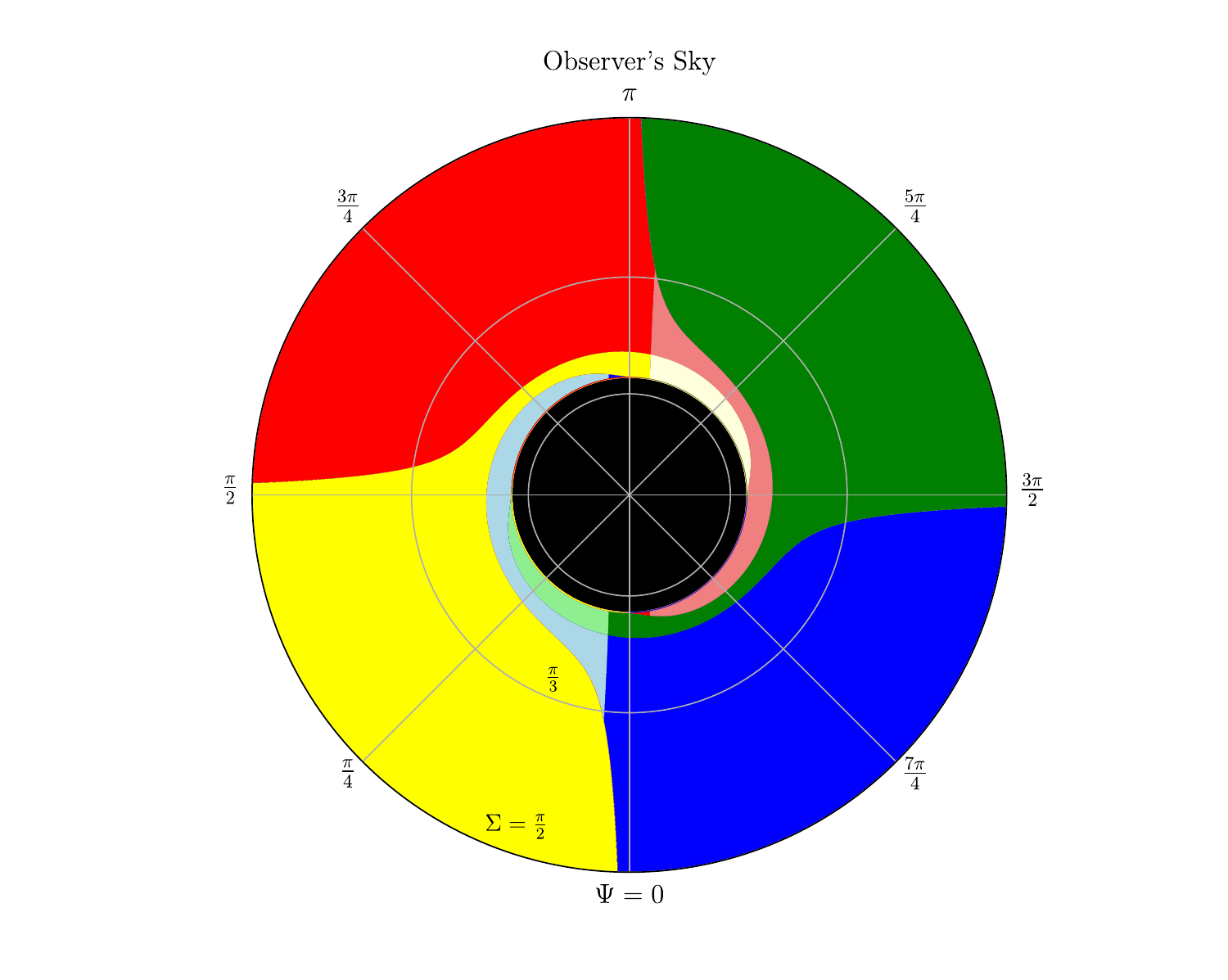} \\
  Reissner-Nordstr\"{o}m-de Sitter Metric& Charged NUT-de Sitter metric\\[6pt]
  \includegraphics[width=89mm]{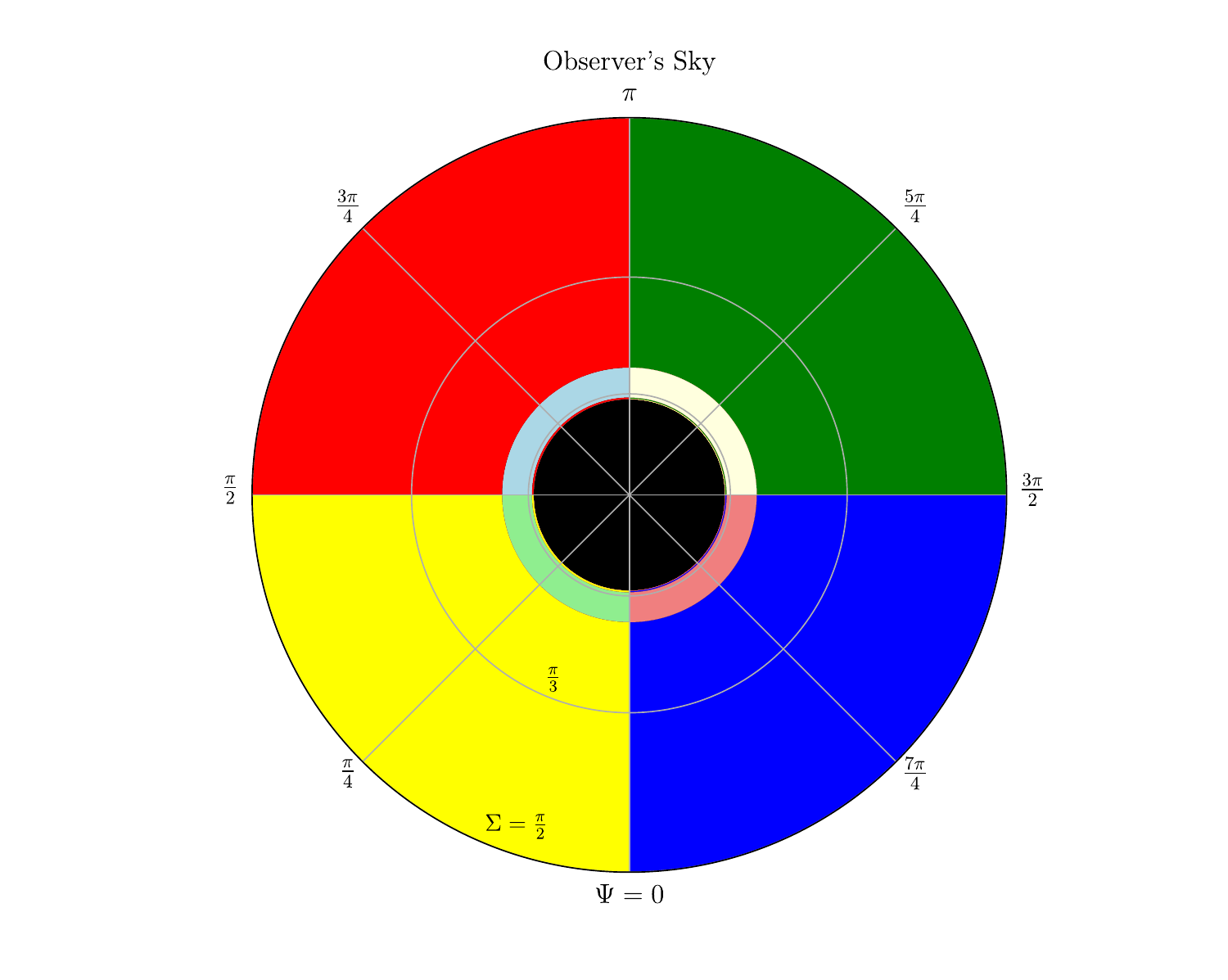} &   \includegraphics[width=89mm]{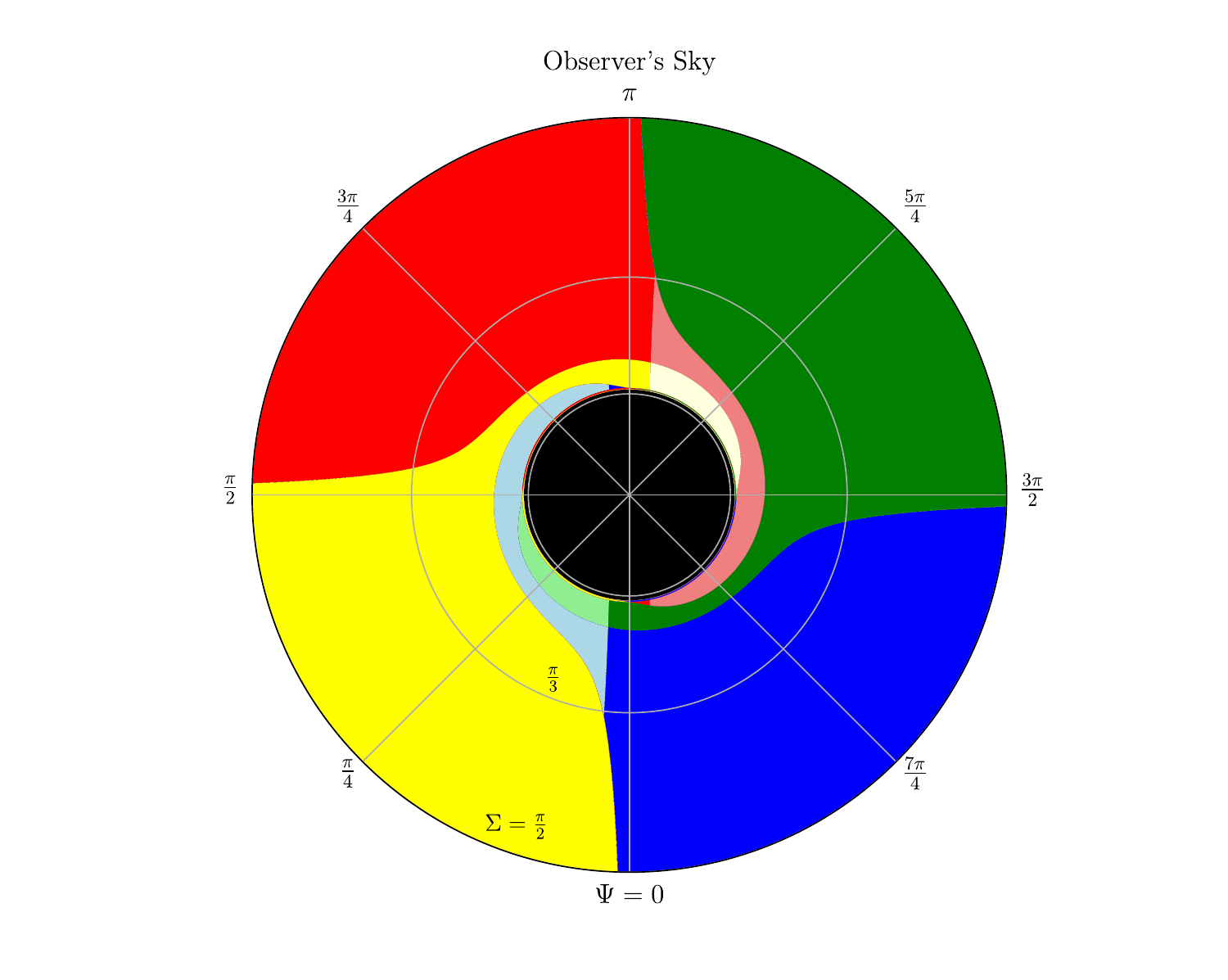} \\
\end{tabular}
\caption{Lens maps for light rays emitted by light sources located on the
two-sphere $S_{L}^2$ at the radius coordinate $r_{L}=9m$ and detected by an observer located at $r_{O}=8m$,
$\vartheta_{O}=\pi/2$, in the Reissner-Nordstr\"{o}m metric (top left), the charged
NUT metric (top right), the Schwarzschild--de Sitter metric (middle left), the
NUT--de Sitter metric (middle right), the Reissner-Nordstr\"{o}m--de Sitter metric
(bottom left), and the charged NUT--de Sitter metric (bottom right). The cosmological
constant $\Lambda$, the electric charge $e$ and the gravitomagnetic charge $n$ are
$\Lambda=1/(200m^2)$, $e=3m/4$, and $n=m/2$, respectively. The Misner string is located at $\vartheta=0$ ($C=1$).}
\end{figure*}

\subsection{Redshift}\label{Sec:Redshift}
\begin{figure*}\label{fig:Redshift}
  \begin{tabular}{cc}
    NUT Metric & Charged NUT Metric\\[6pt]
    \includegraphics[width=90mm]{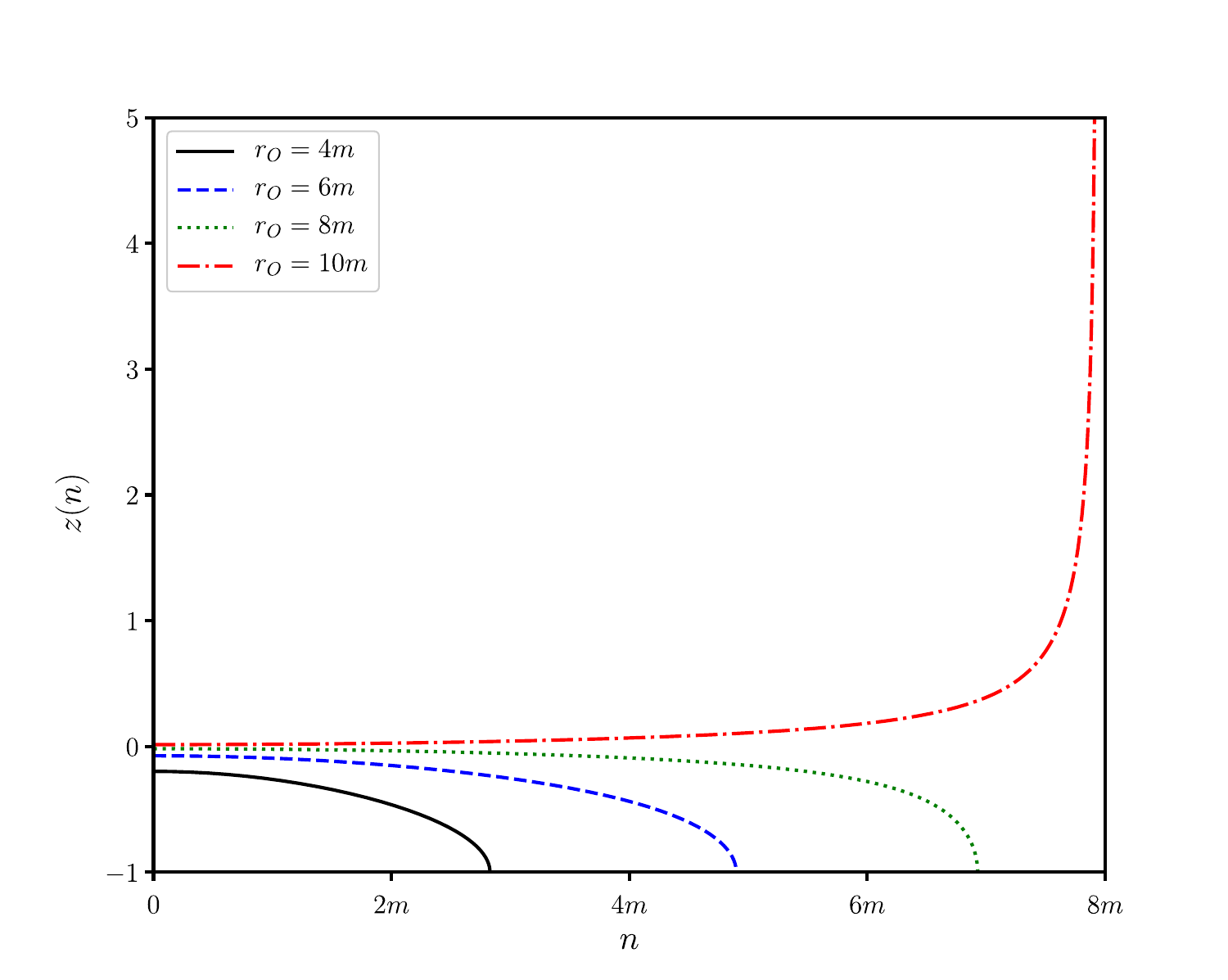} &   \includegraphics[width=90mm]{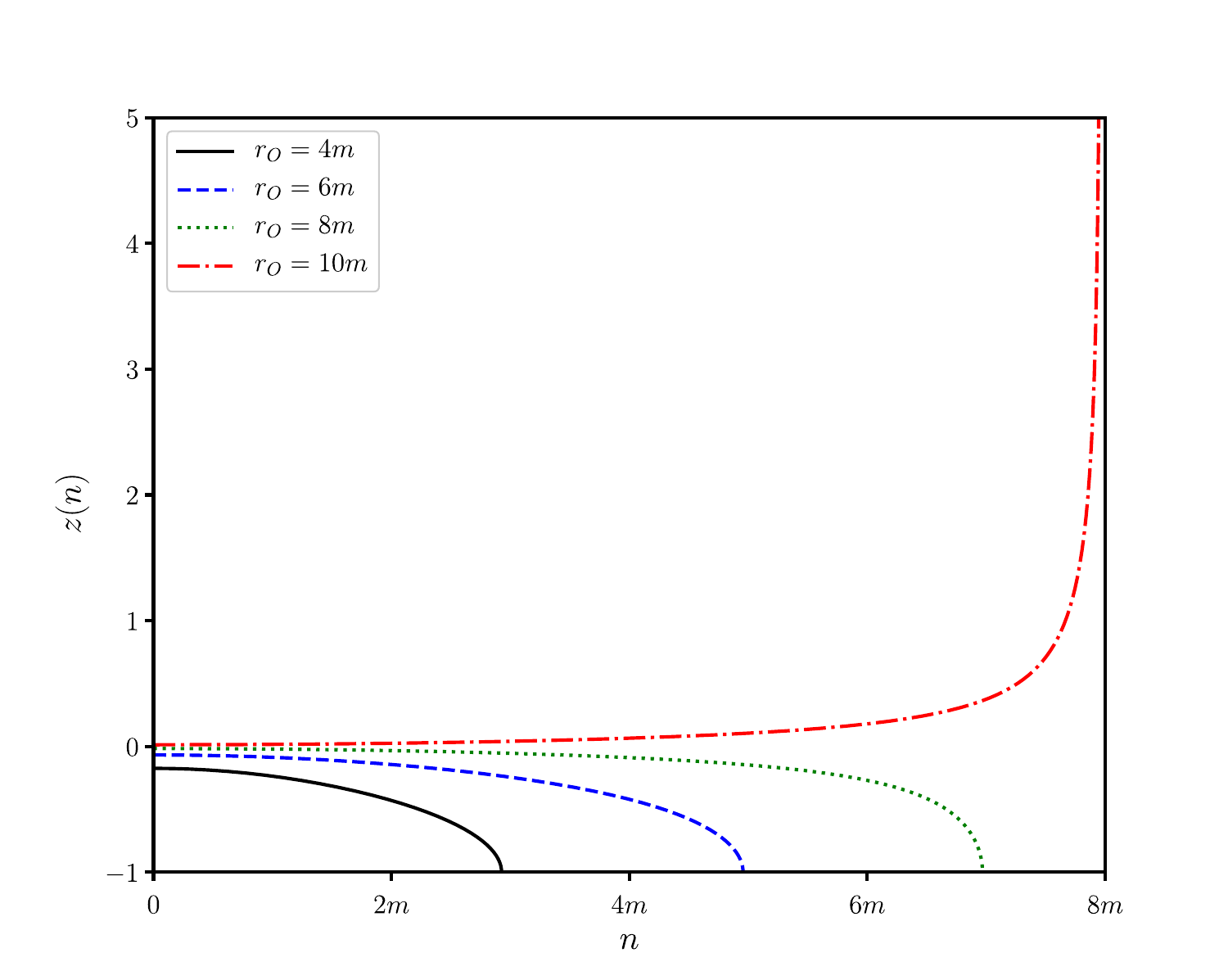} \\
    NUT-de Sitter Metric& Charged NUT-de Sitter Metric\\[6pt]
    \includegraphics[width=90mm]{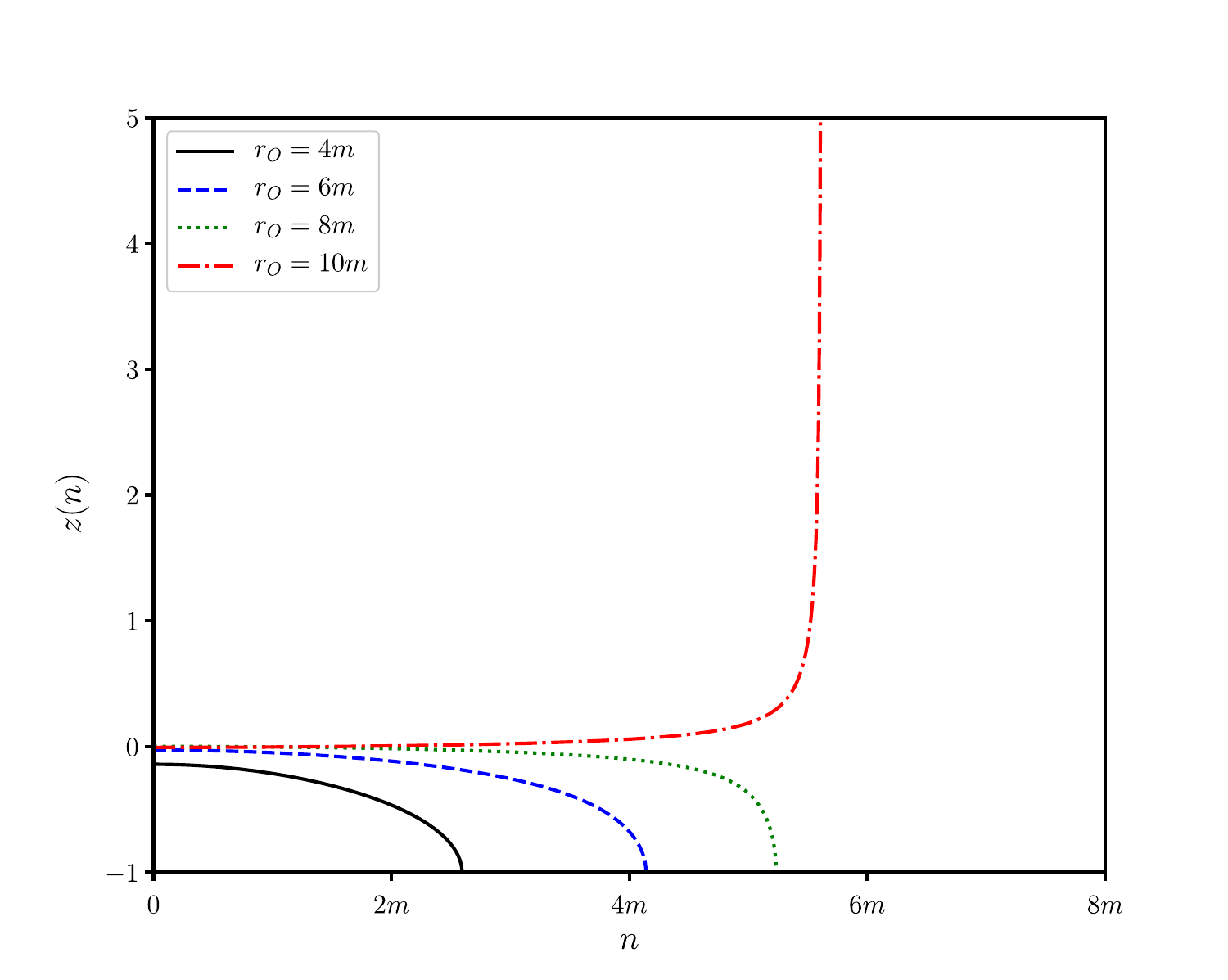} &   \includegraphics[width=90mm]{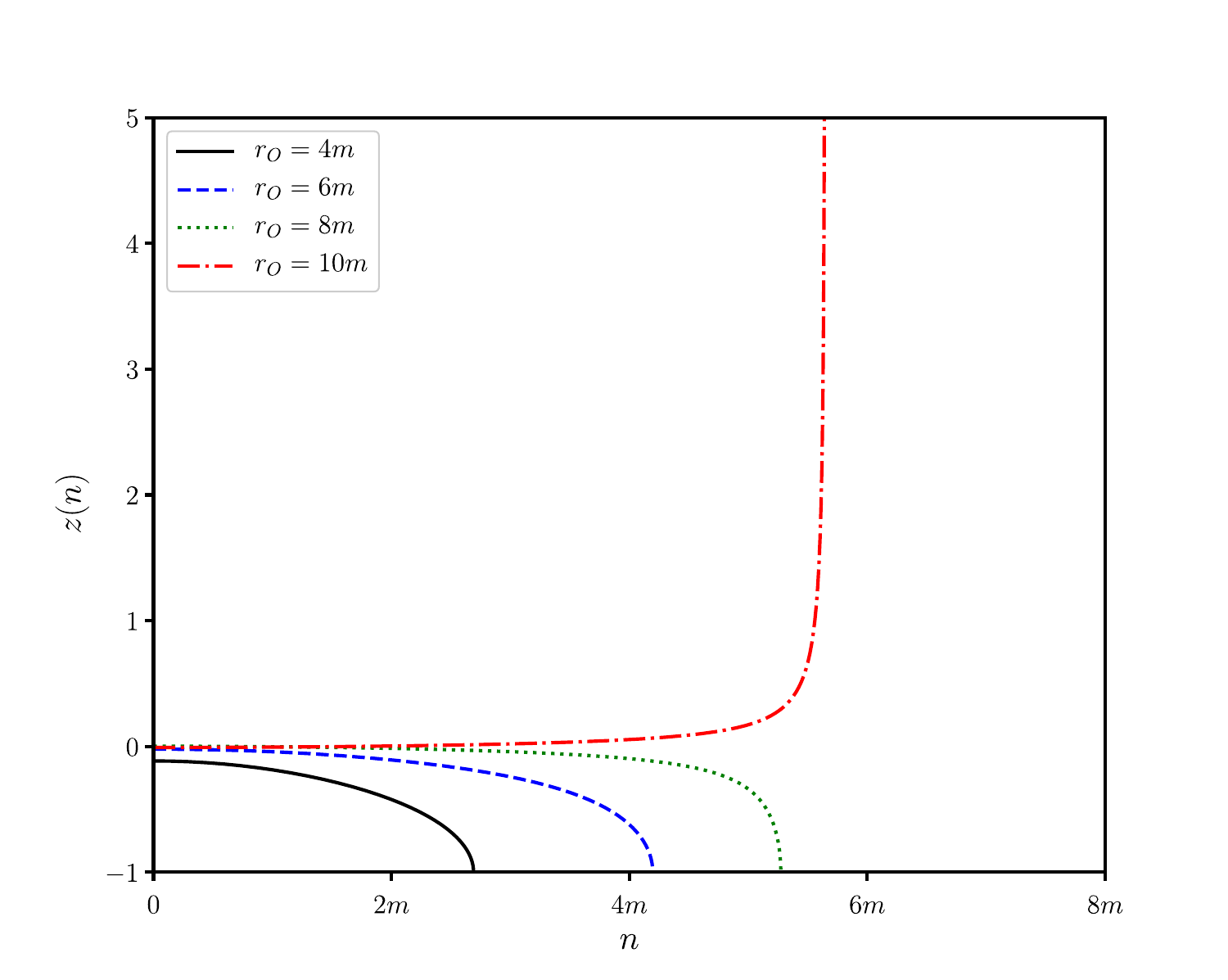} \\
  \end{tabular}
  \caption{Redshift for observers at radii $r_{O}=4m$ (black), $r_{O}=6m$ (blue dashed),
  $r_{O}=8m$ (green dotted), and $r_{O}=10m$ (red dashed-dotted) and a light source at $r_{L}=9m$
  for the NUT metric (top left), the charged NUT metric with $e=3m/4$ (top right),
  the NUT--de Sitter metric with $\Lambda=1/(200m^2)$ (bottom left) and the charged
  NUT--de Sitter metric with $\Lambda=1/(200m^2)$, $e=3m/4$ (bottom right).}
\end{figure*}
The redshift factor $z$ measures the relative energy shift that a light ray experiences on its
way from the light source by which it was emitted to the observer by whom it is
detected. It is one of the few observables that is directly accessible to observations
and can be determined by comparing the measured frequencies of known emission lines
in the emission spectrum of, e.g., a star, to their unshifted frequencies from, e.g.,
laboratory measurements. In our case the observer as well as the light source are stationary
and since we do not consider spinning black holes both move on $t$ lines. For this
emitter-observer constellation the redshift factor $z$ is thoroughly derived, e.g., in the book of
Straumann \cite{Straumann2013}, pp. 45. In terms of the metric coefficients it reads
\begin{eqnarray}\label{eq:rsGen}
z=\sqrt{\frac{\left.g_{tt}\right|_{x_{O}}}{\left.g_{tt}\right|_{x_{L}}}}-1.
\end{eqnarray}
Now we insert the metric coefficient $g_{tt}=-Q(r)/\rho(r)$ and get $z$ in terms
of the spacetime coordinates:
\begin{eqnarray}\label{eq:rsNUT}
z=\sqrt{\frac{\rho(r_{L})Q(r_{O})}{\rho(r_{O})Q(r_{L})}}-1.
\end{eqnarray}
For the charged NUT--de Sitter spacetimes $z$ only depends on the radius coordinates
$r_{O}$ of the observer and $r_{L}$ of the light source and the four parameters $m$, $\Lambda$, $e$ and $n$. Figure~11 shows the redshift
factor $z$ for observers at radius coordinates $r_{O}=4m$, $r_{O}=6m$, $r_{O}=8m$,
and $r_{O}=10m$ and a light source at the radius coordinate
$r_{L}=9m$ as function of the gravitomagnetic charge $n$ for the NUT metric (top
left), the charged NUT metric (top right), the NUT--de Sitter metric (bottom left)
and the charged NUT--de Sitter metric (bottom right). The cosmological constant and
the electric charge are $\Lambda=1/(200m^2)$ and $e=3m/4$, respectively. For $n=0$ the redshift
factor $z$ reduces to the redshift factors in the Schwarzschild metric (top left),
the Reissner-Nordstr\"{o}m metric (top right), the Schwarzschild--de Sitter metric
(bottom left) and the Reissner-Nordstr\"{o}m--de Sitter metric (bottom right), respectively. \\
For $r_{O}<r_{L}$ we mainly have blueshifts while for $r_{L}<r_{O}$ we mainly have redshifts.
When we now turn on the gravitomagnetic charge $n$ in the former case with growing gravitomagnetic charge the outer black hole horizon
approaches the observer $r_{\mathrm{H},\mathrm{o}}\rightarrow r_{O}$ and thus light
rays emitted by the light source are infinitely blueshifted leading to $z\rightarrow -1$.
In the latter case with growing gravitomagnetic charge $n$ the outer black hole
horizon approaches the light source $r_{\mathrm{H},\mathrm{o}}\rightarrow r_{L}$ and thus light rays emitted by this
source become infinitely redshifted and we have $z\rightarrow \infty$.\\
When we now turn on the electric charge $e$ for small $n\approx 0$ the blueshifts and the redshifts slightly
decrease. The outer black hole horizon is
originally located at a smaller radius coordinate and thus
we have $z\rightarrow -1$ and $z\rightarrow \infty$
for slightly larger gravitomagnetic charges $n$, respectively. Turning on the cosmological
constant $\Lambda$ has a similar effect. The outer black hole horizon
is originally located at a slightly larger radius coordinate $r_{\mathrm{H},\mathrm{o}}$ and expands faster with increasing $n$. Therefore
we have $z\rightarrow -1$ and $z\rightarrow \infty$ for much smaller gravitomagnetic
charges $n$.\\
For observations it is rather unfortunate that in addition to the four parameters $m$, $\Lambda$, $e$ and $n$ the redshift factor $z$ only depends
on the radius coordinates of the observer $r_{O}$ and of the light source $r_{L}$.
While the redshift factor $z$ is also affected by the gravitomagnetic charge $n$ this
information is useless as long as we do not \emph{a priori} know the distances between
observer and black hole and light source and black hole. Therefore, similar to the angular radius of the shadow, we have a degeneracy between
the redshift factors in spherically symmetric spacetimes and the charged NUT--de Sitter
metrics for different cosmological constants $\Lambda$, electric charges $e$,
gravitomagnetic charges $n$, $r_{O}$, and $r_{L}$. However, combined with information
about the angular radius of the shadow $\Sigma_{\mathrm{ph}}$, from the lens equation
and travel-time differences (these will be discussed in the next section) there
is a chance that we can lift this degeneracy and determine $\Lambda$, $e$, and $n$.

\subsection{Travel time}\label{Sec:TravTime}
The travel time $T$ measures in terms of the time coordinate $t$ the time a light
ray needs to travel from the light source by which it was emitted to an observer
by whom it is detected. For a light ray that is emitted at the time coordinate $t_{L}$
and detected by an observer at the time coordinate $t_{O}$ it reads
\begin{eqnarray}
T=t_{O}-t_{L}.
\end{eqnarray}
The travel time is not directly measurable; however, in the case that we can identify
multiple images of the same light source, e.g., a quasar (see Fohlmeister \emph{et al.}
\cite{Fohlmeister2013} or Koptelova \emph{et al.} \cite{Koptelova2012}) we can record light curves for each image and compare their variability. When we are able to identify similar structures we can now determine the time delay between the images.\\
We now want to construct travel-time maps for the charged NUT--de Sitter spacetimes.
For this purpose we now insert Eqs.~(\ref{eq:CoME})--(\ref{eq:CoMK}) in Eq.~(\ref{eq:EoMtpart})
and rewrite it with the help of Eq.~(\ref{eq:EoMr}) as (remember that we set $t_{O}=0$)
\begin{eqnarray}
&T(\Sigma,\Psi)=\int_{r_{O}...}^{...r_{L}}\frac{\sqrt{Q(r_{O})}\rho(r')^2\mathrm{d}r'}{Q(r')\sqrt{\rho(r')^2 Q(r_{O})-Q(r')\rho(r_{O})^2\sin^2\Sigma}}\\
&-2n\int_{0}^{\lambda_{L}}(\cos\vartheta(\lambda')+C)\frac{\left(\sqrt{\rho(r_{O})}\sin\vartheta_{O}\sin\Sigma\sin\Psi+2n\left(\cos\vartheta(\lambda')-\cos\vartheta_{O}\right)\sqrt{\frac{Q(r_{O})}{\rho(r_{O})}}\right)\mathrm{d}\lambda'}{1-\cos^2\vartheta(\lambda')}.\nonumber
\end{eqnarray}
The dots in the limits of the integral of the first term shall indicate that we
have to split the integral at the turning point. For observers between photon sphere
and infinity ($\Lambda=0$) or the cosmological horizon ($0<\Lambda<\Lambda_{\mathrm{C}}$) this is always
a minimum. In the same term the sign of the root has to be chosen such that it agrees
with the direction of the $r$ motion along the geodesic. We now rewrite the term in terms of elementary
functions and Legendre's elliptic integrals of the first, second and third kind as described in
Sec.~\ref{Sec:tr}. Analogously we integrate the second term on the right-hand
side following the steps described in Sec.~\ref{Sec:ttheta}. \\
For a fast and efficient evaluation the calculation of the travel time was implemented
in JULIA using the same set of program routines as for the lens equation.\\
Figure~12 shows the travel time in the Schwarzschild metric (top left) and the NUT
metric with a Misner string at $\vartheta=0$ ($C=1$) and $n=m/100$ (top right),
$n=m/10$ (middle left) and $n=m/2$ (middle right) for an observer located at $r_{O}=8m$
and $\vartheta_{O}=\pi/2$. In addition it also shows travel-time maps for observers located
at $r_{O}=8m$ and $\vartheta_{O}=\pi/4$ (bottom left) and $\vartheta_{O}=3\pi/4$
(bottom right) in the NUT spacetime with $n=m/2$. The light sources are located
on the two-sphere $S_{L}^{2}$ at the radius coordinate $r_{L}=9m$. The travel time
increases towards the shadow as the
light ray makes more and more turns around the black hole. For the Schwarzschild
metric (top left) the travel time is rotationally symmetric under arbitrary rotations
about the axis $\Sigma=0$. For $n=m/100$ (top right) the travel time shows still a high degree
of apparent rotational symmetry. When we look closer, however, we can recognize an apparent
sharp discontinuity at $\Psi=\pi$. When we increase the gravitomagnetic charge $n$ this
discontinuity becomes more and more pronounced. When we start at the discontinuity
and go in clockwise direction along a constant latitude $\Sigma$ the travel time
decreases. In the travel-time maps this decrease forms the shape of a spiral. In addition with increasing $n$ a second discontinuity
starts to become visible on the right-hand side of $\Psi=0$ close to the shadow. When we zoom in
on the middle right panel of Fig.~12 we recognize that the first discontinuity consists of very
narrow steps and thus from this map alone it is unclear if this is a real sharp discontinuity
or if the travel time simply shows a very steep increase. In all three panels these
discontinuities appear exactly for lightlike geodesics crossing the Misner string.
Figure~13 shows an enlarged view of the discontinuity close to $\Psi=\pi$ between $\Psi=\pi$ and $\Psi=9\pi/8$ for $n=m/2$ with a
16 times higher $\Psi$ resolution than in the middle right panel of Fig.~12. The figure clearly shows that
the travel time has a real discontinuity for lightlike geodesics crossing the Misner string.
From the observer's perspective lightlike geodesics passing to the left of the Misner string have a shorter travel
time than light rays passing to the right of the Misner string. When the observer
moves to lower spacetime latitudes $\vartheta$ the discontinuity of the
travel time close to $\Psi=\pi$ stretches out to higher latitudes $\Sigma$ on the
observer's celestial sphere while the discontinuity close to $\Psi=0$ is confined
to a much more narrow region close to the shadow.
In addition compared to an observer at $\vartheta_{O}=\pi/2$ for the observer at
$\vartheta_{O}=\pi/4$ they appear closer to $\Psi=\pi$ and $\Psi=0$, respectively.
For an observer at $\vartheta_{O}=3\pi/4$ the situation is reversed. The discontinuity
at $\Psi=\pi$ becomes more confined to the shadow while the discontinuity
at $\Psi=0$ can already be observed at higher latitudes $\Sigma$. In addition both
discontinuities can be found at longitudes $\Psi$ further away from $\Psi=\pi$ and $\Psi=0$,
respectively. \\
Figures~14--16 show the travel-time maps for observers in the Reissner-Nordstr\"{o}m
metric (Fig.~14, top left), the Schwarzschild--de Sitter metric (Fig.~15, top left),
the Reissner-Nordstr\"{o}m--de Sitter metric (Fig.~16, top left), the charged NUT metric (Fig.~14,
top right and bottom row), the NUT--de Sitter metric (Fig.~15, top right and bottom row)
and the charged NUT--de Sitter metric (Fig.~16, top right and bottom row) for observers
located at the radius coordinate $r_{O}=8m$ and the spacetime latitudes $\vartheta_{O}=\pi/4$ (only for $n>0$),
$\vartheta_{O}=\pi/2$ and $\vartheta_{O}=3\pi/4$ (only for $n>0$). The two-sphere $S_{L}^{2} $ is
located at the radius coordinate $r_{L}=9m$. The electric charge and the cosmological
constant are $e=3m/4$ and $\Lambda=1/(200m^2)$, respectively. When we turn on the
electric charge $e$ (Fig.~14) the shadow shrinks and the travel time shows roughly the same pattern just
shifted to lower latitudes. When we turn on the cosmological constant (Figs.~15
and 16) the area of the shadow shrinks while we observe an overall increase of the
travel time. However, like after turning on the electric charge except for some minor
details the overall patterns on the travel-time maps remain the same. \\
The travel time just provided us with a second unique pattern that indicates the
presence of a gravitomagnetic charge. When a black hole has a gravitomagnetic charge
and when it is described by one of the charged NUT--de Sitter metrics we will observe
a discontinuity whenever light rays cross the Misner strings. While the Misner strings
are very likely only mathematical idealizations of a real physical effect and thus
in reality it is more likely that we will observe a transition from shorter to longer
travel times (or vice versa) this effect may still be observable. \\
As stated above we \emph{cannot} observe absolute travel times of light rays but
only travel-time differences. Considering the uniqueness of the discontinuity the
best chance to observe it would be the use of quadruply lensed stars or quasars. When we
observe lensed images of these sources more or less forming a cross around the
lens (see, e.g., Suyu \emph{et al.} \cite{Suyu2017}) we can determine travel-time differences
between the images. In the case they are at roughly the same distance from the black
hole and have roughly the same angular distance from each other (like, e.g., for
HE 0435-1223 in Fig.~1 of Suyu \emph{et al.} \cite{Suyu2017}) the observed discontinuity in the travel time
will lead to a high travel-time difference between the images with the smallest
angular distances to the discontinuity while the travel-time difference between
the other images will be much smaller. The travel-time difference may allow
to distinguish black holes with gravitomagnetic charge $n$ from black holes without
gravitomagnetic charge but it does not allow to lift the degeneracy with respect to
$\Lambda$, $e$, $r_{O}$, and $r_{L}$.\\
Unfortunately so far we did not observe quadruply imaged stars lensed by black
holes. Indeed, so far light sources multiply imaged by black holes were not observed
at all and thus we will have to wait until the next generations of telescopes
become available that have a resolution that is high enough to address this challenge.\\
\begin{figure*}[h]\label{fig:TravelTime1}
\begin{tabular}{cc}
   Schwarzschild Metric $\vartheta_{O}=\pi/2$ & NUT Metric $n=m/100$ and $\vartheta_{O}=\pi/2$\\[6pt]
   \includegraphics[width=90mm]{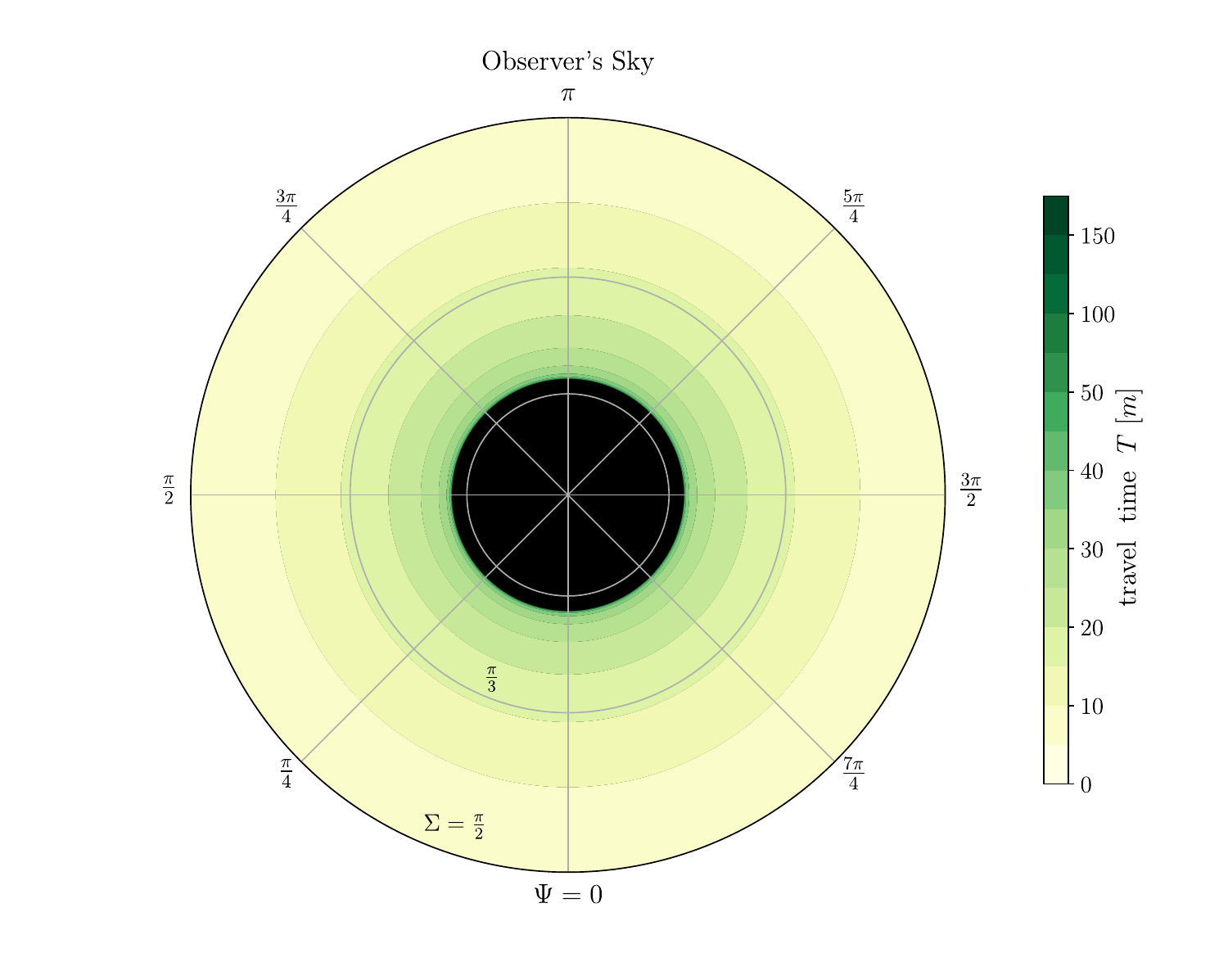} & \includegraphics[width=90mm]{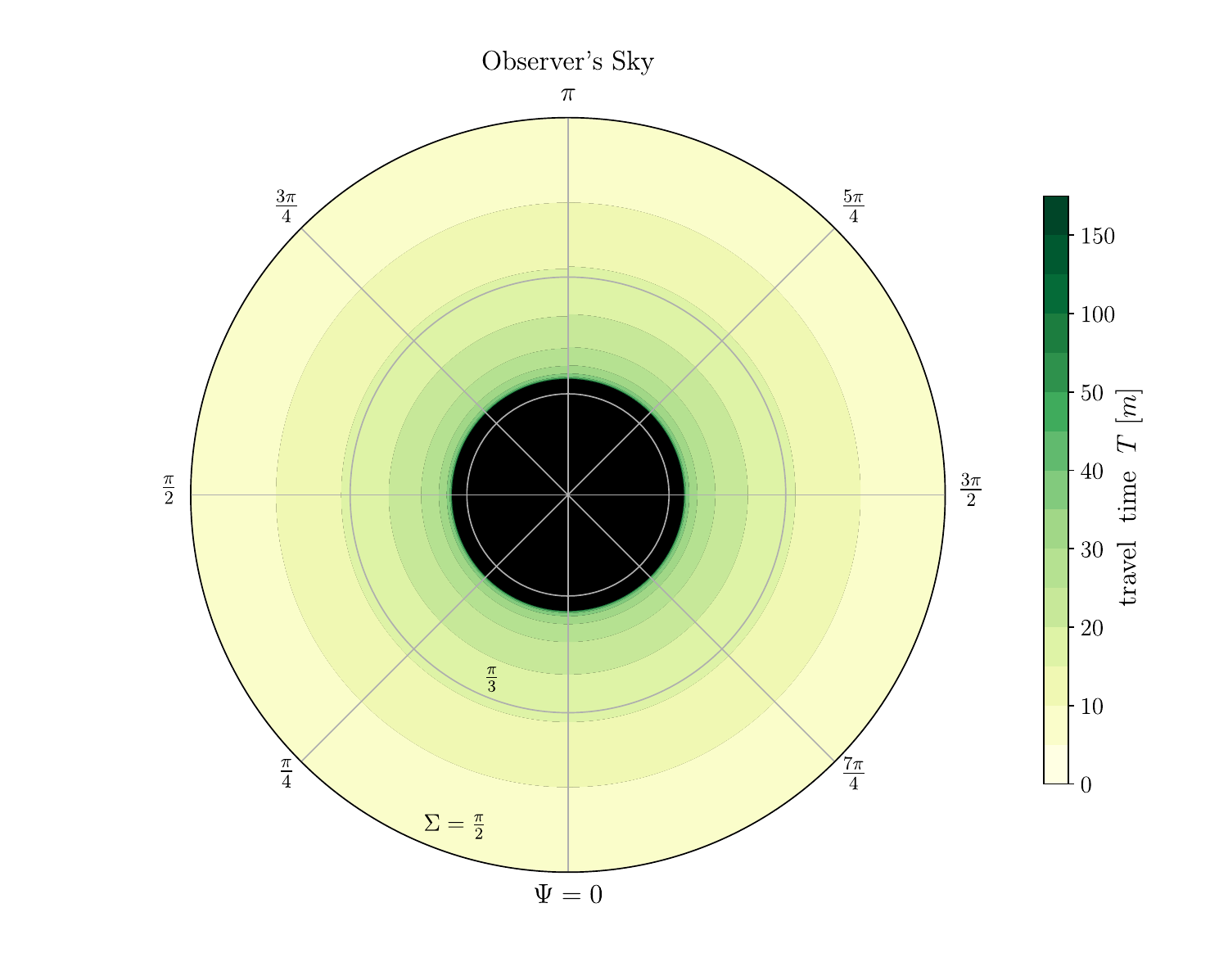} \\
   NUT Metric $n=m/10$ and $\vartheta_{O}=\pi/2$ & NUT Metric $n=m/2$ and $\vartheta_{O}=\pi/2$\\[6pt]
   \includegraphics[width=90mm]{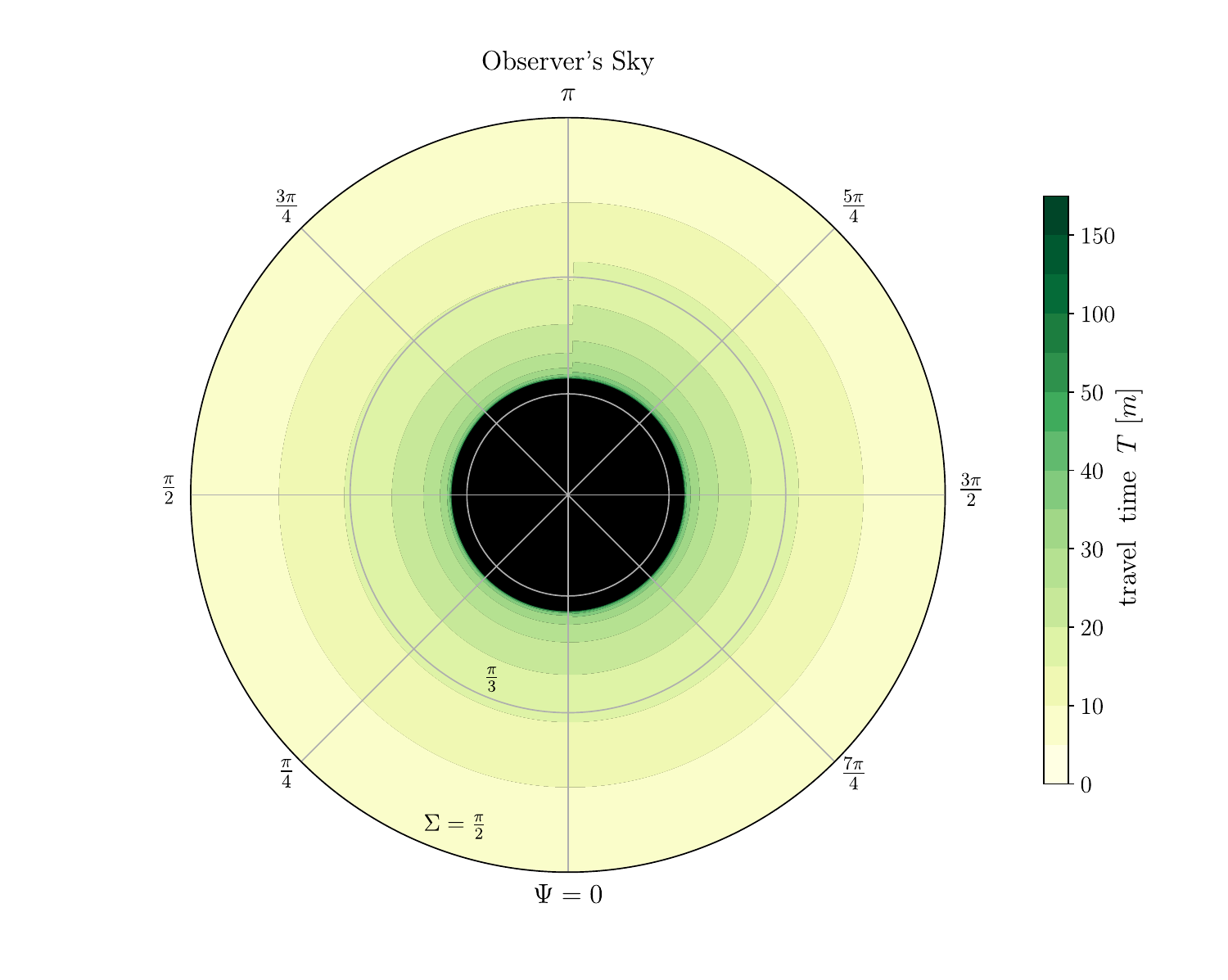} & \includegraphics[width=90mm]{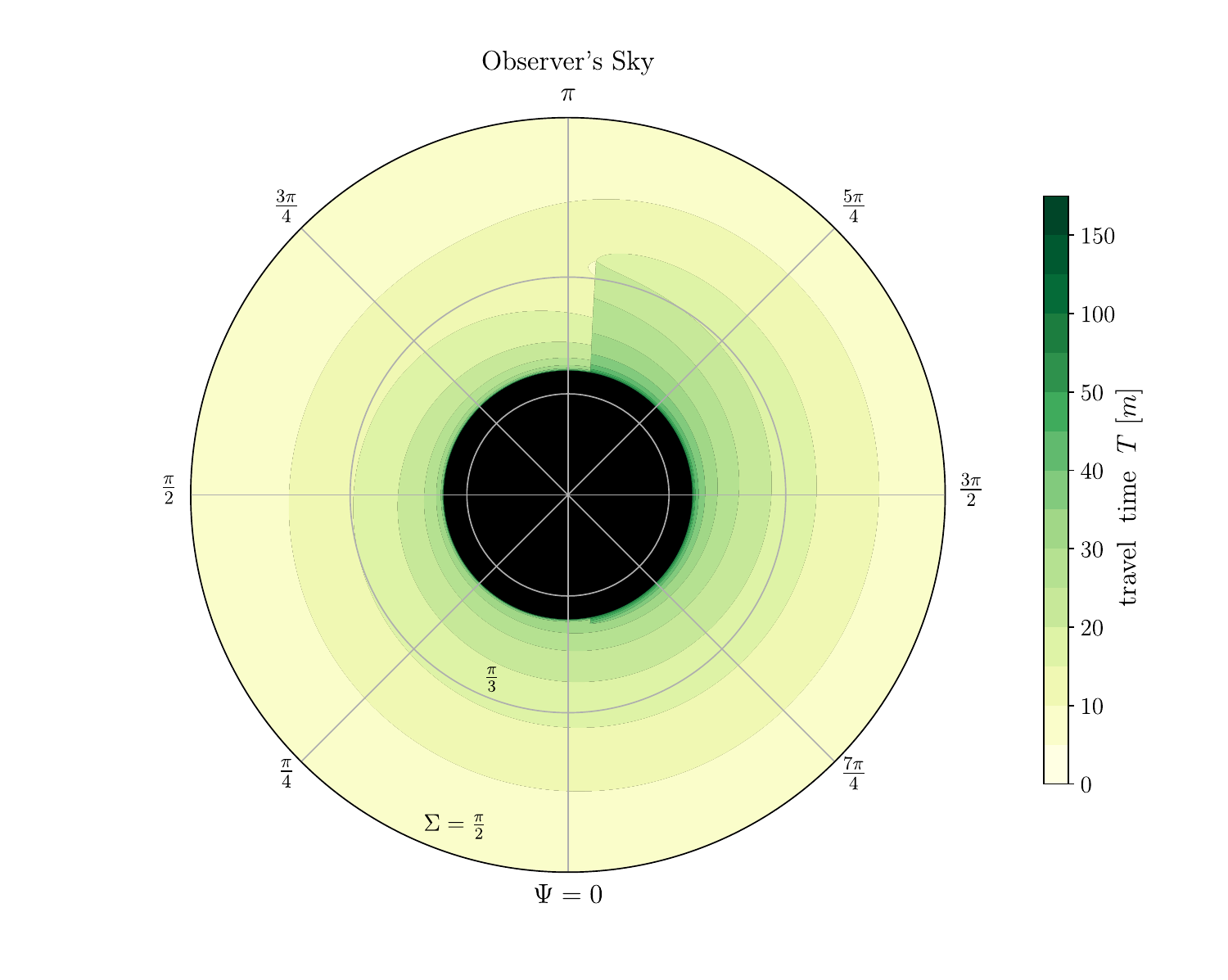}\\
   NUT Metric $n=m/2$ and $\vartheta_{O}=\pi/4$& NUT Metric $n=m/2$ and $\vartheta_{O}=3\pi/4$\\[6pt]
   \includegraphics[width=90mm]{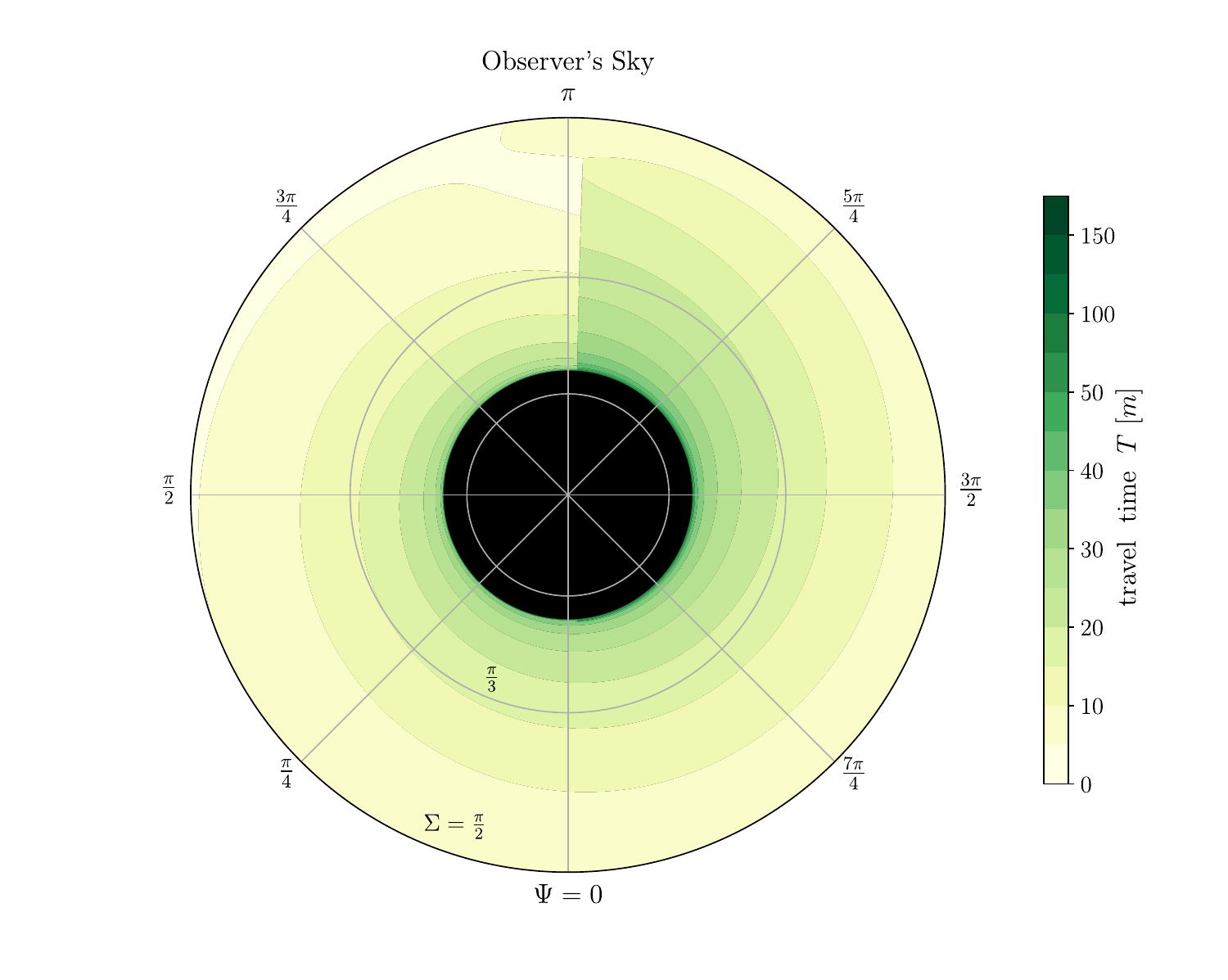} & \includegraphics[width=90mm]{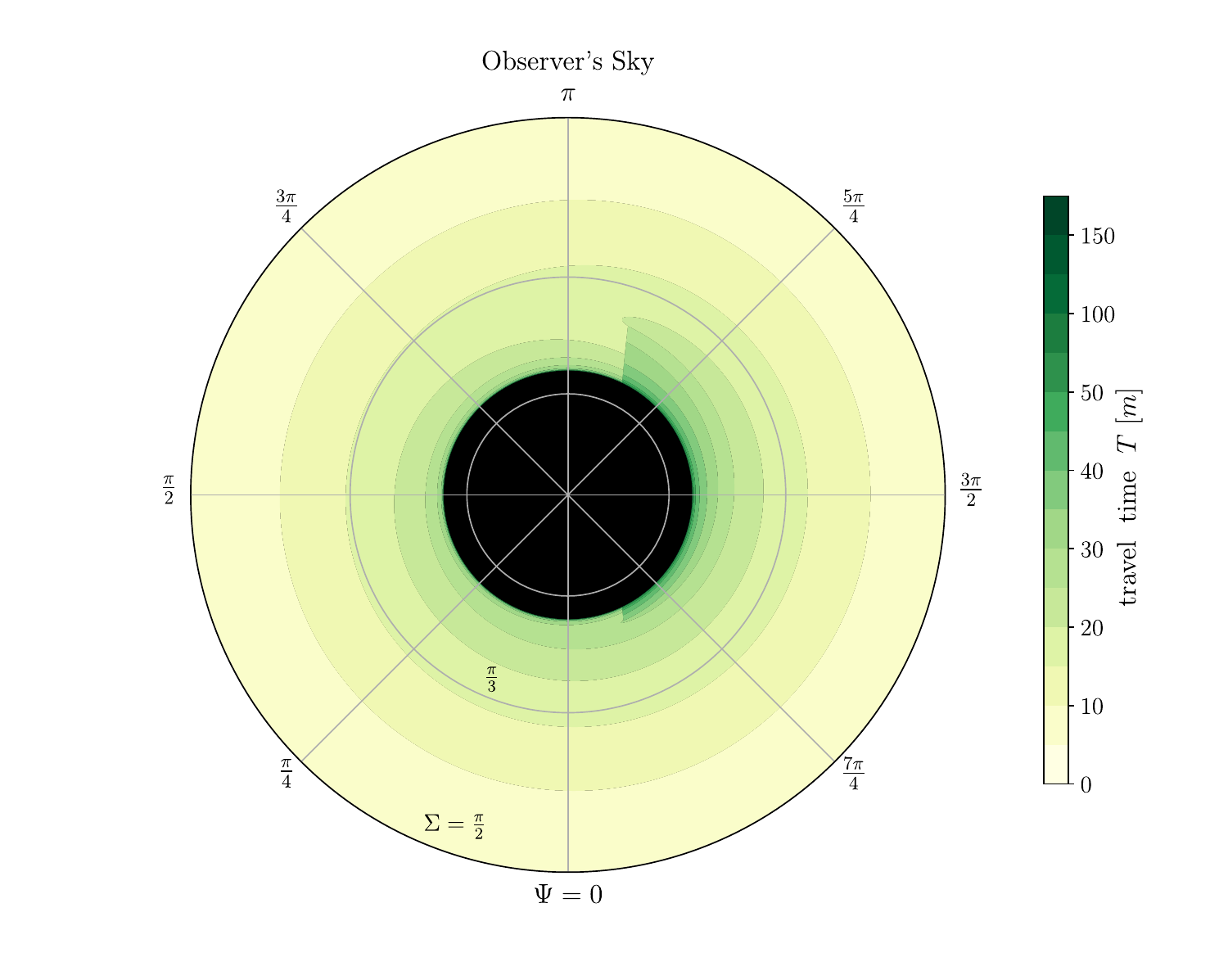}
\end{tabular}
\caption{Travel-time maps for light rays emitted by light sources located on the
two-sphere $S_{L}^{2}$ at the radius coordinate $r_{L}=9m$ and detected by an observer located
at $r_{O}=8m$ and $\vartheta_{O}=\pi/2$ in the Schwarzschild metric (top left) and
the NUT metric with $n=m/100$ (top right), $n=m/10$ (middle left) and $n=m/2$ (middle
right) and two observers located at $\vartheta_{O}=\pi/4$ (bottom left) and $\vartheta_{O}=3\pi/4$
(bottom right) in the NUT metric with $n=m/2$. The Misner string is located at $\vartheta=0$ ($C=1$).}
\end{figure*}
\begin{figure*}[h]\label{fig:TravelTime2}
  \includegraphics[width=90mm]{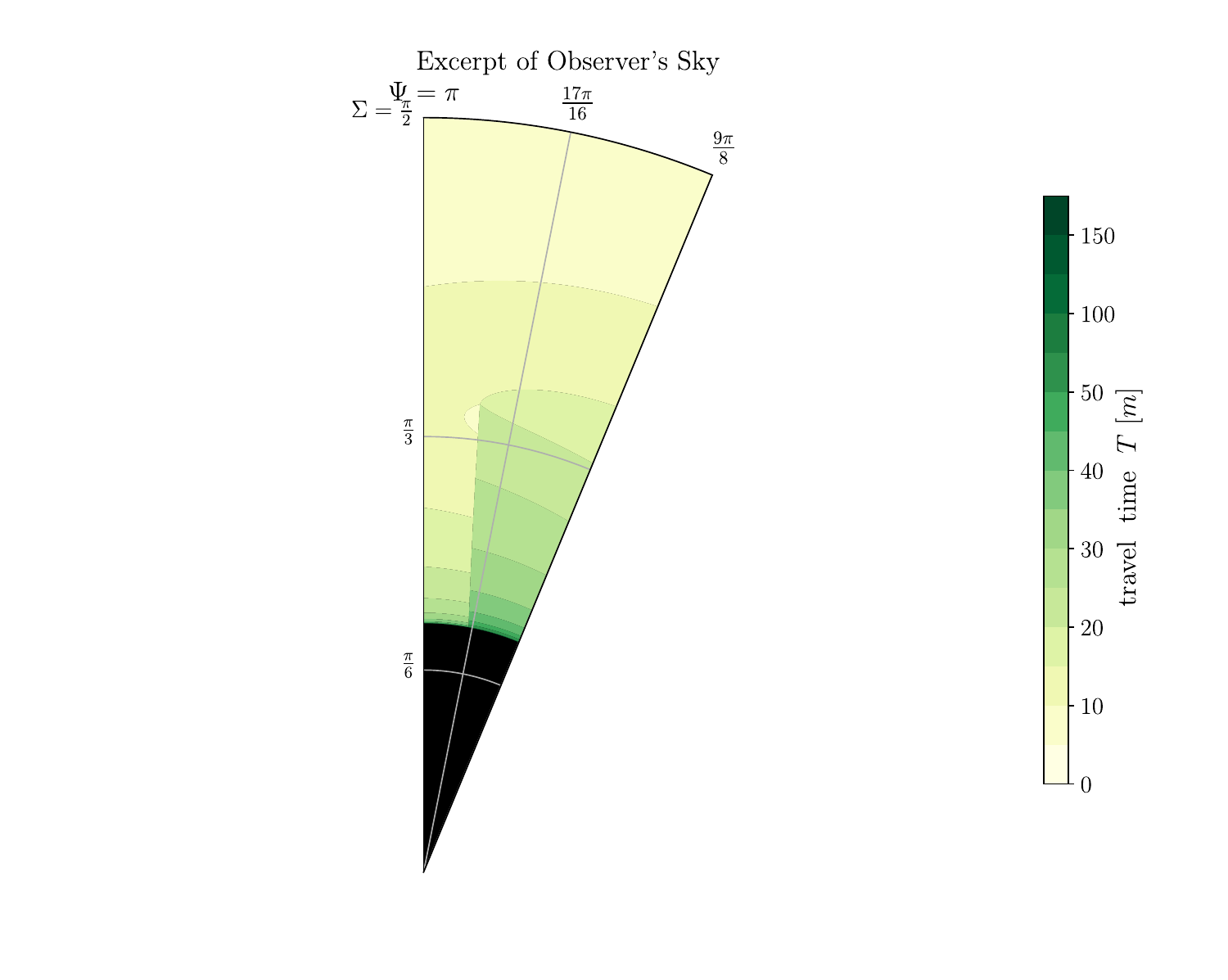} \\
  \caption{Enlarged view of the travel-time map in Fig.~12 (middle right) between $\Psi=\pi$
  and $\Psi=9\pi/8$ for light rays emitted by light sources located on the two-sphere
  $S_{L}^{2}$ at the radius coordinate $r_{L}=9m$ and detected by an observer located
  at $r_{O}=8m$, $\vartheta_{O}=\pi/2$ in the NUT metric with $n=m/2$. The Misner
  string is located at $\vartheta=0$ ($C=1$).}
\end{figure*}
\begin{figure*}\label{fig:TravelTime3}
\begin{tabular}{cc}
  Reissner-Nordstr\"{o}m Metric $\vartheta_{O}=\pi/2$& Charged NUT Metric $\vartheta_{O}=\pi/4$ \\[6pt]
  \includegraphics[width=90mm]{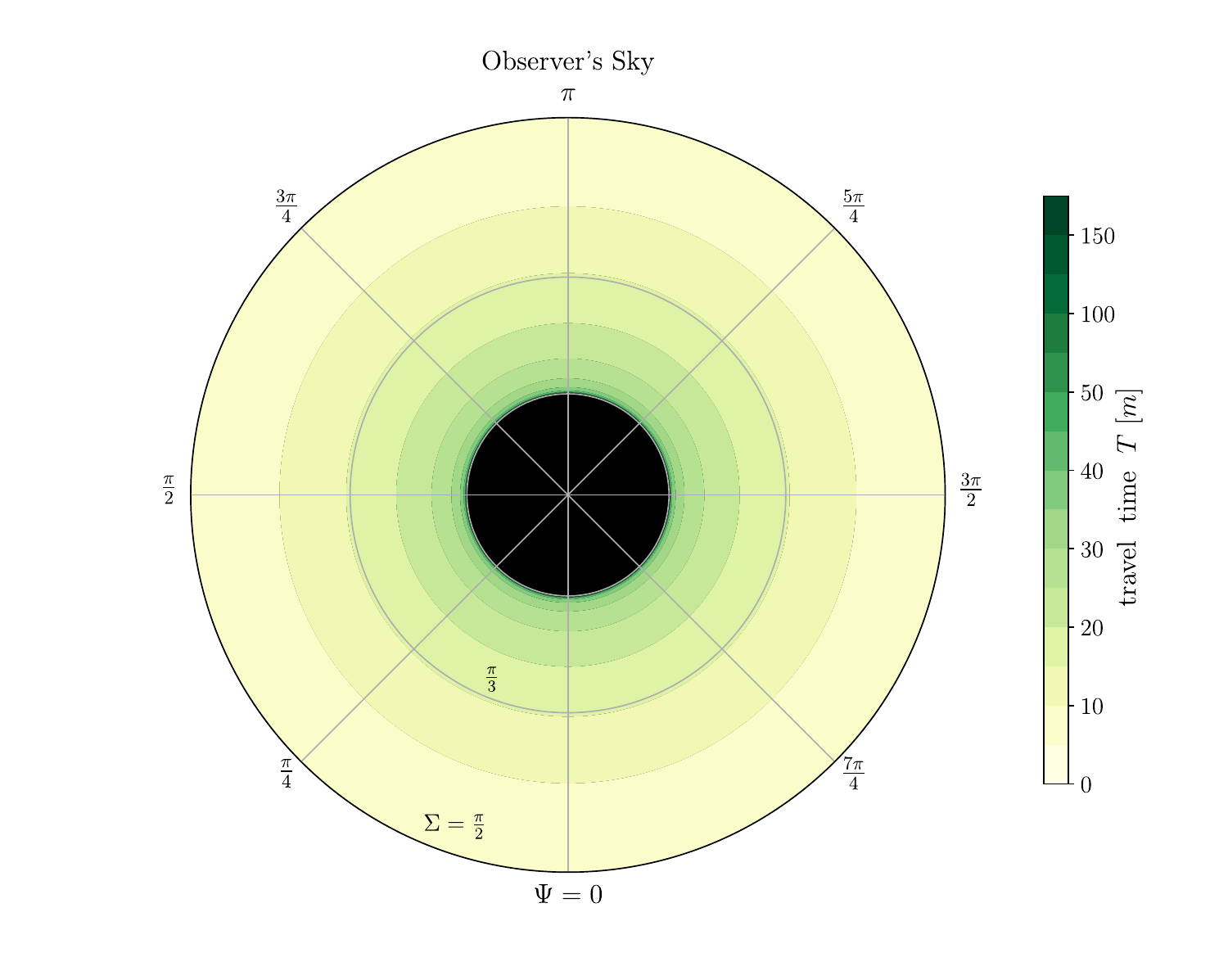} &   \includegraphics[width=90mm]{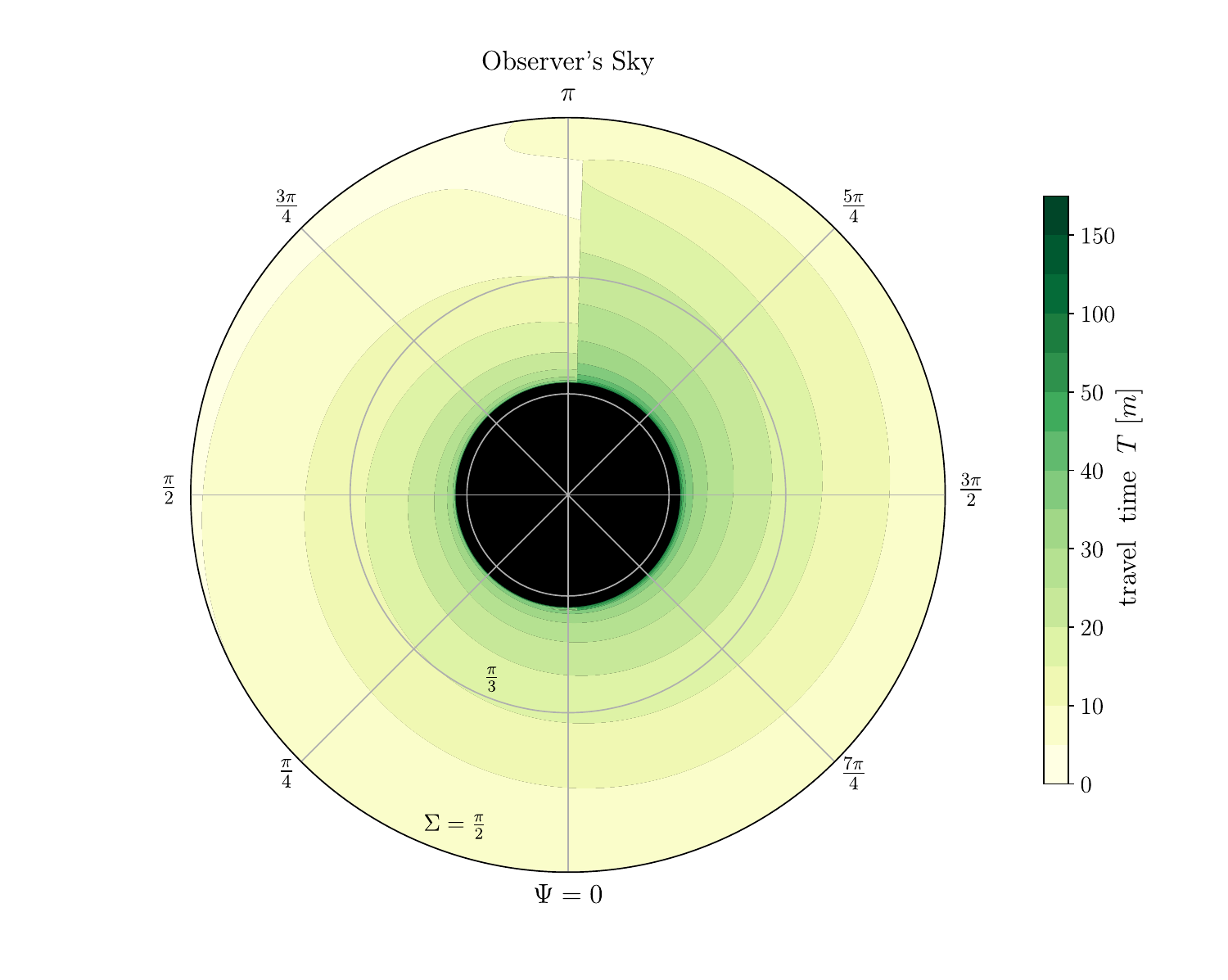} \\
  Charged NUT Metric $\vartheta_{O}=\pi/2$& Charged NUT Metric $\vartheta_{O}=3\pi/4$ \\[6pt]
  \includegraphics[width=90mm]{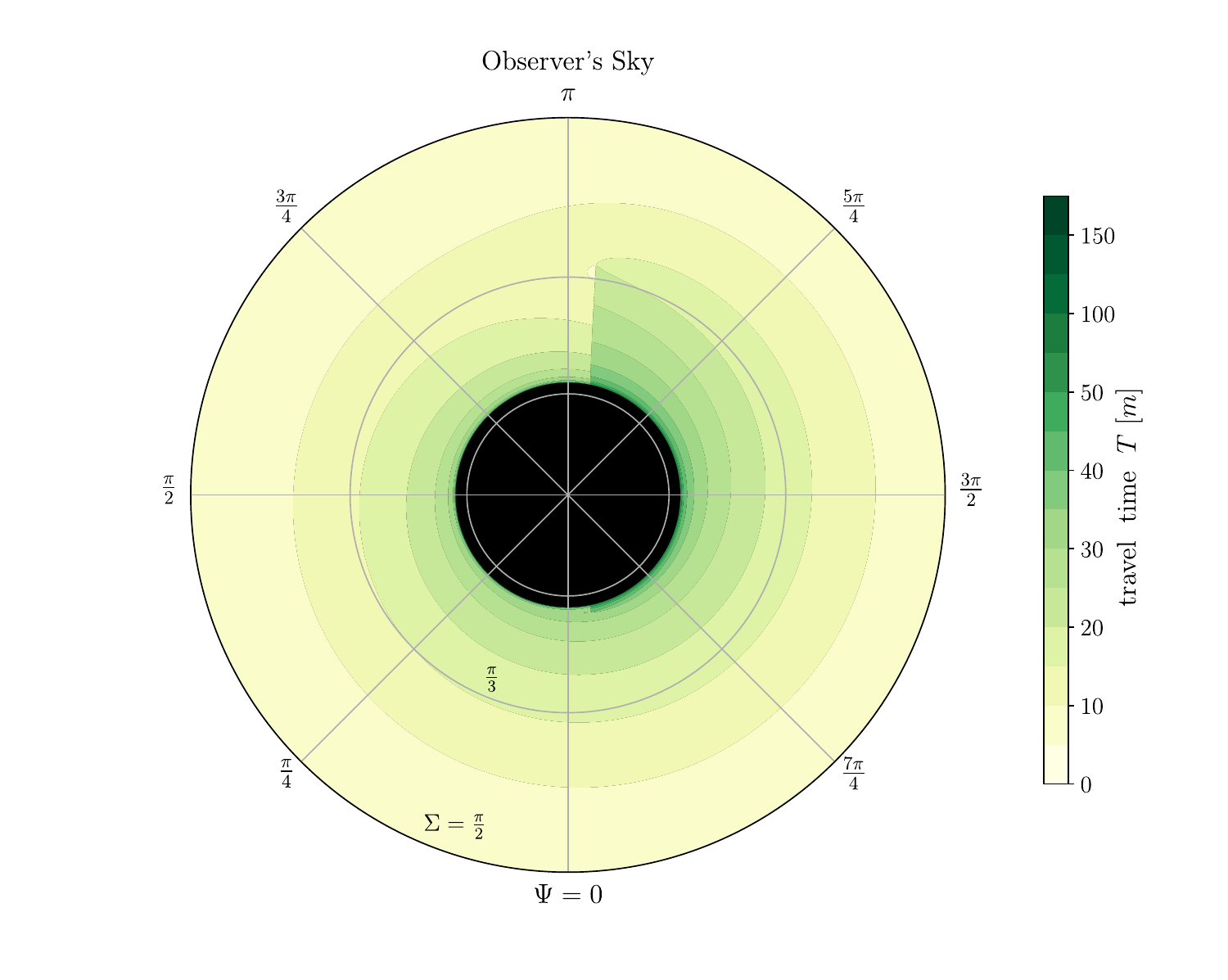} & \includegraphics[width=90mm]{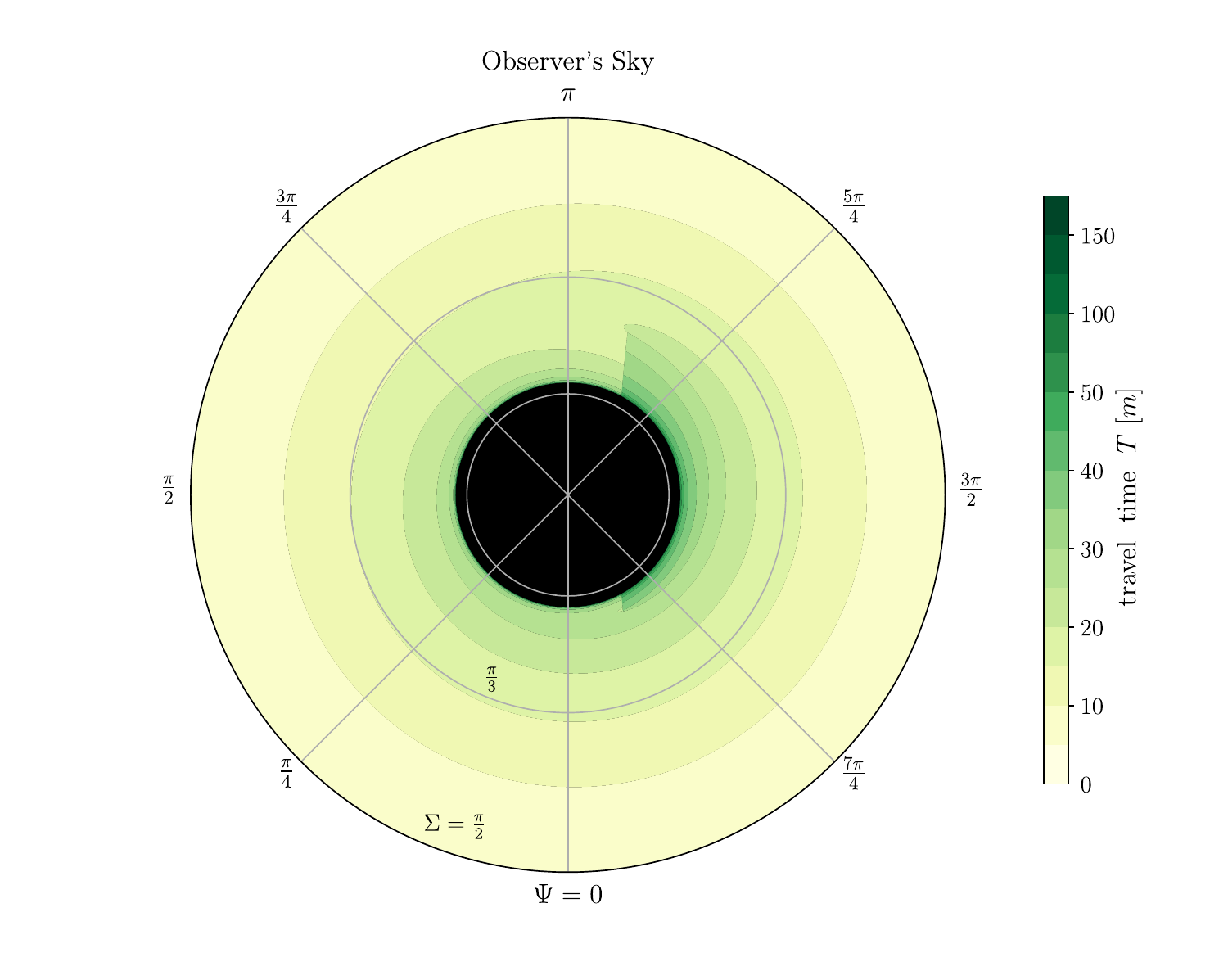}\\
\end{tabular}
\caption{Travel-time maps for light rays emitted by light sources located on the
two-sphere $S_{L}^{2}$ at the radius coordinate $r_{L}=9m$ and detected by an observer located at $r_{O}=8m$
and $\vartheta_{O}=\pi/2$ in the Reissner-Nordstr\"{o}m metric (top left),
$\vartheta_{O}=\pi/4$ (top right), $\vartheta_{O}=\pi/2$ (bottom left), and $\vartheta_{O}=3\pi/4$
(bottom right) in the charged NUT metric with $n=m/2$. The electric charge is $e=3m/4$.
The Misner string is located at $\vartheta=0$ ($C=1$).}
\end{figure*}
\begin{figure*}\label{fig:TravelTime4}
\begin{tabular}{cc}
  Schwarzschild-de Sitter Metric $\vartheta_{O}=\pi/2$& NUT-de Sitter Metric $\vartheta_{O}=\pi/4$ \\[6pt]
  \includegraphics[width=90mm]{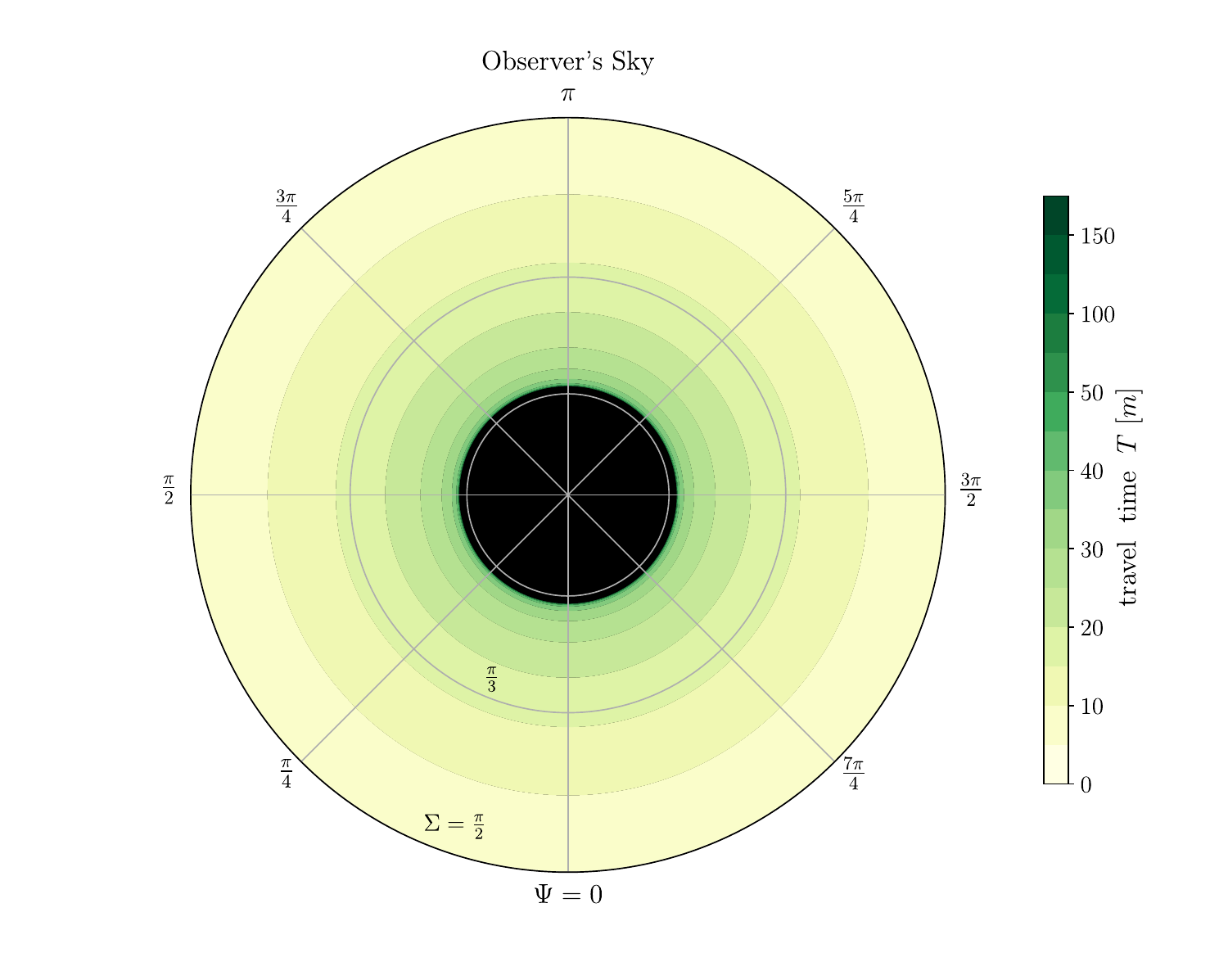} &   \includegraphics[width=90mm]{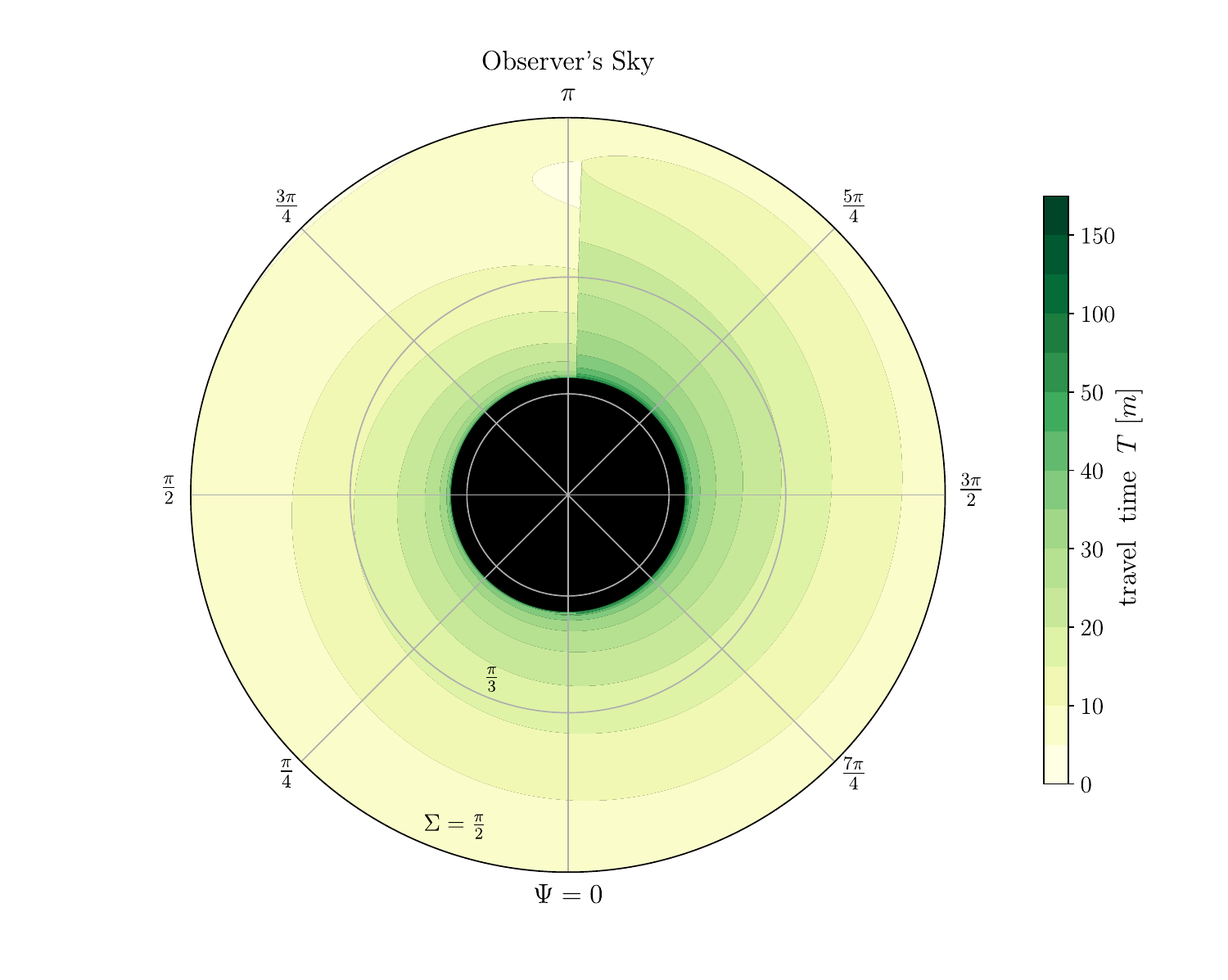} \\
  NUT-de Sitter Metric $\vartheta_{O}=\pi/2$& NUT-de Sitter Metric $\vartheta_{O}=3\pi/4$ \\[6pt]
  \includegraphics[width=90mm]{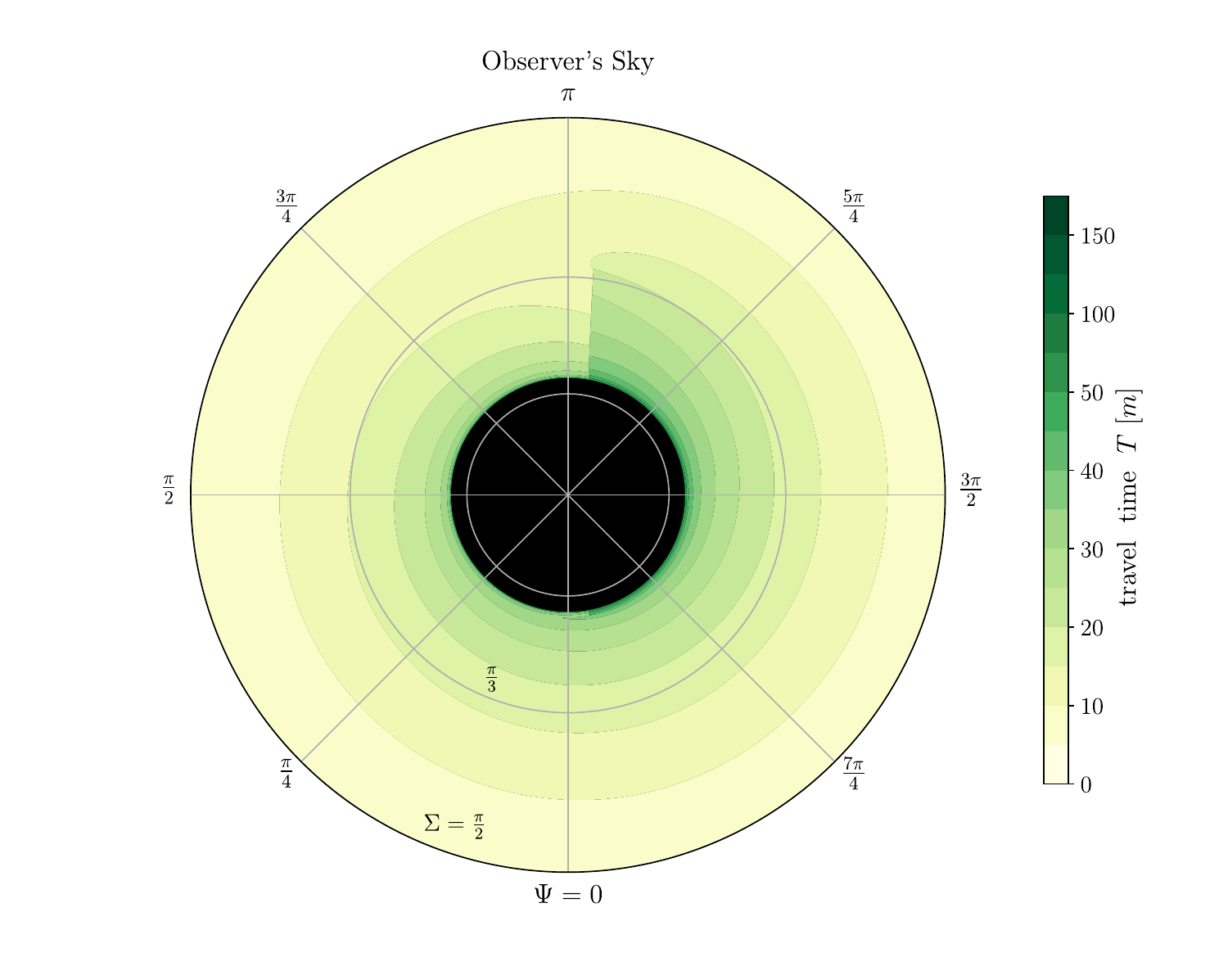} & \includegraphics[width=90mm]{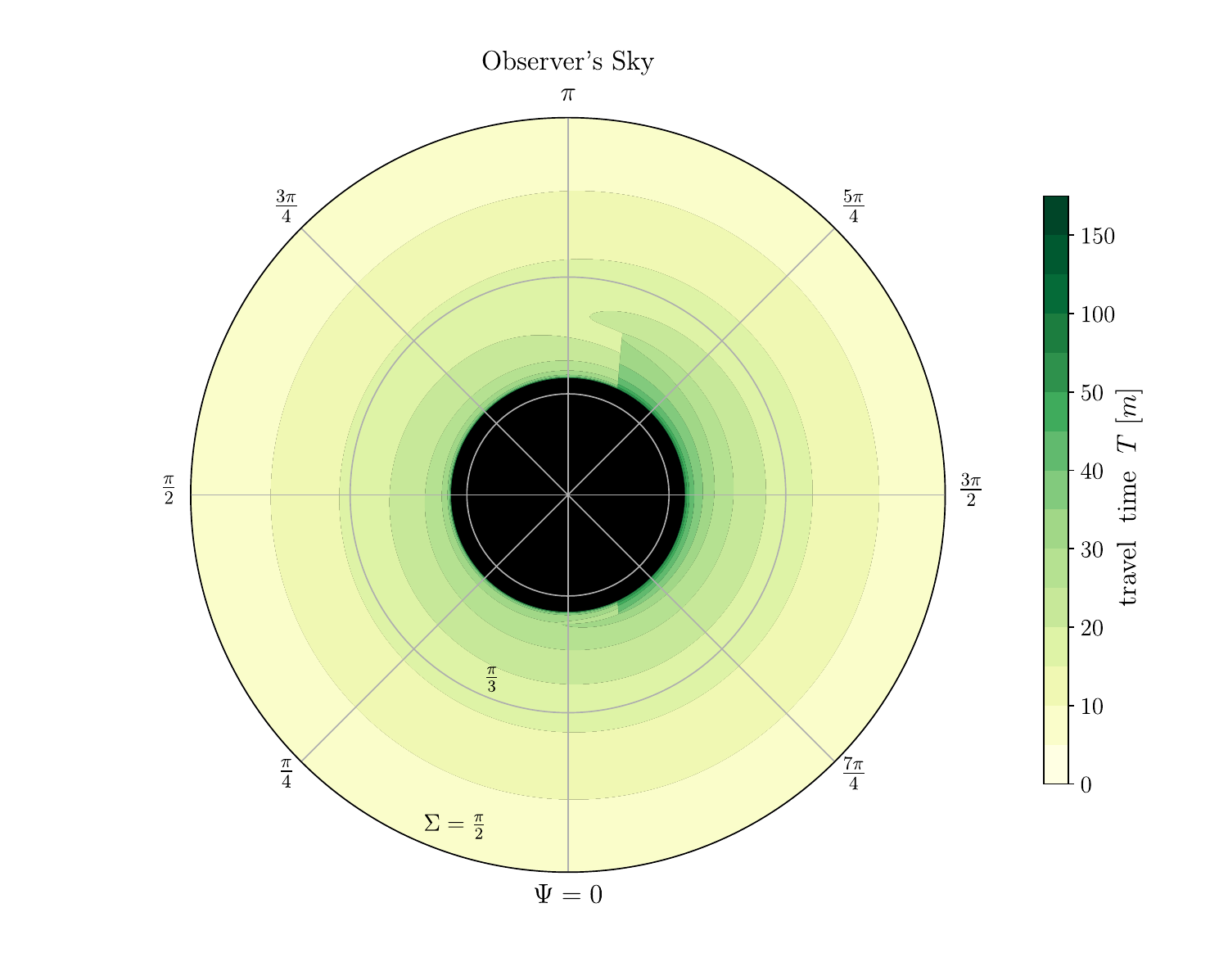}\\
\end{tabular}
\caption{Travel-time maps for light rays emitted by light sources located on the
two-sphere $S_{L}^{2}$ at the radius coordinate $r_{L}=9m$ and detected by an observer located at $r_{O}=8m$
and $\vartheta_{O}=\pi/2$ in the Schwarzschild--de Sitter metric (top left),
$\vartheta_{O}=\pi/4$ (top right), $\vartheta_{O}=\pi/2$ (bottom left), and $\vartheta_{O}=3\pi/4$
(bottom right) in the NUT--de Sitter metric with $n=m/2$. The cosmological constant is $\Lambda=1/(200m^2)$.
The Misner string is located at $\vartheta=0$ ($C=1$).}
\end{figure*}
\begin{figure*}\label{fig:TravelTime5}
\begin{tabular}{cc}
  Reissner-Nordstr\"{o}m-de Sitter Metric $\vartheta_{O}=\pi/2$& Charged NUT-de Sitter Metric $\vartheta_{O}=\pi/4$ \\[6pt]
  \includegraphics[width=90mm]{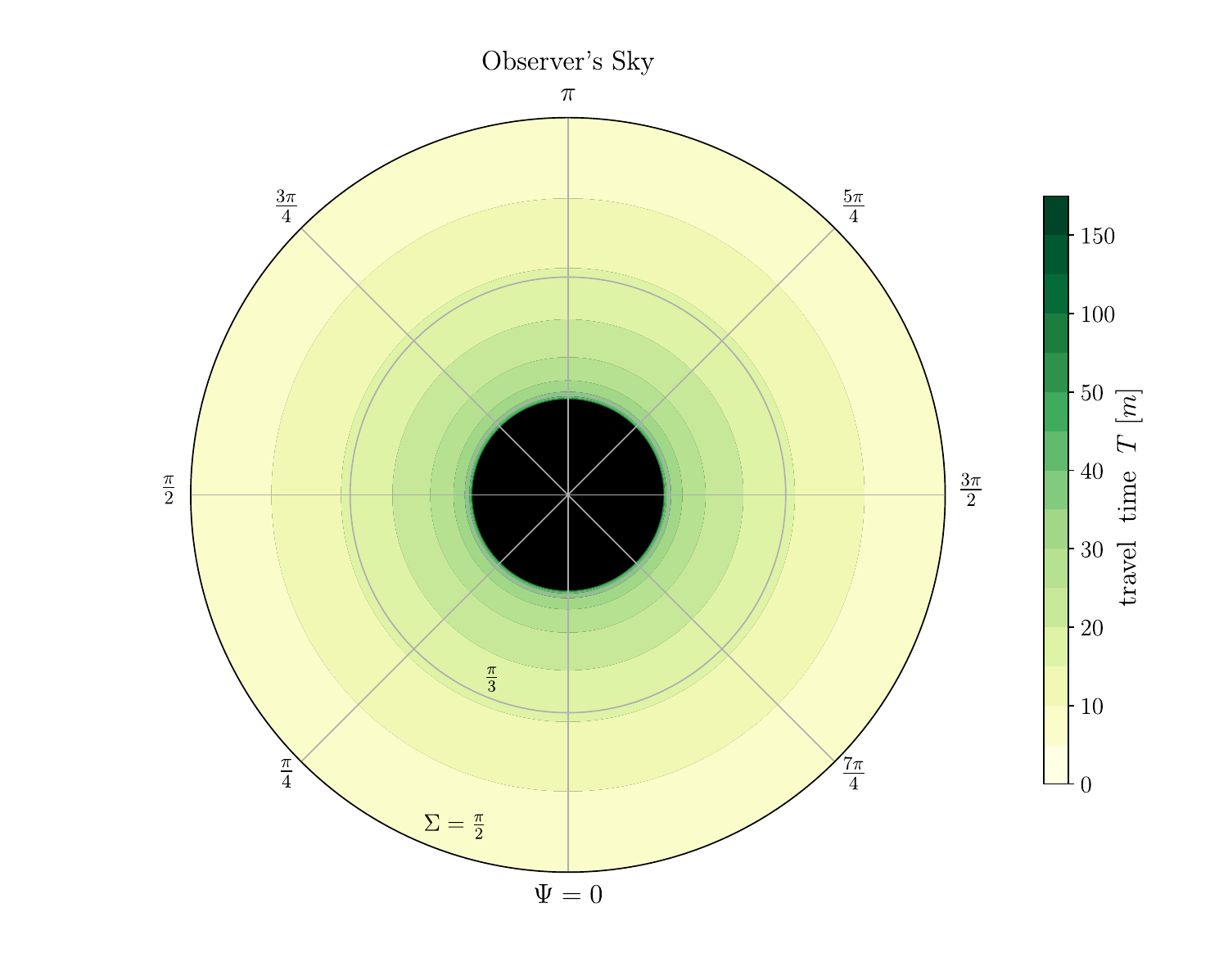} &   \includegraphics[width=90mm]{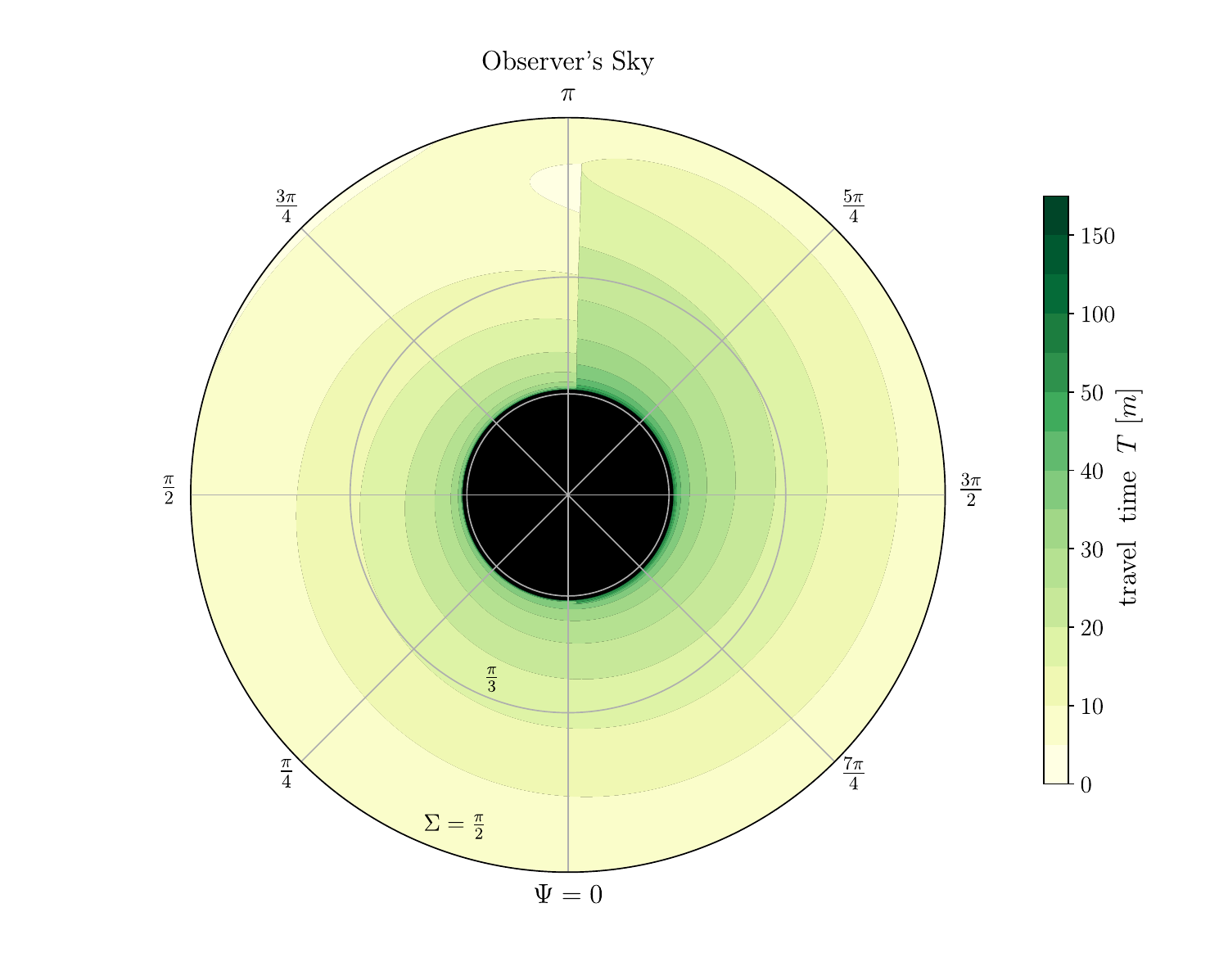} \\
  Charged NUT-de Sitter Metric $\vartheta_{O}=\pi/2$& Charged NUT-de Sitter Metric $\vartheta_{O}=3\pi/4$ \\[6pt]
  \includegraphics[width=90mm]{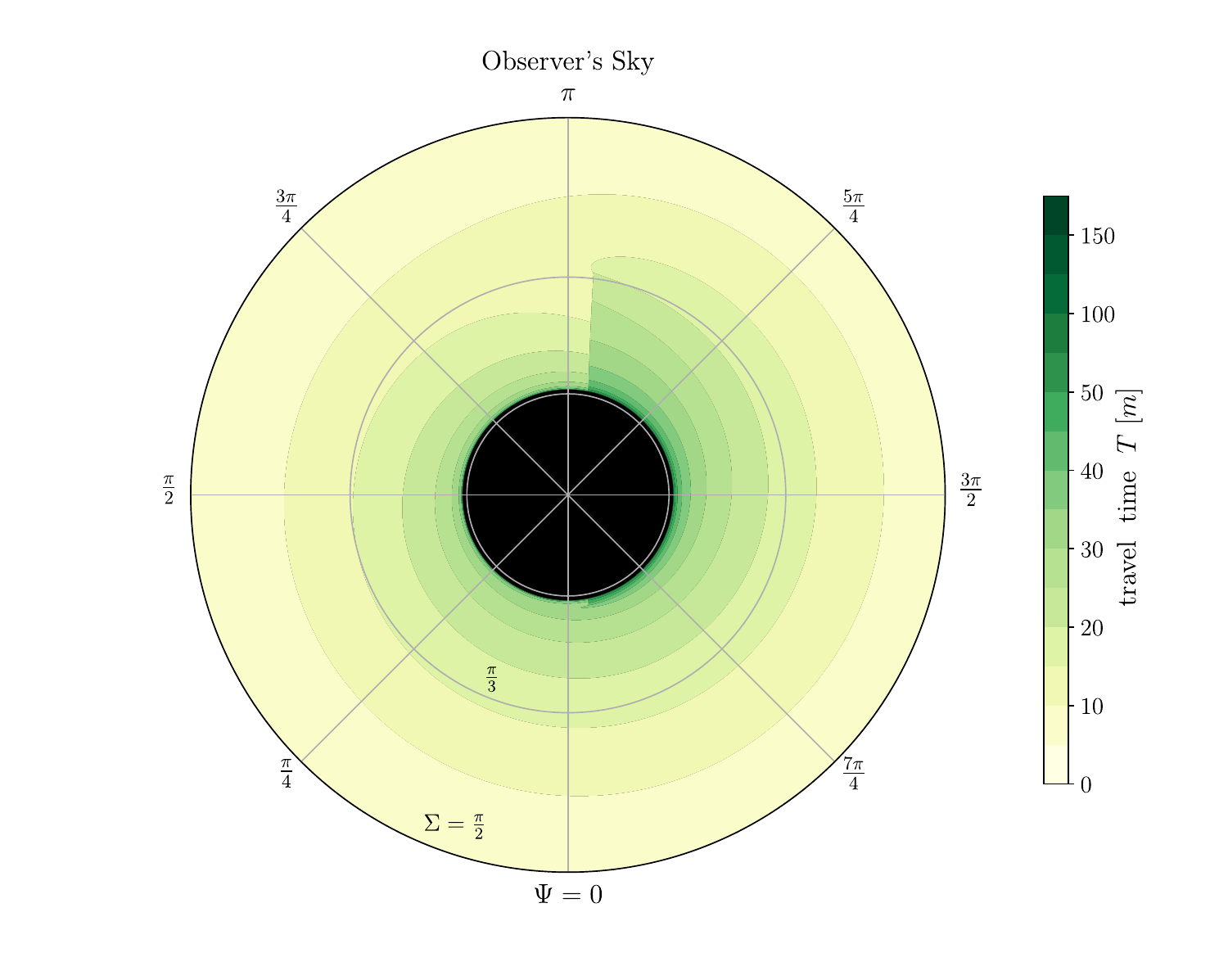} & \includegraphics[width=90mm]{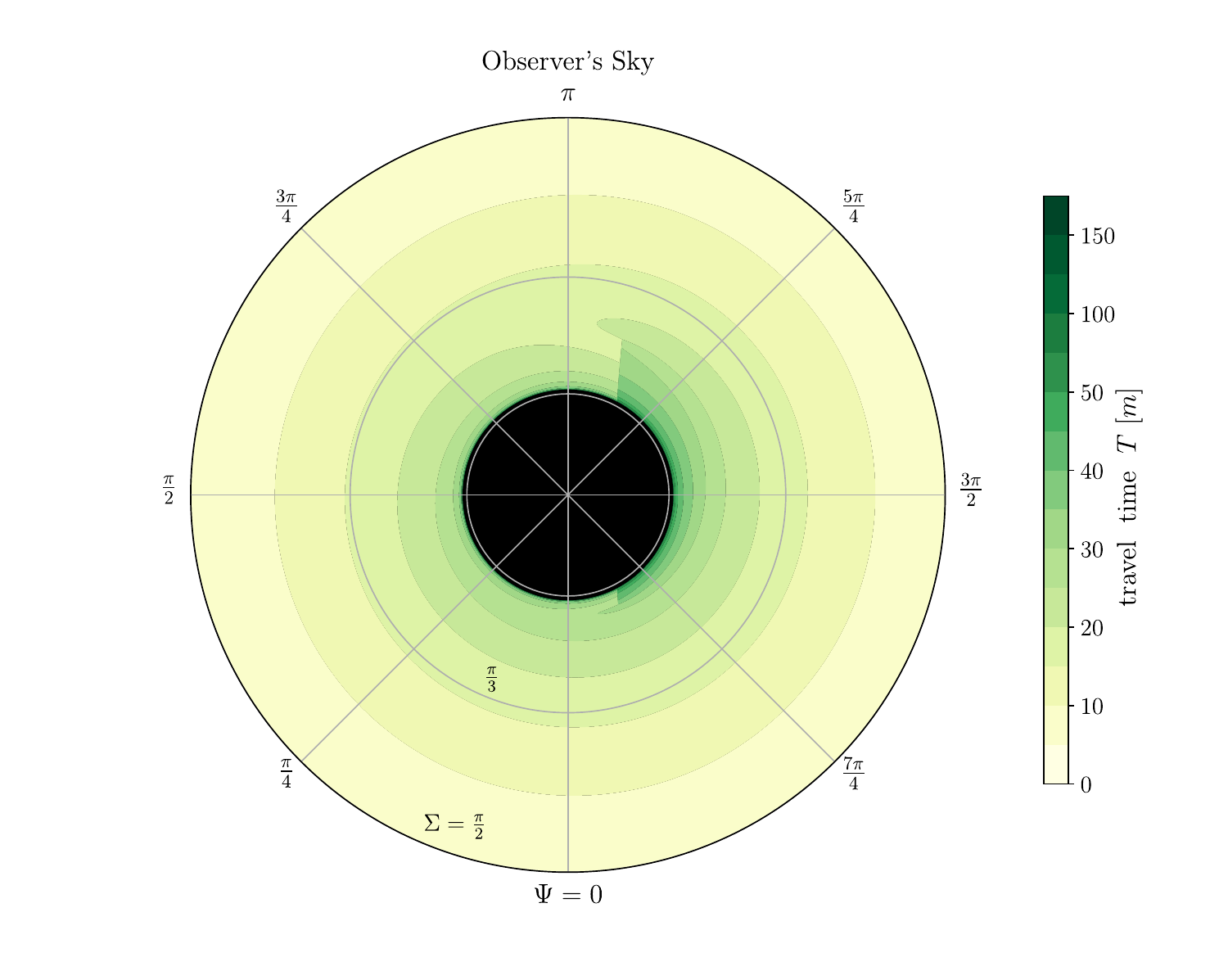}\\
\end{tabular}
\caption{Travel-time maps for light rays emitted by light sources located on the
two-sphere $S_{L}^{2}$ at the radius coordinate $r_{L}=9m$ and detected by an observer located at $r_{O}=8m$
and $\vartheta_{O}=\pi/2$ in the Reissner-Nordstr\"{o}m--de Sitter metric (top left),
$\vartheta_{O}=\pi/4$ (top right), $\vartheta_{O}=\pi/2$ (bottom left), and $\vartheta_{O}=3\pi/4$
(bottom right) in the charged NUT--de Sitter metric with $n=m/2$. The cosmological constant and the electric charge are $\Lambda=1/(200m^2)$
and $e=3m/4$, respectively. The Misner string is located at $\vartheta=0$ ($C=1$).}
\end{figure*}

\section{Summary and Conclusions}\label{Sec:Summary}
In this paper we first discussed and solved the equations of motion in the domain
of outer communication of the charged NUT--de Sitter metrics using Legendre's canonical
forms of the elliptic integrals and Jacobi's elliptic functions. While for $\vartheta$
and $\varphi$ our results are not particularly new we believe that our representation
makes them easily accessible without any further rescalings; see, e.g., Kagramanova \emph{et al.} \cite{Kagramanova2010},
or conventions using Killing vector fields, see, e.g., Cl\'{e}ment \emph{et al.} \cite{Clement2015}. It is true that we can also use Weierstrass' elliptic
$\wp$ function and Weierstrass' $\zeta$ and $\sigma$ functions to solve the equations
of motion for $r$ and $t$; see Kagramanova \emph{et al.} \cite{Kagramanova2010}. However, using
Legendre's canonical form of the elliptic integrals has the clear advantage that
we do not have to consider and manually adjust the branches of the $ln$
that occur in the equations for $t$ in Kagramanova \emph{et al.} \cite{Kagramanova2010}.
Along the way we also derived and discussed the properties of the photon
sphere and the \emph{individual photon cones}. The radius coordinate of the photon
sphere in the NUT metric was already well known for quite some time; see Jefremov
and Perlick \cite{Jefremov2016}. For the charged NUT--de Sitter metrics it is also
included as special case in the results of Grenzebach \emph{et al.}
\cite{Grenzebach2014}. However, we believe that the approach to derive it using
the potential $V_{r}(r)$ makes it particularly easy to access and to understand
the related classification of the different types of lightlike geodesic motion. \\
In the second part of the paper we employed the derived solutions to the equations of motion to thoroughly
investigate gravitational lensing in the charged NUT--de Sitter spacetimes. For this
purpose we introduced a stationary observer at the radius coordinate $r_{O}$ and a two-sphere $S_{L}^{2}$ of light sources at the radius coordinate $r_{L}$, both measured
in units of $m$, in the domain of outer communication between photon sphere and
infinity ($\Lambda=0$) or the cosmological horizon ($0<\Lambda<\Lambda_{\mathrm{C}}$). We introduced
an orthonormal tetrad to parametrize the constants of motion using latitude-longitude
coordinates on the observer's celestial sphere following the approach of Grenzebach \emph{et al.} \cite{Grenzebach2015a}. In this parametrization we derived
the angular radius of the shadow, set up a lens equation, defined the redshift, and
the travel time. \\
For the charged NUT--de Sitter metrics we found that the shadow is always circular.
Although the charged NUT--de Sitter metrics are only axisymmetric this result is not
really surprising because the spatial component of the metrics maintains
a rotational $SO(3,\mathbb{R})$ symmetry. The angular radius of the shadow is a
function of the gravitomagnetic charge $n$ and, for a fixed $r_{O}$, grows when we
increase the gravitomagnetic charge. Unfortunately, as long as we do not know $r_{O}$,
$\Lambda$ and $e$, for the latter two the shadow shrinks compared to the NUT metric
as soon as we turn them on, we have a degeneracy with respect to $\Lambda$, $e$, $n$
and also $r_{O}$.\\
As first main result of this paper we wrote down an exact lens equation for the
charged NUT--de Sitter metrics. Here, we have to stress that we did not derive it
using numerical ray tracing but the exact analytic solutions to the equations of
motion. The lens map shows images up to fourth order. We found that unlike in static
and spherically symmetric spacetimes the images of first and second orders from
the same quadrant on the two-sphere $S_{L}^{2}$ connect and are twisted. In addition
we found two regions with images of first order. The first region appears relatively
far away from the shadow while the second region appears very close to
the shadow. In the second region the direction of the $\varphi$ motion reverses and thus lightlike geodesics
do not perform a full orbit about the axes $\vartheta=0$ or $\vartheta=\pi$. The
images of first and second order are separated by very clean-cut lines which mark
lightlike geodesics crossing the axes. We found that for these geodesics all three
spatial coordinates are regular confirming the results of Cl\'{e}ment \emph{et al.} \cite{Clement2015}. In addition we found that when we turn on the
cosmological constant $\Lambda$ and the electric charge $e$ the lens map maintains
its basic structure.\\
We also discussed the potential location of the critical curves. We argued that
it is unlikely that the boundaries between images of different orders from the same
region on the two-sphere $S_{L}^{2}$ are part of the critical curves because they
immediately occur when we turn on the gravitomagnetic charge. We came to the
conclusion that it is very likely that the critical curves still form circles because
(i) the spacetime maintains the spatial rotational symmetry of the static and spherically
symmetric spacetimes and (ii) the boundary between images of first and second order
and images of third and fourth order are still circles. However, for confirming
our claims and for finding the exact position of the critical curves we need a
much more detailed and thorough investigation of lightlike geodesic motion
in the charged NUT--de Sitter metrics, in particular the Jacobian of the lens equation, which was beyond
the scope of this paper.\\
We also derived the redshift and plotted it as function of $n$ for observer constellations
$r_{O}<r_{L}$ and $r_{L}<r_{O}$. For the former we mainly observed blueshifts while
for the latter we mainly observed redshifts. We found that for these observers the observed blueshift and the observed redshift
of light rays emitted by a light source located at $r_{L}$ increase with growing
gravitomagnetic charge $n$, respectively. Adding the electric charge has only a very small effect
while adding the cosmological constant $\Lambda$ shifts the limits $z\rightarrow -1$ and $z\rightarrow \infty$
to much lower $n$. \\
As the second main result of this paper we derived the travel time $T(\Sigma,\Psi)$
and plotted it as a function of $\Sigma$ and $\Psi$ on the observer's celestial
sphere. When we compared the travel-time maps of the charged NUT--de Sitter metrics
to their spherically symmetric and static counterparts two very distinct differences
immediately caught our eye. First, for the charged NUT--de Sitter metrics the travel
time shows a discontinuity when light rays cross the Misner string at least once
(in our case we have $C=1$ and thus it is located at $\vartheta=0$). In addition,
when we go from the first crossing in clockwise direction along a constant latitude
$\Sigma$ the travel time decreases, resulting in a spiral pattern. In addition we
found that turning on the electric charge $e$ did not significantly affect the travel
time. However, in the presence of a positive cosmological constant the travel time gets significantly
longer.\\
From the astrophysical point of view it is unfortunate that the shadow and the redshift
factor $z$ are degenerate with respect to $\Lambda$, $e$, $n$, $r_{O}$, and $r_{L}$
(the latter is only true for the redshift factor $z$). However, the lens equation
and the travel time very beautifully demonstrate that the presence of a gravitomagnetic
charge is always connected with a twist in the lens map and a discontinuity in the
travel time. The former was already observed in the weak-field limit by Nouri-Zonoz
and Lynden-Bell \cite{NouriZonoz1997,LyndenBell1998} and our results confirm it
for the exact lens map. The discontinuity in the travel time, and as a consequence
of the time coordinate, confirms Misner's conclusion that the time coordinate has
a singularity at the Misner string \cite{Misner1963}. \\
The twist and the discontinuity of the travel time are unique features caused by
the gravitomagnetic charge $n$ and therefore if they are observed for a black hole
they will be strong indicators for the presence of a gravitomagnetic charge. In addition
the strength of the twist and the discontinuity will also allow to draw conclusions
on the magnitude of the gravitomagnetic charge $n$. However, for lifting the degeneracy
with respect to the cosmological constant $\Lambda$, the electric charge $e$, the
distances between observer and black hole and light source and black hole, $r_{O}$
and $r_{L}$, we have to combine observations of the angular radius of the shadow,
the redshift, the positions of multiple images of the same light source on the observer's celestial sphere and the travel-time differences between these images.

\section*{ACKNOWLEDGMENTS}
I would like to thank Volker Perlick for the helpful discussions. I acknowledge
financial support from the Cluster of Excellence QuantumFrontiers. I also acknowledge support from Deutsche
Forschungsgemeinschaft within the Research Training Group 1620 Models of Gravity.
I also would like to express my gratitude to all contributors to the JULIA project
and in particular the authors of the packages ELLIPTIC, BLOSC, HDF5 and PYPLOT.

\appendix
\section{ELEMENTARY AND ELLIPTIC INTEGRALS}\label{Sec:Integrals}
While integrating the equations of motion for $r$ in Sec.~\ref{Sec:EoMr} and
the time coordinate $t$ in Sec.~\ref{Sec:EoMt} we encountered several elementary
and elliptic integrals. In this appendix we will demonstrate how to calculate them.
\subsection{Elementary Integrals}\label{Sec:EMIntegrals}
We start with the elementary integrals required to calculate the solutions for $r(\lambda)$
in Sec.~\ref{Sec:Solr} and the $r$-dependent part $t_{r}(\lambda)$ of the time
coordinate $t$ in Sec.~\ref{Sec:tr} for cases 3 and 5.
\subsubsection{$r$ motion and time coordinate $t$: Case 3}\label{Sec:EMInt1}
In Secs.~\ref{Sec:Solr} and \ref{Sec:tr} we encountered in total five
different elementary integrals associated with the geodesic motion of light rays
with $E^2/K=V_{r}(r_{\mathrm{ph}-})$. These geodesics have a double root at $r_{1}=r_{2}=r_{\mathrm{ph-}}$
and a pair of complex conjugate roots at $r_{3}=\bar{r}_{4}=R_{3}+iR_{4}$. The first
two integrals $I_{1}$ and $I_{2}$ are given by Eqs.~(\ref{eq:EMInt1}) and (\ref{eq:EMInt2})
and are easy to calculate
\begin{eqnarray}\label{eq:EMInt1}
I_{1}=\int\frac{r\mathrm{d}r}{\sqrt{(R_{3}-r)^2+R_{4}^2}}=\sqrt{(R_{3}-r)^2+R_{4}^2}+R_{3}\mathrm{arsinh}\left(\frac{r-R_{3}}{R_{4}}\right),
\end{eqnarray}
\begin{eqnarray}\label{eq:EMInt2}
I_{2}=\int\frac{\mathrm{d}r}{\sqrt{(R_{3}-r)^2+R_{4}^2}}=\mathrm{arsinh}\left(\frac{r-R_{3}}{R_{4}}\right).
\end{eqnarray}
The other three integrals $I_{3}$, $I_{4}$ and $I_{5}$ are given by Eqs.~(\ref{eq:EMInt3})--(\ref{eq:EMInt5}).
In $I_{3}$ and $I_{4}$ we always have $a<r$. Here $a$ can take the values $r_{1}$,
$r_{\mathrm{C}-}$, $r_{\mathrm{H},\mathrm{i}}$ or $r_{\mathrm{H},\mathrm{o}}$. In $I_{5}$
on the other hand we only have $a=r_{\mathrm{C}+}$ and thus $r<a$. Now we substitute $x=r-a$
in $I_{3}$ and $I_{4}$ and $x=a-r$ in $I_{5}$ and integrate. After integration and
resubstitution $I_{3}$, $I_{4}$ and $I_{5}$ read
\begin{eqnarray}\label{eq:EMInt3}
I_{3}=\int\frac{\mathrm{d}r}{(r-a)\sqrt{(R_{3}-r)^2+R_{4}^2}}=-\frac{1}{\sqrt{(R_{3}-a)^2+R_{4}^2}}\mathrm{arsinh}\left(\frac{(a-R_{3})(r-a)+(R_{3}-a)^2+R_{4}^2}{(r-a)R_{4}}\right),
\end{eqnarray}
\begin{eqnarray}\label{eq:EMInt4}
&I_{4}=\int\frac{\mathrm{d}r}{(r-a)^2\sqrt{(R_{3}-r)^2+R_{4}^2}}=-\frac{\sqrt{(R_{3}-r)^2+R_{4}^2}}{((R_{3}-a)^2+R_{4}^2)(r-a)}\\ \nonumber
&+\frac{a-R_{3}}{\left((R_{3}-a)^2+R_{4}^2\right)^{\frac{3}{2}}}\mathrm{arsinh}\left(\frac{(a-R_{3})(r-a)+(R_{3}-a)^2+R_{4}^2}{(r-a)R_{4}}\right),
\end{eqnarray}
\begin{eqnarray}\label{eq:EMInt5}
I_{5}=\int\frac{\mathrm{d}r}{(a-r)\sqrt{(R_{3}-r)^2+R_{4}^2}}=\frac{1}{\sqrt{(R_{3}-a)^2+R_{4}^2}}\mathrm{arsinh}\left(\frac{(R_{3}-a)^2+R_{4}^2-(a-R_{3})(a-r)}{(a-r)R_{4}}\right).
\end{eqnarray}
\subsubsection{$r$ motion and time coordinate $t$: Case 5}\label{Sec:EMInt2}
In addition to the five integrals discussed in the last subsection of this appendix
in Sec.~\ref{Sec:Solr} and Sec.~\ref{Sec:tr} we also encountered four elementary
integrals associated with lightlike geodesics asymptotically coming from or going
to the photon sphere. In their most general form these integrals are given by $I_{6}$--$I_{9}$ [Eqs.~(\ref{eq:EMInty1})--(\ref{eq:EMInty4})].
In $I_{6}$ and $I_{7}$ we always have $y>a$, where $a$ is either $y_{\mathrm{ph}}$,
$y_{\mathrm{C}-}$, $y_{\mathrm{C}+}$ or $a_{2,r}/12$. $y_{\mathrm{ph}}$, $y_{\mathrm{C}-}$ and $y_{\mathrm{C}+}$ are related to $r_{\mathrm{ph}}$,
$r_{\mathrm{C}-}$ and $r_{\mathrm{C}+}$ via Eq.~(\ref{eq:Sub3}), respectively. Now we substitute
$z=y-a$ and integrate. After integration and resubstitution $I_{6}$ and $I_{7}$
read
\begin{eqnarray}\label{eq:EMInty1}
I_{6}=\int\frac{\mathrm{d}y}{(y-a)\sqrt{y-y_{1}}}=-\frac{2}{\sqrt{a-y_{1}}}\mathrm{arcoth}\left(\sqrt{\frac{y-y_{1}}{a-y_{1}}}\right),
\end{eqnarray}
\begin{eqnarray}\label{eq:EMInty2}
I_{7}=\int\frac{\mathrm{d}y}{(y-a)^2\sqrt{y-y_{1}}}=-\frac{\sqrt{y-y_{1}}}{(a-y_{1})(y-a)}+\frac{1}{(a-y_{1})^{\frac{3}{2}}}\mathrm{arcoth}\left(\sqrt{\frac{y-y_{1}}{a-y_{1}}}\right).
\end{eqnarray}
In $I_{8}$ and $I_{9}$ we always have $y<a$, where $a$ is either $y_{\mathrm{ph}}$,
$y_{\mathrm{H},\mathrm{i}}$ or $y_{\mathrm{H},\mathrm{o}}$. $y_{\mathrm{ph}}$,
$y_{\mathrm{H},\mathrm{i}}$ and $y_{\mathrm{H},\mathrm{o}}$ are related to $r_{\mathrm{ph}}$,
$r_{\mathrm{H},\mathrm{i}}$ and $r_{\mathrm{H},\mathrm{o}}$ via Eq.~(\ref{eq:Sub3}),
respectively. Now we substitute $z=y-y_{1}$ and integrate. After integration and resubstitution
$I_{8}$ and $I_{9}$ read
\begin{eqnarray}\label{eq:EMInty3}
I_{8}=\int\frac{\mathrm{d}y}{(a-y)\sqrt{y-y_{1}}}=\frac{2}{\sqrt{a-y_{1}}}\mathrm{artanh}\left(\sqrt{\frac{y-y_{1}}{a-y_{1}}}\right),
\end{eqnarray}
\begin{eqnarray}\label{eq:EMInty4}
I_{9}=\int\frac{\mathrm{d}y}{(a-y)^2\sqrt{y-y_{1}}}=\frac{\sqrt{y-y_{1}}}{(a-y_{1})(a-y)}+\frac{1}{(a-y_{1})^{\frac{3}{2}}}\mathrm{artanh}\left(\sqrt{\frac{y-y_{1}}{a-y_{1}}}\right).
\end{eqnarray}

\subsection{Elliptic Integrals}
In Sec.~\ref{Sec:tr} we encountered several general elliptic integrals. The
main purpose of this section is to demonstrate how to rewrite them in terms of
elementary functions and Legendre's canonical forms of the elliptic integrals of the first,
second and third kind. Let us start by defining Legendre's elliptic integrals of the first,
second and third kind. In their canonical form they read
\begin{eqnarray}\label{eq:FL}
F_{L}(\chi,k)=\int_{0}^{\chi}\frac{\mathrm{d}\chi'}{\sqrt{1-k\sin^2\chi'}},
\end{eqnarray}
\begin{eqnarray}\label{eq:EL}
E_{L}(\chi,k)=\int_{0}^{\chi}\sqrt{1-k\sin^2\chi'}\mathrm{d}\chi',
\end{eqnarray}
\begin{eqnarray}\label{eq:PiL}
\hspace*{-0.5cm}\Pi_{L}(\chi,k,n_{i})=\int_{0}^{\chi}\frac{\mathrm{d}\chi'}{(1-n_{i}\sin^2\chi')\sqrt{1-k\sin^2\chi'}},
\end{eqnarray}
where $\chi$ is called the argument of the elliptic functions, $k$ is the square
of the elliptic modulus and $n_{i}\in \mathbb{R}$ is an arbitrary parameter. In the case $\chi=\pi/2$
we refer to them as complete elliptic integrals. For the complete elliptic integrals
one commonly omits $\chi$ in the arguments and writes the complete elliptic integral
of the first kind as $K_{L}(k)$. The integrand of Eq.~(\ref{eq:PiL}) becomes singular
whenever we integrate over a horizon. We can alleviate this problem by rewriting
it as \cite{MilneThomson1972}
\begin{eqnarray}\label{eq:Ellsing}
\hspace*{-0.5cm}\Pi_{L}(\chi,k,n_{i})=F_{L}(\chi,k)-\Pi_{L}\left(\chi,k,\frac{k}{n_{i}}\right)+\frac{1}{2p}\ln\left(\frac{\cos\chi \sqrt{1-k\sin^2\chi}+p\sin\chi}{\left|\cos\chi \sqrt{1-k\sin^2\chi}-p\sin\chi\right|}\right),
\end{eqnarray}
where
\begin{eqnarray}
p=\sqrt{\frac{(n_{i}-1)(n_{i}-k)}{n_{i}}}.
\end{eqnarray}
While integrating the radial part of the time coordinate $t_{r}(\lambda)$ in Sec.~\ref{Sec:tr}
we also encountered in total five elliptic integrals that do not immediately take one of Legendre's
canonical forms given by Eqs.~(\ref{eq:FL})--(\ref{eq:PiL}). In the following we
demonstrate how to rewrite them as elementary functions and Legendre's elliptic integrals
of the first, second and third kind.

\subsubsection{Time coordinate $t$: Case 2} \label{Sec:Ell1}
In this case we have two pairs of complex conjugate roots. Employing the notation from Sec.~\ref{sec:ToMr}
we write them as $r_{1}=\bar{r}_{2}=R_{1}+iR_{2}$ and $r_{3}=\bar{r}_{4}=R_{3}+iR_{4}$,
where $R_{1}<R_{3}$, $0<R_{2}$, and $0<R_{4}$. In this notation the integrals take
the following two general forms:
\begin{eqnarray}\label{eq:T1}
t_{r,1}(r_{i},r)=\int_{r_{i}}^{r}\frac{r'^{m_{k}}\mathrm{d}r'}{\sqrt{((R_{1}-r')^2+R_{2}^2)((R_{3}-r')^2+R_{4}^2)}},
\end{eqnarray}
\begin{eqnarray}\label{eq:T2}
\hspace*{-0.5cm}t_{r,2}(r_{i},r)=\int_{r_{i}}^{r}\frac{\mathrm{d}r'}{(r'-r_{h})^{m_{k}}\sqrt{((R_{1}-r')^2+R_{2}^2)((R_{3}-r')^2+R_{4}^2)}},
\end{eqnarray}
where in our case $r_{h}$ always corresponds to the radius coordinate of one of the horizons.
Applying the coordinate transformation Eq.~(\ref{eq:Subr1}) and defining two new
constants of motion following Byrd and Friedman \cite{Byrd1954}
\begin{eqnarray}
n_{1}=\frac{R_{2}+g_{0}R_{1}}{R_{1}-g_{0}R_{2}}~~~\mathrm{and}~~~n_{2}=\frac{R_{2}+g_{0}(R_{1}-r_{h})}{R_{1}-g_{0}R_{2}-r_{h}}
\end{eqnarray}
then transforms the integrals Eqs.~(\ref{eq:T1}) and (\ref{eq:T2}) to
\begin{eqnarray}\label{eq:TE1}
t_{r,1}(r_{i},r)=\frac{2(R_{1}-g_{0}R_{2})^{m_{k}}}{(S+\bar{S})g_{0}^{m_{k}}}\sum_{j=0}^{m_{k}}\frac{m_{k}!n_{1}^{m_{k}-j}(g_{0}-n_{1})^j}{(m_{k}-j)!j!}\int_{\chi_{i}}^{\chi}\frac{\mathrm{d}\chi'}{(1+g_{0}\tan\chi')^{j}\sqrt{1-k_{1}\sin^2\chi'}},
\end{eqnarray}
\begin{eqnarray}\label{eq:TE2}
t_{r,2}(r_{i},r)=\frac{2}{(S+\bar{S})(R_{2}+g_{0}\left(R_{1}-r_{h}\right))^{m_{k}}}\sum_{j=0}^{m_{k}}\frac{m_{k}!g_{0}^{m_{k}-j}(n_{2}-g_{0})^j}{(m_{k}-j)!j!}\int_{\chi_{i}}^{\chi}\frac{\mathrm{d}\chi'}{(1+n_{2}\tan\chi')^{j}\sqrt{1-k_{1}\sin^2\chi'}},
\end{eqnarray}
where $S$, $\bar{S}$, and $g_{0}$ are defined by Eqs.~(\ref{eq:S}), (\ref{eq:Sbar}),
and (\ref{eq:g0}), respectively, the square of the elliptic modulus $k_{1}$ is given by Eq.~(\ref{eq:EM1})
and $\chi_{i}$ and $\chi$ are related to $r_{i}$
and $r$ by Eq.~(\ref{eq:chi1}), respectively. Equations (\ref{eq:TE1}) and (\ref{eq:TE2})
contain elliptic integrals that do not immediately take one of Legendre's canonical
forms. Thus they have to be calculated separately. In our case we always have either
$m_{k}=0$, $m_{k}=1$ or $m_{k}=2$. For $m_{k}=0$ Eqs.~(\ref{eq:TE1}) and (\ref{eq:TE2})
reduce to the same term containing two elliptic integrals of the first kind. It is related
to the Mino parameter $\lambda$ by
\begin{eqnarray}
\lambda-\lambda_{i}=\frac{i_{r_{i}}2(F_{L}(\chi,k_{1})-F_{L}(\chi_{i},k_{1})}{(S+\bar{S})\sqrt{E^2+\frac{\Lambda}{3}K}}.
\end{eqnarray}
For $m_{k}=1$ and $m_{k}=2$ Eqs.~(\ref{eq:TE1}) and (\ref{eq:TE2}) contain two elliptic
integrals not immediately taking one of Legendre's canonical forms. The two integrals
have $j=1$ and $j=2$ and read in their most general form
\begin{eqnarray}\label{eq:GL}
G_{L}(\chi_{i},\chi,k_{1},n_{k})=\int_{\chi_{i}}^{\chi}\frac{\mathrm{d}\chi'}{\left(1+n_{k}\tan\chi'\right)\sqrt{1-k_{1}\sin^2\chi'}},
\end{eqnarray}
and
\begin{eqnarray}\label{eq:HL}
H_{L}(\chi_{i},\chi,k_{1},n_{k})=\int_{\chi_{i}}^{\chi}\frac{\mathrm{d}\chi'}{\left(1+n_{k}\tan\chi'\right)^2\sqrt{1-k_{1}\sin^2\chi'}},
\end{eqnarray}
where $n_{k}=g_{0}$ or $n_{k}=n_{2}$. For brevity we will now drop $\chi_{i}$ in the argument.
Following Gralla and Lupsasca \cite{Gralla2020} we can now rewrite $G_{L}(\chi,k_{1},n_{k})$
and $H_{L}(\chi,k_{1},n_{k})$ in terms of elementary functions and Legendre's elliptic integrals of the
first, second and third kind
\begin{eqnarray}\label{eq:GLint}
G_{L}(\chi,k_{1},n_{k})=\frac{F_{L}(\chi,k_{1})+n_{k}^2\Pi_{L}(\chi,k_{1},1+n_{k}^2)}{1+n_{k}^2}+\frac{n_{k} \tilde{G}_{L}(\chi,k_{1},n_{k})}{2\sqrt{(1+n_{k}^2)(1-k_{1}+n_{k}^2)}},
\end{eqnarray}
\begin{eqnarray}\label{eq:HLint}
H_{L}(\chi,k_{1},n_{k})=&\frac{F_{L}(\chi,k_{1})}{(1+n_{k}^2)^2}+\frac{n_{k}^2}{(1+n_{k}^2)(1-k_{1}+n_{k}^2)}\left(n_{k}+\frac{\sin\chi-n_{k}\cos\chi}{\cos\chi+n_{k}\sin\chi}\sqrt{1-k_{1}\sin^2\chi}-E_{L}(\chi,k_{1})\right)\\ \nonumber
&+\frac{2(1-k_{1}+n_{k}^2)-n_{k}^2k_{1}}{(1+n_{k}^2)(1-k_{1}+n_{k}^2)}\left(\frac{n_{k}^2\Pi_{L}(\chi,k_{1},1+n_{k}^2)}{1+n_{k}^2}+\frac{n_{k} \tilde{G}_{L}(\chi,k_{1},n_{k})}{2\sqrt{(1+n_{k}^2)(1-k_{1}+n_{k}^2)}}\right),
\end{eqnarray}
where
\begin{eqnarray}\label{eq:GTL}
\tilde{G}_{L}(\chi,k_{1},n_{k})=\ln\left(\left|\frac{\left(1+\sqrt{\frac{1+n_{k}^2}{1-k_{1}+n_{k}^2}}\right)\left(1-\sqrt{\frac{1+n_{k}^2}{1-k_{1}+n_{k}^2}}\sqrt{1-k_{1}\sin^2\chi}\right)}{\left(1-\sqrt{\frac{1+n_{k}^2}{1-k_{1}+n_{k}^2}}\right)\left(1+\sqrt{\frac{1+n_{k}^2}{1-k_{1}+n_{k}^2}}\sqrt{1-k_{1}\sin^2\chi}\right)}\right|\right).
\end{eqnarray}
In addition, because we always have $0<n_{k}^2$, we evoke Eq.~(\ref{eq:Ellsing}) to
avoid the divergence of $\Pi_{L}(\chi,k_{1},1+n_{k}^2)$.
\subsubsection{Time coordinate $t$: Case 4} \label{Sec:Ell2}
In Sec.~\ref{Sec:tr} we also encountered the two elliptic integrals $I_{L}(\chi_{i},\chi,k_{2},n_{k})$
and $J_{L}(\chi_{i},\chi,k_{2},n_{k})$ that do not immediately take one of Legendre's canonical
forms [$k_{2}$ is the square of the elliptic modulus given by Eq.~(\ref{eq:MOD2})
and $\chi_{i}$ and $\chi$ are related to $r_{i}$ and $r$ by Eq.~(\ref{eq:Ichi1}), respectively].
We will now demonstrate how to rewrite them in terms of elementary functions
and Legendre's elliptic integrals of the first, second and third kind. For this purpose let
us first write them down in their general forms:
\begin{eqnarray}\label{eq:IL}
I_{L}(\chi_{i},\chi,k_{2},n_{k})=\int_{\chi_{i}}^{\chi}\frac{\mathrm{d}\chi'}{\left(1+n_{k}\cos\chi'\right)\sqrt{1-k_{2}\sin^2\chi'}},
\end{eqnarray}
\begin{eqnarray}\label{eq:JL}
J_{L}(\chi_{i},\chi,k_{2},n_{k})=\int_{\chi_{i}}^{\chi}\frac{\mathrm{d}\chi'}{\left(1+n_{k}\cos\chi'\right)^2\sqrt{1-k_{2}\sin^2\chi'}}.
\end{eqnarray}
We start by integrating $I_{L}(\chi_{i},\chi,k_{2},n_{k})$. For this purpose we first omit,
for brevity, $\chi_{i}$ in the argument and then expand by $1-n_{k}\cos\chi'$:
\begin{eqnarray}\label{eq:ILim}
I_{L}(\chi,k_{2},n_{k})=&\int_{0}^{\chi}\frac{\mathrm{d}\chi'}{\left(1+n_{k}\cos\chi'\right)\sqrt{1-k_{2}\sin^2\chi'}}=\frac{1}{1-n_{k}^2}\left(\int_{0}^{\chi}\frac{\mathrm{d}\chi'}{\left(1-\frac{n_{k}^2}{n_{k}^2-1}\sin^2\chi'\right)\sqrt{1-k_{2}\sin^2\chi'}}\right. \\
&\left.-n_{k}\int_{0}^{\chi}\frac{\cos\chi'\mathrm{d}\chi'}{\left(1-\frac{n_{k}^2}{n_{k}^2-1}\sin^2\chi'\right)\sqrt{1-k_{2}\sin^2\chi'}}\right). \nonumber
\end{eqnarray}
Now we rewrite the first term as Legendre's elliptic integral of the third kind. The second term
is an elementary integral. Its calculation involves several case-by-case analyses
which are too long to be reproduced here. After the integration $I_{L}(\chi,k_{2},n_{k})$ becomes \cite{Frost2021a} [see also Eqs.~(B61), (B62) and (B65) in Gralla and Lupsasca \cite{Gralla2020} for an alternative formulation]
\begin{eqnarray}\label{eq:ILint}
I_{L}(\chi,k_{2},n_{k})=\frac{\Pi_{L}\left(\chi,k_{2},\frac{n_{k}^2}{n_{k}^2-1}\right)}{1-n_{k}^2}+\frac{n_{k}\tilde{I}_{L}(\chi,k_{2},n_{k})}{2\sqrt{(n_{k}^2-1)(n_{k}^2(1-k_{2})+k_{2})}},
\end{eqnarray}
where
\begin{eqnarray}\label{eq:ILTint}
\tilde{I}_{L}(\chi,k_{2},n_{k})=\ln\left(\frac{\sin\chi\sqrt{\frac{n_{k}^2(1-k_{2})+k_{2}}{n_{k}^2-1}}+\sqrt{1-k_{2}\sin^2\chi}}{\left|\sin\chi\sqrt{\frac{n_{k}^2(1-k_{2})+k_{2}}{n_{k}^2-1}}-\sqrt{1-k_{2}\sin^2\chi}\right|}\right).
\end{eqnarray}
For $J_{L}(\chi_{i},\chi,k_{2},n_{k})$ we proceed analogously. We first omit $\chi_{i}$
in the argument and then expand by $(1-n_{k}\cos\chi')^2$ and write the third term as
Legendre's elliptic integral of the third kind:
\begin{eqnarray}\label{eq:JLint}
J_{L}(\chi,k_{2},n_{k})=&\int_{0}^{\chi}\frac{\mathrm{d}\chi'}{\left(1+n_{k}\cos\chi'\right)^2\sqrt{1-k_{2}\sin^2\chi'}}=\frac{2}{(n_{k}^2-1)^2}\left(\int_{0}^{\chi}\frac{\mathrm{d}\chi'}{\left(1-\frac{n_{k}^2}{n_{k}^2-1}\sin^2\chi'\right)^2\sqrt{1-k_{2}\sin^2\chi'}}\right.\\
&\left.-n_{k}\int_{0}^{\chi}\frac{\cos\chi'\mathrm{d}\chi'}{\left(1-\frac{n_{k}^2}{n_{k}^2-1}\sin^2\chi'\right)^2\sqrt{1-k_{2}\sin^2\chi'}}\right)+\frac{\Pi_{L}\left(\chi,k_{2},\frac{n_{k}^2}{n_{k}^2-1}\right)}{n_{k}^2-1}.\nonumber
\end{eqnarray}
The first term is again an elliptic integral. It is given by Eq.~(\ref{eq:MLint})
in Sec.~\ref{Sec:Ell3} and its evaluation will be discussed there. The second
term is, again, an elementary integral. Together their evaluation requires several
case-by-case analyses. After the integration and simplifying all terms $J_{L}(\chi,k_{2},n_{k})$
reads [see also Eqs.~(B61)--(B65) in Gralla and Lupsasca \cite{Gralla2020} for an alternative formulation]
\begin{eqnarray}\label{eq:JLTint}
J_{L}(\chi,k_{2},n_{k})=&\frac{n_{k}^3\sin\chi\sqrt{1-k_{2}\sin^2\chi}}{(n_{k}^2-1)(n_{k}^2(1-k_{2})+k_{2})(1+n_{k}\cos\chi)}-\frac{n_{k}(n_{k}^2(1-2k_{2})+2k_{2})\tilde{I}_{L}(\chi,k_{2},n_{k})}{2\left((n_{k}^2-1)(n_{k}^2(1-k_{2})+k_{2})\right)^{\frac{3}{2}}}+\frac{F_{L}(\chi,k_{2})}{n_{k}^2-1}\\
&-\frac{n_{k}^2E_{L}(\chi,k_{2})}{(n_{k}^2-1)(n_{k}^2(1-k_{2})+k_{2})}+\frac{(n_{k}^2(1-2k_{2})+2k_{2})\Pi_{L}\left(\chi,k_{2},\frac{n_{k}^2}{n_{k}^2-1}\right)}{(n_{k}^2-1)^2(n_{k}^2(1-k_{2})+k_{2})}.\nonumber
\end{eqnarray}
Note that in $I_{L}(\chi,k_{2},n_{k})$ and $J_{L}(\chi,k_{2},n_{k})$ we always have
$n_{k}^2/(n_{k}^2-1)>1$ and thus we again evoke Eq.~(\ref{eq:Ellsing}) to avoid the
divergence of $\Pi_{L}(\chi,k_{2},n_{k}^2/(n_{k}^2-1))$.

\subsubsection{Time coordinate $t$: Case 6}\label{Sec:Ell3}
In Sec.~\ref{Sec:tr} and Appendix~\ref{Sec:Ell2} we encountered the elliptic
integral $M_{L}(\chi_{i},\chi,k_{i},n_{k})$ in two different forms [$k_{i}=k_{2}$
or $k_{i}=k_{3}$ is the square of the elliptic modulus given by Eqs.~(\ref{eq:MOD2})
or (\ref{eq:MOD3}) and $\chi_{i}$ and $\chi$ are related to $r_{i}$ and $r$ by
Eqs.~(\ref{eq:Ichi1}), (\ref{eq:Ichi3}) or (\ref{eq:Ichi4}), respectively]. In its
explicit form it reads
\begin{eqnarray}\label{eq:ML}
M_{L}(\chi_{i},\chi,k_{i},n_{k})=\int_{\chi_{i}}^{\chi}\frac{\mathrm{d}\chi'}{\left(1-n_{k}\sin^2\chi'\right)^2\sqrt{1-k_{i}\sin^2\chi'}}.
\end{eqnarray}
We can now rewrite this integral in terms of elementary functions and Legendre's
elliptic integrals of the first, second and third kind (again we omit the first
argument $\chi_{i}$):
\begin{eqnarray}\label{eq:MLint}
M_{L}(\chi,k_{i},n_{k})&=\int_{0}^{\chi}\frac{\mathrm{d}\chi'}{\left(1-n_{k}\sin^2\chi'\right)^2\sqrt{1-k_{i}\sin^2\chi'}}=\frac{n_{k}^2\sin(2\chi)\sqrt{1-k_{i}\sin^2\chi}}{4(n_{k}-k_{i})(n_{k}-1)(1-n_{k}\sin^2\chi)}+\frac{F_{L}(\chi,k_{i})}{2(n_{k}-1)}\\
&-\frac{n_{k}E_{L}(\chi,k_{i})}{2(n_{k}-k_{i})(n_{k}-1)}+\frac{n_{k}(n_{k}-2)-(2n_{k}-3)k_{i}}{2(n_{k}-k_{i})(n_{k}-1)}\Pi_{L}(\chi,k_{i},n_{k}). \nonumber
\end{eqnarray}
Note that for the integral in Sec.~\ref{Sec:Ell2} we have to replace $n_{k}\rightarrow n_{k}^2/(n_{k}^2-1)$.
For lightlike geodesics with turning points at $r_{\mathrm{min}}=r_{1}$
and $r_{\mathrm{max}}=r_{2}$ we always chose the coordinate transformations Eq.~(\ref{eq:Sub4})
and Eq.~(\ref{eq:Sub5}) such that Legendre's elliptic integral of the third kind does not diverge.
Therefore, in these two cases we can use Eq.~(\ref{eq:MLint}) directly.

\section{ELLIPTIC FUNCTIONS}\label{Sec:Ellfunc}
In this appendix we demonstrate how to solve the differential equation associated
with the equation of motion for $r$ given by Eq.~(\ref{eq:EoMr}) for case 2, case 4 and case 6 in Sec.~\ref{Sec:Solr}
using Jacobi's elliptic functions. Before we turn to explicitly solving the differential
equation we will give a brief introduction to Jacobi's elliptic functions and their
properties. For a thorough introduction we refer the interested reader to the book
of Hancock \cite{Hancock1917}. \\
The theory of elliptic functions after Jacobi defines three elementary elliptic
functions. These are Jacobi's $\mathrm{sn}$, $\mathrm{cn}$ and $\mathrm{dn}$ functions.
Starting from the sine and the cosine they are defined by
\begin{eqnarray}
\mathrm{sn}(\lambda,k)=\sin\mathrm{am}\lambda=\sin\chi,
\end{eqnarray}
\begin{eqnarray}
\mathrm{cn}(\lambda,k)=\cos\mathrm{am}\lambda=\cos\chi,
\end{eqnarray}
\begin{eqnarray}
\mathrm{dn}(\lambda,k)=\sqrt{1-k\sin^2\text{am}\lambda}=\sqrt{1-k\sin^2\chi},
\end{eqnarray}
where for now $\lambda$ is an arbitrary independent variable, $k$ is the square
of the elliptic modulus and $\chi=\mathrm{am}\lambda$ is called the amplitude
of $\lambda$. In addition one can also define six associated elliptic functions.
In this paper we only need one, Jacobi's elliptic $\mathrm{sc}$ function. It is
defined by
\begin{eqnarray}
\mathrm{sc}(\lambda,k)=\frac{\mathrm{sn}(\lambda,k)}{\mathrm{cn}(\lambda,k)}.
\end{eqnarray}
Jacobi's elliptic functions are periodic with respect to the complete elliptic integral
of the first kind $K_{L}(k)$ and fulfill the following periodicity relations:
\begin{eqnarray}
\mathrm{sn}(\lambda\pm4K_{L}(k),k)=\mathrm{sn}(\lambda,k),
\end{eqnarray}
\begin{eqnarray}
\mathrm{cn}(\lambda\pm4K_{L}(k),k)=\mathrm{cn}(\lambda,k),
\end{eqnarray}
\begin{eqnarray}
\mathrm{dn}(\lambda\pm2K_{L}(k),k)=\mathrm{dn}(\lambda,k),
\end{eqnarray}
\begin{eqnarray}
\mathrm{sc}(\lambda\pm2K_{L}(k),k)=\mathrm{sc}(\lambda,k).
\end{eqnarray}
Jacobi's elliptic functions have the characteristic that they solve the differential
equation
\begin{eqnarray}\label{eq:elldiff}
\left(\frac{\mathrm{d}\chi}{\mathrm{d}\lambda}\right)^2=a\left(1-k\sin^2\chi\right).
\end{eqnarray}
Although Eq.~(\ref{eq:EoMr}) does not immediately take the Legendre form of Eq.~(\ref{eq:elldiff})
using an appropriate coordinate transformation $z=f(\sin\chi)$, $z=f(\cos\chi)$ or
$z=f(\tan\chi)$ we can transform any differential equation of the form
\begin{eqnarray}
\left(\frac{\mathrm{d}z}{\mathrm{d}\lambda}\right)^2=a_{4}z^4+a_{3}z^3+a_{2}z^2+a_{1}z+a_{0}
\end{eqnarray}
into the form of Eq.~(\ref{eq:elldiff}). Now we separate variables and integrate:
\begin{eqnarray}
\int_{\lambda_{i}}^{\lambda}\mathrm{d}\lambda'=\frac{i_{\chi_{i}}}{\sqrt{a}}\int_{\chi_{i}}^{\chi}\frac{\mathrm{d}\chi'}{\sqrt{1-k\sin^2\chi'}},
\end{eqnarray}
where $1/\sqrt{a}=c/\sqrt{a_{4}}$, $i_{\chi_{i}}=\mathrm{sgn}\left(\left.\mathrm{d}\chi/\mathrm{d}\lambda\right|_{\chi=\chi_{i}}\right)$
and $c$ is a new constant that is specific to the chosen coordinate transformation. We can
now rewrite this equation as
\begin{eqnarray}
\tilde{\lambda}=i_{\chi_{i}}\frac{\sqrt{a_{4}}}{c}\left(\lambda-\lambda_{i}\right)+F_{L}\left(\chi_{i},k\right)=\int_{0}^{\chi}\frac{\mathrm{d}\chi'}{\sqrt{1-k\sin^2\chi'}}.
\end{eqnarray}
With $\chi=\mathrm{am}\tilde{\lambda}$ we can now write the solution $z(\lambda)$
to Eq.~(\ref{eq:elldiff}) in terms of Jacobi's elliptic $\mathrm{sn}$, $\mathrm{cn}$,
and $\mathrm{sc}$ functions.
\bibliography{Charged_NUT_de_Sitter_Lensing.bib}

\begin{thebibliography}{56}%
\makeatletter
\providecommand \@ifxundefined [1]{%
 \@ifx{#1\undefined}
}%
\providecommand \@ifnum [1]{%
 \ifnum #1\expandafter \@firstoftwo
 \else \expandafter \@secondoftwo
 \fi
}%
\providecommand \@ifx [1]{%
 \ifx #1\expandafter \@firstoftwo
 \else \expandafter \@secondoftwo
 \fi
}%
\providecommand \natexlab [1]{#1}%
\providecommand \enquote  [1]{``#1''}%
\providecommand \bibnamefont  [1]{#1}%
\providecommand \bibfnamefont [1]{#1}%
\providecommand \citenamefont [1]{#1}%
\providecommand \href@noop [0]{\@secondoftwo}%
\providecommand \href [0]{\begingroup \@sanitize@url \@href}%
\providecommand \@href[1]{\@@startlink{#1}\@@href}%
\providecommand \@@href[1]{\endgroup#1\@@endlink}%
\providecommand \@sanitize@url [0]{\catcode `\\12\catcode `\$12\catcode
  `\&12\catcode `\#12\catcode `\^12\catcode `\_12\catcode `\%12\relax}%
\providecommand \@@startlink[1]{}%
\providecommand \@@endlink[0]{}%
\providecommand \url  [0]{\begingroup\@sanitize@url \@url }%
\providecommand \@url [1]{\endgroup\@href {#1}{\urlprefix }}%
\providecommand \urlprefix  [0]{URL }%
\providecommand \Eprint [0]{\href }%
\providecommand \doibase [0]{https://doi.org/}%
\providecommand \selectlanguage [0]{\@gobble}%
\providecommand \bibinfo  [0]{\@secondoftwo}%
\providecommand \bibfield  [0]{\@secondoftwo}%
\providecommand \translation [1]{[#1]}%
\providecommand \BibitemOpen [0]{}%
\providecommand \bibitemStop [0]{}%
\providecommand \bibitemNoStop [0]{.\EOS\space}%
\providecommand \EOS [0]{\spacefactor3000\relax}%
\providecommand \BibitemShut  [1]{\csname bibitem#1\endcsname}%
\let\auto@bib@innerbib\@empty
\bibitem [{\citenamefont {Plebanski}\ and\ \citenamefont
  {Demianski}(1976)}]{Plebanski1976}%
  \BibitemOpen
  \bibfield  {author} {\bibinfo {author} {\bibfnamefont {J.~F.}\ \bibnamefont
  {Plebanski}}\ and\ \bibinfo {author} {\bibfnamefont {M.}~\bibnamefont
  {Demianski}},\ }\bibfield  {title} {\bibinfo {title} {Rotating, charged, and
  uniformly accelerating mass in general relativity},\ }\href
  {https://doi.org/10.1016/0003-4916(76)90240-2} {\bibfield  {journal}
  {\bibinfo  {journal} {Ann. {P}hys. ({N}. {Y}.)}\ }\textbf {\bibinfo {volume}
  {98}},\ \bibinfo {pages} {98} (\bibinfo {year} {1976})}\BibitemShut {NoStop}%
\bibitem [{\citenamefont {Manko}\ and\ \citenamefont {Ruiz}(2005)}]{Manko2005}%
  \BibitemOpen
  \bibfield  {author} {\bibinfo {author} {\bibfnamefont {V.~S.}\ \bibnamefont
  {Manko}}\ and\ \bibinfo {author} {\bibfnamefont {E.}~\bibnamefont {Ruiz}},\
  }\bibfield  {title} {\bibinfo {title} {Physical interpretation of the {NUT}
  family of solutions},\ }\href {https://doi.org/10.1088/0264-9381/22/17/014}
  {\bibfield  {journal} {\bibinfo  {journal} {Classical {Q}uantum {G}ravity}\
  }\textbf {\bibinfo {volume} {22}},\ \bibinfo {pages} {3555} (\bibinfo {year}
  {2005})}\BibitemShut {NoStop}%
\bibitem [{\citenamefont {Taub}(1951)}]{Taub1951}%
  \BibitemOpen
  \bibfield  {author} {\bibinfo {author} {\bibfnamefont {A.~H.}\ \bibnamefont
  {Taub}},\ }\bibfield  {title} {\bibinfo {title} {Empty space-times admitting
  a three parameter group of motions},\ }\href
  {https://doi.org/10.2307/1969567} {\bibfield  {journal} {\bibinfo  {journal}
  {Ann. {M}ath.}\ }\textbf {\bibinfo {volume} {53}},\ \bibinfo {pages} {472}
  (\bibinfo {year} {1951})}\BibitemShut {NoStop}%
\bibitem [{\citenamefont {Newman}\ \emph {et~al.}(1963)\citenamefont {Newman},
  \citenamefont {Tamburino},\ and\ \citenamefont {Unti}}]{Newman1963}%
  \BibitemOpen
  \bibfield  {author} {\bibinfo {author} {\bibfnamefont {E.}~\bibnamefont
  {Newman}}, \bibinfo {author} {\bibfnamefont {L.}~\bibnamefont {Tamburino}},\
  and\ \bibinfo {author} {\bibfnamefont {T.}~\bibnamefont {Unti}},\ }\bibfield
  {title} {\bibinfo {title} {Empty-space generalization of the {S}chwarzschild
  metric},\ }\href {https://doi.org/10.1063/1.1704018} {\bibfield  {journal}
  {\bibinfo  {journal} {J. {M}ath. {P}hys. ({N}.{Y}.)}\ }\textbf {\bibinfo
  {volume} {4}},\ \bibinfo {pages} {915} (\bibinfo {year} {1963})}\BibitemShut
  {NoStop}%
\bibitem [{\citenamefont {Misner}(1963)}]{Misner1963}%
  \BibitemOpen
  \bibfield  {author} {\bibinfo {author} {\bibfnamefont {C.~W.}\ \bibnamefont
  {Misner}},\ }\bibfield  {title} {\bibinfo {title} {The flatter regions of
  {N}ewman, {U}nti, and {T}amburino's generalized {S}chwarzschild space},\
  }\href {https://doi.org/10.1063/1.1704019} {\bibfield  {journal} {\bibinfo
  {journal} {J. {M}ath. {P}hys. ({N}.{Y}.)}\ }\textbf {\bibinfo {volume} {4}},\
  \bibinfo {pages} {924} (\bibinfo {year} {1963})}\BibitemShut {NoStop}%
\bibitem [{\citenamefont {Bonnor}(1969)}]{Bonnor1969}%
  \BibitemOpen
  \bibfield  {author} {\bibinfo {author} {\bibfnamefont {W.~B.}\ \bibnamefont
  {Bonnor}},\ }\bibfield  {title} {\bibinfo {title} {A new interpretation of
  the {NUT} metric in general relativity},\ }\href
  {https://doi.org/10.1017/S0305004100044807} {\bibfield  {journal} {\bibinfo
  {journal} {Proc. {C}ambridge {P}hilos. {S}oc.}\ }\textbf {\bibinfo {volume}
  {66}},\ \bibinfo {pages} {145} (\bibinfo {year} {1969})}\BibitemShut
  {NoStop}%
\bibitem [{\citenamefont {Sackfield}(1971)}]{Sackfield1971}%
  \BibitemOpen
  \bibfield  {author} {\bibinfo {author} {\bibfnamefont {A.}~\bibnamefont
  {Sackfield}},\ }\bibfield  {title} {\bibinfo {title} {Physical interpretation
  of {N}.{U}.{T}. metric},\ }\href {https://doi.org/10.1017/S0305004100049707}
  {\bibfield  {journal} {\bibinfo  {journal} {Proc. {C}ambridge {P}hilos.
  {S}oc.}\ }\textbf {\bibinfo {volume} {70}},\ \bibinfo {pages} {89} (\bibinfo
  {year} {1971})}\BibitemShut {NoStop}%
\bibitem [{\citenamefont {Griffiths}\ and\ \citenamefont {Podolsk{\'
  y}}(2009)}]{Griffiths2009}%
  \BibitemOpen
  \bibfield  {author} {\bibinfo {author} {\bibfnamefont {J.~B.}\ \bibnamefont
  {Griffiths}}\ and\ \bibinfo {author} {\bibfnamefont {J.}~\bibnamefont
  {Podolsk{\' y}}},\ }\href {https://doi.org/10.1017/CBO9780511635397} {\emph
  {\bibinfo {title} {Exact {S}pace-{T}imes in {E}instein's {G}eneral
  {R}elativity}}}\ (\bibinfo  {publisher} {Cambridge {U}niversity {P}ress},\
  \bibinfo {address} {Cambridge, England},\ \bibinfo {year} {2009})\ pp.\
  \bibinfo {pages} {213--237}\BibitemShut {NoStop}%
\bibitem [{\citenamefont {Brill}(1964)}]{Brill1964}%
  \BibitemOpen
  \bibfield  {author} {\bibinfo {author} {\bibfnamefont {D.~R.}\ \bibnamefont
  {Brill}},\ }\bibfield  {title} {\bibinfo {title} {Electromagnetic fields in a
  homogeneous, nonisotropic universe},\ }\href
  {https://doi.org/10.1103/PhysRev.133.B845} {\bibfield  {journal} {\bibinfo
  {journal} {Phys. {R}ev.}\ }\textbf {\bibinfo {volume} {133}},\ \bibinfo
  {pages} {{B}845} (\bibinfo {year} {1964})}\BibitemShut {NoStop}%
\bibitem [{\citenamefont {Cl{\' e}ment}\ \emph {et~al.}(2016)\citenamefont
  {Cl{\' e}ment}, \citenamefont {Gal'tsov},\ and\ \citenamefont
  {Guenouche}}]{Clement2016}%
  \BibitemOpen
  \bibfield  {author} {\bibinfo {author} {\bibfnamefont {G.}~\bibnamefont
  {Cl{\' e}ment}}, \bibinfo {author} {\bibfnamefont {D.}~\bibnamefont
  {Gal'tsov}},\ and\ \bibinfo {author} {\bibfnamefont {M.}~\bibnamefont
  {Guenouche}},\ }\bibfield  {title} {\bibinfo {title} {{NUT} wormholes},\
  }\href {https://doi.org/10.1103/PhysRevD.93.024048} {\bibfield  {journal}
  {\bibinfo  {journal} {Phys. {R}ev. {D}}\ }\textbf {\bibinfo {volume} {93}},\
  \bibinfo {pages} {024048} (\bibinfo {year} {2016})}\BibitemShut {NoStop}%
\bibitem [{\citenamefont {Ruban}(1972)}]{Ruban1972}%
  \BibitemOpen
  \bibfield  {author} {\bibinfo {author} {\bibfnamefont {V.~A.}\ \bibnamefont
  {Ruban}},\ }\bibfield  {title} {\bibinfo {title} {Non-singular metrics of
  {T}aub–{N}ewman–{U}nti–{T}amburino type with an electromagnetic
  field},\ }\href
  {http://www.mathnet.ru/php/archive.phtml?wshow=paper&jrnid=dan&apperid=36957&option_lang=eng}
  {\bibfield  {journal} {\bibinfo  {journal} {Dokl. {A}kad. {N}auk {SSSR}}\
  }\textbf {\bibinfo {volume} {204}},\ \bibinfo {pages} {1086} (\bibinfo {year}
  {1972})}\BibitemShut {NoStop}%
\bibitem [{\citenamefont {Misner}\ and\ \citenamefont
  {Taub}(1969)}]{Misner1969}%
  \BibitemOpen
  \bibfield  {author} {\bibinfo {author} {\bibfnamefont {C.~W.}\ \bibnamefont
  {Misner}}\ and\ \bibinfo {author} {\bibfnamefont {A.~H.}\ \bibnamefont
  {Taub}},\ }\bibfield  {title} {\bibinfo {title} {A singularity-free empty
  universe},\ }\href
  {https://ui.adsabs.harvard.edu/abs/1969JETP...28..122M/abstract} {\bibfield
  {journal} {\bibinfo  {journal} {Sov. {P}hys. {JETP}}\ }\textbf {\bibinfo
  {volume} {28}},\ \bibinfo {pages} {122} (\bibinfo {year} {1969})}\BibitemShut
  {NoStop}%
\bibitem [{\citenamefont {Miller}\ \emph {et~al.}(1971)\citenamefont {Miller},
  \citenamefont {Kruskal},\ and\ \citenamefont {Godfrey}}]{Miller1971}%
  \BibitemOpen
  \bibfield  {author} {\bibinfo {author} {\bibfnamefont {J.~G.}\ \bibnamefont
  {Miller}}, \bibinfo {author} {\bibfnamefont {M.~D.}\ \bibnamefont
  {Kruskal}},\ and\ \bibinfo {author} {\bibfnamefont {B.~B.}\ \bibnamefont
  {Godfrey}},\ }\bibfield  {title} {\bibinfo {title} {Taub-{NUT} ({N}ewman,
  {U}nti, {T}amburino) metric and incompatible extensions},\ }\href
  {https://doi.org/10.1103/PhysRevD.4.2945} {\bibfield  {journal} {\bibinfo
  {journal} {Phys. {R}ev. {D}}\ }\textbf {\bibinfo {volume} {4}},\ \bibinfo
  {pages} {2945} (\bibinfo {year} {1971})}\BibitemShut {NoStop}%
\bibitem [{\citenamefont {Kagramanova}\ \emph {et~al.}(2010)\citenamefont
  {Kagramanova}, \citenamefont {Kunz}, \citenamefont {Hackmann},\ and\
  \citenamefont {L{\"a}mmerzahl}}]{Kagramanova2010}%
  \BibitemOpen
  \bibfield  {author} {\bibinfo {author} {\bibfnamefont {V.}~\bibnamefont
  {Kagramanova}}, \bibinfo {author} {\bibfnamefont {J.}~\bibnamefont {Kunz}},
  \bibinfo {author} {\bibfnamefont {E.}~\bibnamefont {Hackmann}},\ and\
  \bibinfo {author} {\bibfnamefont {C.}~\bibnamefont {L{\"a}mmerzahl}},\
  }\bibfield  {title} {\bibinfo {title} {Analytic treatment of complete and
  incomplete geodesics in {T}aub-{NUT} space-times},\ }\href
  {https://doi.org/10.1103/PhysRevD.81.124044} {\bibfield  {journal} {\bibinfo
  {journal} {Phys. {R}ev. {D}.}\ }\textbf {\bibinfo {volume} {81}},\ \bibinfo
  {pages} {124044} (\bibinfo {year} {2010})}\BibitemShut {NoStop}%
\bibitem [{\citenamefont {Cl{\' e}ment}\ \emph {et~al.}(2015)\citenamefont
  {Cl{\' e}ment}, \citenamefont {Gal'tsov},\ and\ \citenamefont
  {Guenouche}}]{Clement2015}%
  \BibitemOpen
  \bibfield  {author} {\bibinfo {author} {\bibfnamefont {G.}~\bibnamefont
  {Cl{\' e}ment}}, \bibinfo {author} {\bibfnamefont {D.}~\bibnamefont
  {Gal'tsov}},\ and\ \bibinfo {author} {\bibfnamefont {M.}~\bibnamefont
  {Guenouche}},\ }\bibfield  {title} {\bibinfo {title} {Rehabilitating
  space-times with {NUT}s},\ }\href
  {https://doi.org/10.1016/j.physletb.2015.09.074} {\bibfield  {journal}
  {\bibinfo  {journal} {Phys. {L}ett. {B}}\ }\textbf {\bibinfo {volume}
  {750}},\ \bibinfo {pages} {591} (\bibinfo {year} {2015})}\BibitemShut
  {NoStop}%
\bibitem [{\citenamefont {Lynden-Bell}\ and\ \citenamefont
  {Nouri-Zonoz}(1998)}]{LyndenBell1998}%
  \BibitemOpen
  \bibfield  {author} {\bibinfo {author} {\bibfnamefont {D.}~\bibnamefont
  {Lynden-Bell}}\ and\ \bibinfo {author} {\bibfnamefont {M.}~\bibnamefont
  {Nouri-Zonoz}},\ }\bibfield  {title} {\bibinfo {title} {Classical monopoles:
  {N}ewton, {NUT} space, gravomagnetic lensing, and atomic spectra},\ }\href
  {https://doi.org/10.1103/RevModPhys.70.427} {\bibfield  {journal} {\bibinfo
  {journal} {Rev. Mod. Phys.}\ }\textbf {\bibinfo {volume} {70}},\ \bibinfo
  {pages} {427} (\bibinfo {year} {1998})}\BibitemShut {NoStop}%
\bibitem [{\citenamefont {Rahvar}\ and\ \citenamefont
  {Nouri-Zonoz}(2003)}]{Rahvar2003}%
  \BibitemOpen
  \bibfield  {author} {\bibinfo {author} {\bibfnamefont {S.}~\bibnamefont
  {Rahvar}}\ and\ \bibinfo {author} {\bibfnamefont {M.}~\bibnamefont
  {Nouri-Zonoz}},\ }\bibfield  {title} {\bibinfo {title} {Gravitational
  microlensing in {NUT} space},\ }\href
  {https://doi.org/10.1046/j.1365-8711.2003.06137.x} {\bibfield  {journal}
  {\bibinfo  {journal} {Mon. Not. R. Astron. Soc.}\ }\textbf {\bibinfo {volume}
  {338}},\ \bibinfo {pages} {926} (\bibinfo {year} {2003})}\BibitemShut
  {NoStop}%
\bibitem [{\citenamefont {{The LIGO Scientific Collaboration}}\ \emph
  {et~al.}(2015)\citenamefont {{The LIGO Scientific Collaboration}} \emph
  {et~al.}}]{LIGOCollaboration2015}%
  \BibitemOpen
  \bibfield  {author} {\bibinfo {author} {\bibnamefont {{The LIGO Scientific
  Collaboration}}} \emph {et~al.},\ }\bibfield  {title} {\bibinfo {title}
  {Advanced {LIGO}},\ }\href {https://doi.org/10.1088/0264-9381/32/7/074001}
  {\bibfield  {journal} {\bibinfo  {journal} {Classical Quantum Gravity}\
  }\textbf {\bibinfo {volume} {32}},\ \bibinfo {pages} {074001} (\bibinfo
  {year} {2015})}\BibitemShut {NoStop}%
\bibitem [{\citenamefont {Acernese}\ \emph {et~al.}(2015)\citenamefont
  {Acernese} \emph {et~al.}}]{Acernese2015}%
  \BibitemOpen
  \bibfield  {author} {\bibinfo {author} {\bibfnamefont {F.}~\bibnamefont
  {Acernese}} \emph {et~al.},\ }\bibfield  {title} {\bibinfo {title} {Advanced
  {V}irgo: a second-generation interferometric gravitational wave detector},\
  }\href {https://doi.org/10.1088/0264-9381/32/2/024001} {\bibfield  {journal}
  {\bibinfo  {journal} {Classical Quantum Gravity}\ }\textbf {\bibinfo {volume}
  {32}},\ \bibinfo {pages} {024001} (\bibinfo {year} {2015})}\BibitemShut
  {NoStop}%
\bibitem [{\citenamefont {Akutsu}\ \emph {et~al.}(2018)\citenamefont {Akutsu}
  \emph {et~al.}}]{Akutsu2018}%
  \BibitemOpen
  \bibfield  {author} {\bibinfo {author} {\bibfnamefont {T.}~\bibnamefont
  {Akutsu}} \emph {et~al.},\ }\bibfield  {title} {\bibinfo {title}
  {Construction of {KAGRA}: An underground gravitational-wave observatory},\
  }\href {https://doi.org/10.1093/ptep/ptx180} {\bibfield  {journal} {\bibinfo
  {journal} {Prog. Theor. Exp. Phys.}\ }\textbf {\bibinfo {volume} {2018}},\
  \bibinfo {pages} {013F01} (\bibinfo {year} {2018})}\BibitemShut {NoStop}%
\bibitem [{\citenamefont {{The Event Horizon Telescope Collaboration}}\ \emph
  {et~al.}(2019{\natexlab{a}})\citenamefont {{The Event Horizon Telescope
  Collaboration}} \emph {et~al.}}]{EHTCollaboration2019a}%
  \BibitemOpen
  \bibfield  {author} {\bibinfo {author} {\bibnamefont {{The Event Horizon
  Telescope Collaboration}}} \emph {et~al.},\ }\bibfield  {title} {\bibinfo
  {title} {First {M}87 {E}vent {H}orizon {T}elescope results. {I}. {T}he shadow
  of the supermassive black hole},\ }\href
  {https://doi.org/10.3847/2041-8213/ab0ec7} {\bibfield  {journal} {\bibinfo
  {journal} {{A}strophys. {J}. {L}ett.}\ }\textbf {\bibinfo {volume} {875}},\
  \bibinfo {pages} {L1} (\bibinfo {year} {2019}{\natexlab{a}})}\BibitemShut
  {NoStop}%
\bibitem [{\citenamefont {{The Event Horizon Telescope Collaboration}}\ \emph
  {et~al.}(2019{\natexlab{b}})\citenamefont {{The Event Horizon Telescope
  Collaboration}} \emph {et~al.}}]{EHTCollaboration2019b}%
  \BibitemOpen
  \bibfield  {author} {\bibinfo {author} {\bibnamefont {{The Event Horizon
  Telescope Collaboration}}} \emph {et~al.},\ }\bibfield  {title} {\bibinfo
  {title} {First {M}87 {E}vent {H}orizon {T}elescope results. {II}. {A}rray and
  instrumentation},\ }\href {https://doi.org/10.3847/2041-8213/ab0c96}
  {\bibfield  {journal} {\bibinfo  {journal} {{A}strophys. {J}. {L}ett.}\
  }\textbf {\bibinfo {volume} {875}},\ \bibinfo {pages} {L2} (\bibinfo {year}
  {2019}{\natexlab{b}})}\BibitemShut {NoStop}%
\bibitem [{\citenamefont {Kardashev}\ \emph {et~al.}(2013)\citenamefont
  {Kardashev} \emph {et~al.}}]{Kardashev2013}%
  \BibitemOpen
  \bibfield  {author} {\bibinfo {author} {\bibfnamefont {N.~S.}\ \bibnamefont
  {Kardashev}} \emph {et~al.},\ }\bibfield  {title} {\bibinfo {title}
  {"{R}adio{A}stron"--{A} telescope with a size of 300 000 km: {M}ain
  parameters and first observational results},\ }\href
  {https://doi.org/10.1134/S1063772913030025} {\bibfield  {journal} {\bibinfo
  {journal} {Astronomy {R}eports}\ }\textbf {\bibinfo {volume} {57}},\ \bibinfo
  {pages} {153} (\bibinfo {year} {2013})}\BibitemShut {NoStop}%
\bibitem [{\citenamefont {Kardashev}\ \emph {et~al.}(2017)\citenamefont
  {Kardashev} \emph {et~al.}}]{Kardashev2017}%
  \BibitemOpen
  \bibfield  {author} {\bibinfo {author} {\bibfnamefont {N.~S.}\ \bibnamefont
  {Kardashev}} \emph {et~al.},\ }\bibfield  {title} {\bibinfo {title}
  {Radio{A}stron science program five years after launch: {M}ain science
  results},\ }\href {https://doi.org/10.1134/S0038094617070085} {\bibfield
  {journal} {\bibinfo  {journal} {Solar {S}ystem {R}esearch}\ }\textbf
  {\bibinfo {volume} {51}},\ \bibinfo {pages} {535} (\bibinfo {year}
  {2017})}\BibitemShut {NoStop}%
\bibitem [{\citenamefont {Zimmerman}\ and\ \citenamefont
  {Shahir}(1989)}]{Zimmerman1989}%
  \BibitemOpen
  \bibfield  {author} {\bibinfo {author} {\bibfnamefont {R.~L.}\ \bibnamefont
  {Zimmerman}}\ and\ \bibinfo {author} {\bibfnamefont {B.~Y.}\ \bibnamefont
  {Shahir}},\ }\bibfield  {title} {\bibinfo {title} {Geodesics for the {NUT}
  metric and gravitational monopoles},\ }\href
  {https://doi.org/10.1007/BF00758986} {\bibfield  {journal} {\bibinfo
  {journal} {Gen. Relativ. Gravit.}\ }\textbf {\bibinfo {volume} {21}},\
  \bibinfo {pages} {821} (\bibinfo {year} {1989})}\BibitemShut {NoStop}%
\bibitem [{\citenamefont {Nouri-Zonoz}\ and\ \citenamefont
  {Lynden-Bell}(1997)}]{NouriZonoz1997}%
  \BibitemOpen
  \bibfield  {author} {\bibinfo {author} {\bibfnamefont {M.}~\bibnamefont
  {Nouri-Zonoz}}\ and\ \bibinfo {author} {\bibfnamefont {D.}~\bibnamefont
  {Lynden-Bell}},\ }\bibfield  {title} {\bibinfo {title} {Gravomagnetic lensing
  by {NUT} space},\ }\href {https://doi.org/10.1093/mnras/292.3.714} {\bibfield
   {journal} {\bibinfo  {journal} {Mon. Not. R. Astron. Soc.}\ }\textbf
  {\bibinfo {volume} {292}},\ \bibinfo {pages} {714} (\bibinfo {year}
  {1997})}\BibitemShut {NoStop}%
\bibitem [{\citenamefont {Halla}\ and\ \citenamefont
  {Perlick}(2020)}]{Halla2020}%
  \BibitemOpen
  \bibfield  {author} {\bibinfo {author} {\bibfnamefont {M.}~\bibnamefont
  {Halla}}\ and\ \bibinfo {author} {\bibfnamefont {V.}~\bibnamefont
  {Perlick}},\ }\bibfield  {title} {\bibinfo {title} {Application of the
  {G}auss-{B}onnet theorem to lensing in the {NUT} metric},\ }\href
  {https://doi.org/10.1007/s10714-020-02766-z} {\bibfield  {journal} {\bibinfo
  {journal} {Gen. Relativ. Gravit.}\ }\textbf {\bibinfo {volume} {52}},\
  \bibinfo {pages} {112} (\bibinfo {year} {2020})}\BibitemShut {NoStop}%
\bibitem [{\citenamefont {Werner}(2012)}]{Werner2012}%
  \BibitemOpen
  \bibfield  {author} {\bibinfo {author} {\bibfnamefont {M.~C.}\ \bibnamefont
  {Werner}},\ }\bibfield  {title} {\bibinfo {title} {Gravitational lensing in
  the {K}err-{R}anders optical geometry},\ }\href
  {https://doi.org/10.1007/s10714-012-1458-9} {\bibfield  {journal} {\bibinfo
  {journal} {Gen. Relativ. Gravit.}\ }\textbf {\bibinfo {volume} {44}},\
  \bibinfo {pages} {3047} (\bibinfo {year} {2012})}\BibitemShut {NoStop}%
\bibitem [{\citenamefont {Wei}\ \emph {et~al.}(2012)\citenamefont {Wei},
  \citenamefont {Liu}, \citenamefont {Fu},\ and\ \citenamefont
  {Yang}}]{Wei2012}%
  \BibitemOpen
  \bibfield  {author} {\bibinfo {author} {\bibfnamefont {S.-W.}\ \bibnamefont
  {Wei}}, \bibinfo {author} {\bibfnamefont {Y.-X.}\ \bibnamefont {Liu}},
  \bibinfo {author} {\bibfnamefont {C.-E.}\ \bibnamefont {Fu}},\ and\ \bibinfo
  {author} {\bibfnamefont {K.}~\bibnamefont {Yang}},\ }\bibfield  {title}
  {\bibinfo {title} {Strong field limit analysis of gravitational lensing in
  {K}err-{T}aub-{NUT} spacetime},\ }\href
  {https://doi.org/10.1088/1475-7516/2012/10/053} {\bibfield  {journal}
  {\bibinfo  {journal} {J. Cosmol. Astropart. Phys.}\ }\textbf {\bibinfo
  {volume} {2012}}\bibinfo  {number} { (10)},\ \bibinfo {pages}
  {053}}\BibitemShut {NoStop}%
\bibitem [{\citenamefont {Sharif}\ and\ \citenamefont
  {Iftikhar}(2016)}]{Sharif2016}%
  \BibitemOpen
\bibfield  {number} {  }\bibfield  {author} {\bibinfo {author} {\bibfnamefont
  {M.}~\bibnamefont {Sharif}}\ and\ \bibinfo {author} {\bibfnamefont
  {S.}~\bibnamefont {Iftikhar}},\ }\bibfield  {title} {\bibinfo {title}
  {Equatorial gravitational lensing by accelerating and rotating black hole
  with {NUT} parameter},\ }\href {https://doi.org/10.1007/s10509-015-2623-x}
  {\bibfield  {journal} {\bibinfo  {journal} {Astrophys. {S}pace {S}ci.}\
  }\textbf {\bibinfo {volume} {361}},\ \bibinfo {pages} {36} (\bibinfo {year}
  {2016})}\BibitemShut {NoStop}%
\bibitem [{\citenamefont {Grenzebach}\ \emph {et~al.}(2014)\citenamefont
  {Grenzebach}, \citenamefont {Perlick},\ and\ \citenamefont
  {L\"ammerzahl}}]{Grenzebach2014}%
  \BibitemOpen
  \bibfield  {author} {\bibinfo {author} {\bibfnamefont {A.}~\bibnamefont
  {Grenzebach}}, \bibinfo {author} {\bibfnamefont {V.}~\bibnamefont
  {Perlick}},\ and\ \bibinfo {author} {\bibfnamefont {C.}~\bibnamefont
  {L\"ammerzahl}},\ }\bibfield  {title} {\bibinfo {title} {Photon regions and
  shadows of {K}err-{N}ewman-{NUT} black holes with a cosmological constant},\
  }\href {https://doi.org/10.1103/PhysRevD.89.124004} {\bibfield  {journal}
  {\bibinfo  {journal} {Phys. {R}ev. {D}}\ }\textbf {\bibinfo {volume} {89}},\
  \bibinfo {pages} {124004} (\bibinfo {year} {2014})}\BibitemShut {NoStop}%
\bibitem [{\citenamefont {Grenzebach}(2016)}]{Grenzebach2016}%
  \BibitemOpen
  \bibfield  {author} {\bibinfo {author} {\bibfnamefont {A.}~\bibnamefont
  {Grenzebach}},\ }\href {https://doi.org/10.1007/978-3-319-30066-5} {\emph
  {\bibinfo {title} {The {S}hadow of {B}lack {H}oles}}},\ Springer Briefs in
  Physics\ (\bibinfo  {publisher} {Springer},\ \bibinfo {address} {Cham},\
  \bibinfo {year} {2016})\BibitemShut {NoStop}%
\bibitem [{\citenamefont {Forsyth}(1920)}]{Forsyth1920}%
  \BibitemOpen
  \bibfield  {author} {\bibinfo {author} {\bibfnamefont {A.~R.}\ \bibnamefont
  {Forsyth}},\ }\bibfield  {title} {\bibinfo {title} {Note on the central
  differential equation in the relativity theory of gravitation},\ }\href
  {https://doi.org/10.1098/rspa.1920.0019} {\bibfield  {journal} {\bibinfo
  {journal} {Proc. R. Soc. A}\ }\textbf {\bibinfo {volume} {97}},\ \bibinfo
  {pages} {145} (\bibinfo {year} {1920})}\BibitemShut {NoStop}%
\bibitem [{\citenamefont {Morton}(1921)}]{Morton1921}%
  \BibitemOpen
  \bibfield  {author} {\bibinfo {author} {\bibfnamefont {W.~B.}\ \bibnamefont
  {Morton}},\ }\bibfield  {title} {\bibinfo {title} {The forms of planetary
  orbits on the theory of relativity},\ }\href
  {https://doi.org/10.1080/14786442108633793} {\bibfield  {journal} {\bibinfo
  {journal} {{L}ondon, {E}dinburgh, {D}ublin {P}hilos. {M}ag. J. Sci.}\
  }\textbf {\bibinfo {volume} {42}},\ \bibinfo {pages} {511} (\bibinfo {year}
  {1921})}\BibitemShut {NoStop}%
\bibitem [{\citenamefont {Darwin}(1959)}]{Darwin1959}%
  \BibitemOpen
  \bibfield  {author} {\bibinfo {author} {\bibfnamefont {C.}~\bibnamefont
  {Darwin}},\ }\bibfield  {title} {\bibinfo {title} {The gravity field of a
  particle},\ }\href {https://www.jstor.org/stable/100508} {\bibfield
  {journal} {\bibinfo  {journal} {Proc. {R}. Soc. {A}}\ }\textbf {\bibinfo
  {volume} {249}},\ \bibinfo {pages} {180} (\bibinfo {year}
  {1959})}\BibitemShut {NoStop}%
\bibitem [{\citenamefont {Yang}\ and\ \citenamefont {Wang}(2013)}]{Yang2013}%
  \BibitemOpen
  \bibfield  {author} {\bibinfo {author} {\bibfnamefont {X.}~\bibnamefont
  {Yang}}\ and\ \bibinfo {author} {\bibfnamefont {J.}~\bibnamefont {Wang}},\
  }\bibfield  {title} {\bibinfo {title} {Y{NOGK}: {A} new public code for
  calculating null geodesics in the {K}err spacetime},\ }\href
  {https://doi.org/10.1088/0067-0049/207/1/6} {\bibfield  {journal} {\bibinfo
  {journal} {{A}strophys. {J}. {S}uppl. {S}er.}\ }\textbf {\bibinfo {volume}
  {207}},\ \bibinfo {pages} {6} (\bibinfo {year} {2013})}\BibitemShut {NoStop}%
\bibitem [{\citenamefont {Gralla}\ and\ \citenamefont
  {Lupsasca}(2020)}]{Gralla2020}%
  \BibitemOpen
  \bibfield  {author} {\bibinfo {author} {\bibfnamefont {S.~E.}\ \bibnamefont
  {Gralla}}\ and\ \bibinfo {author} {\bibfnamefont {A.}~\bibnamefont
  {Lupsasca}},\ }\bibfield  {title} {\bibinfo {title} {Null geodesics of the
  {K}err exterior},\ }\href {https://doi.org/10.1103/PhysRevD.101.044032}
  {\bibfield  {journal} {\bibinfo  {journal} {Phys. {R}ev. {D}}\ }\textbf
  {\bibinfo {volume} {101}},\ \bibinfo {pages} {044032} (\bibinfo {year}
  {2020})}\BibitemShut {NoStop}%
\bibitem [{\citenamefont {Frost}\ and\ \citenamefont
  {Perlick}(2021)}]{Frost2021a}%
  \BibitemOpen
  \bibfield  {author} {\bibinfo {author} {\bibfnamefont {T.~C.}\ \bibnamefont
  {Frost}}\ and\ \bibinfo {author} {\bibfnamefont {V.}~\bibnamefont
  {Perlick}},\ }\bibfield  {title} {\bibinfo {title} {Lightlike geodesics and
  gravitational lensing in the spacetime of an accelerating black hole},\
  }\href {https://doi.org/10.1088/1361-6382/abe0f5} {\bibfield  {journal}
  {\bibinfo  {journal} {Classical {Q}uantum {G}ravity}\ }\textbf {\bibinfo
  {volume} {38}},\ \bibinfo {pages} {085016} (\bibinfo {year}
  {2021})}\BibitemShut {NoStop}%
\bibitem [{\citenamefont {Grenzebach}\ \emph {et~al.}(2015)\citenamefont
  {Grenzebach}, \citenamefont {Perlick},\ and\ \citenamefont
  {L\"ammerzahl}}]{Grenzebach2015a}%
  \BibitemOpen
  \bibfield  {author} {\bibinfo {author} {\bibfnamefont {A.}~\bibnamefont
  {Grenzebach}}, \bibinfo {author} {\bibfnamefont {V.}~\bibnamefont
  {Perlick}},\ and\ \bibinfo {author} {\bibfnamefont {C.}~\bibnamefont
  {L\"ammerzahl}},\ }\bibfield  {title} {\bibinfo {title} {Photon regions and
  shadows of accelerated black holes},\ }\href
  {https://doi.org/10.1142/S0218271815420249} {\bibfield  {journal} {\bibinfo
  {journal} {Int. {J}. {M}od. {P}hys. {D}}\ }\textbf {\bibinfo {volume} {24}},\
  \bibinfo {pages} {1542024} (\bibinfo {year} {2015})}\BibitemShut {NoStop}%
\bibitem [{\citenamefont {Jefremov}\ and\ \citenamefont
  {Perlick}(2016)}]{Jefremov2016}%
  \BibitemOpen
  \bibfield  {author} {\bibinfo {author} {\bibfnamefont {P.~I.}\ \bibnamefont
  {Jefremov}}\ and\ \bibinfo {author} {\bibfnamefont {V.}~\bibnamefont
  {Perlick}},\ }\bibfield  {title} {\bibinfo {title} {Circular motion in {NUT}
  space-time},\ }\href {https://doi.org/10.1088/0264-9381/33/24/245014}
  {\bibfield  {journal} {\bibinfo  {journal} {Classical {Q}uantum {G}ravity}\
  }\textbf {\bibinfo {volume} {33}},\ \bibinfo {pages} {245014} (\bibinfo
  {year} {2016})}\BibitemShut {NoStop}%
\bibitem [{\citenamefont {Mino}(2003)}]{Mino2003}%
  \BibitemOpen
  \bibfield  {author} {\bibinfo {author} {\bibfnamefont {Y.}~\bibnamefont
  {Mino}},\ }\bibfield  {title} {\bibinfo {title} {Perturbative approach to an
  orbital evolution around a supermassive black hole},\ }\href
  {https://doi.org/10.1103/PhysRevD.67.084027} {\bibfield  {journal} {\bibinfo
  {journal} {Phys. {R}ev. {D}}\ }\textbf {\bibinfo {volume} {67}},\ \bibinfo
  {pages} {084027} (\bibinfo {year} {2003})}\BibitemShut {NoStop}%
\bibitem [{\citenamefont {Frost}(2021)}]{Frost2021b}%
  \BibitemOpen
  \bibfield  {author} {\bibinfo {author} {\bibfnamefont {T.~C.}\ \bibnamefont
  {Frost}},\ }\href@noop {} {\bibinfo {title} {Gravitational lensing by charged
  accelerating black holes}},\ \bibinfo {howpublished} {arXiv:2111.00283}
  (\bibinfo {year} {2021})\BibitemShut {NoStop}%
\bibitem [{\citenamefont {Byrd}\ and\ \citenamefont
  {Friedman}(1954)}]{Byrd1954}%
  \BibitemOpen
  \bibfield  {author} {\bibinfo {author} {\bibfnamefont {P.~F.}\ \bibnamefont
  {Byrd}}\ and\ \bibinfo {author} {\bibfnamefont {M.~D.}\ \bibnamefont
  {Friedman}},\ }\href {https://doi.org/10.1007/978-3-642-52803-3} {\emph
  {\bibinfo {title} {Handbook of {E}lliptic {I}ntegrals for {E}ngineers and
  {P}hysicists}}},\ \bibinfo {edition} {1st}\ ed.,\ Die {G}rundlehren der
  {M}athematischen {W}issenschaften\ (\bibinfo  {publisher} {Springer-Verlag,
  Berlin, Heidelberg},\ \bibinfo {year} {1954})\BibitemShut {NoStop}%
\bibitem [{\citenamefont {Hancock}(1917)}]{Hancock1917}%
  \BibitemOpen
  \bibfield  {author} {\bibinfo {author} {\bibfnamefont {H.}~\bibnamefont
  {Hancock}},\ }\href@noop {} {\emph {\bibinfo {title} {Elliptic
  {I}ntegrals}}},\ \bibinfo {edition} {1st}\ ed.,\ edited by\ \bibinfo {editor}
  {\bibfnamefont {M.}~\bibnamefont {Merriman}}\ and\ \bibinfo {editor}
  {\bibfnamefont {R.~S.}\ \bibnamefont {Woodward}},\ Mathematical {M}onographs\
  (\bibinfo  {publisher} {John Wiley \& Sons, New York},\ \bibinfo {year}
  {1917})\BibitemShut {NoStop}%
\bibitem [{\citenamefont {Synge}(1966)}]{Synge1966}%
  \BibitemOpen
  \bibfield  {author} {\bibinfo {author} {\bibfnamefont {J.~L.}\ \bibnamefont
  {Synge}},\ }\bibfield  {title} {\bibinfo {title} {The escape of photons from
  gravitationally intense stars},\ }\href
  {https://doi.org/10.1093/mnras/131.3.463} {\bibfield  {journal} {\bibinfo
  {journal} {Mon. {N}ot. {R}. {A}stron. {S}oc.}\ }\textbf {\bibinfo {volume}
  {131}},\ \bibinfo {pages} {463} (\bibinfo {year} {1966})}\BibitemShut
  {NoStop}%
\bibitem [{\citenamefont {Penrose}(1959)}]{Penrose1959}%
  \BibitemOpen
  \bibfield  {author} {\bibinfo {author} {\bibfnamefont {R.}~\bibnamefont
  {Penrose}},\ }\bibfield  {title} {\bibinfo {title} {The apparent shape of a
  relativistically moving sphere},\ }\href
  {https://doi.org/10.1017/S0305004100033776} {\bibfield  {journal} {\bibinfo
  {journal} {Proc. {C}ambridge {P}hilos. {S}oc.}\ }\textbf {\bibinfo {volume}
  {55}},\ \bibinfo {pages} {137} (\bibinfo {year} {1959})}\BibitemShut
  {NoStop}%
\bibitem [{\citenamefont {Mars}\ \emph {et~al.}(2018)\citenamefont {Mars},
  \citenamefont {Paganini},\ and\ \citenamefont {Oancea}}]{Mars2018}%
  \BibitemOpen
  \bibfield  {author} {\bibinfo {author} {\bibfnamefont {M.}~\bibnamefont
  {Mars}}, \bibinfo {author} {\bibfnamefont {C.~F.}\ \bibnamefont {Paganini}},\
  and\ \bibinfo {author} {\bibfnamefont {M.~A.}\ \bibnamefont {Oancea}},\
  }\bibfield  {title} {\bibinfo {title} {The fingerprints of black
  holes--shadows and their degeneracies},\ }\href
  {https://doi.org/10.1088/1361-6382/aa97ff} {\bibfield  {journal} {\bibinfo
  {journal} {Classical {Q}uantum {G}ravity}\ }\textbf {\bibinfo {volume}
  {35}},\ \bibinfo {pages} {025005} (\bibinfo {year} {2018})}\BibitemShut
  {NoStop}%
\bibitem [{\citenamefont {Frittelli}\ and\ \citenamefont
  {Newman}(1999)}]{Frittelli1999}%
  \BibitemOpen
  \bibfield  {author} {\bibinfo {author} {\bibfnamefont {S.}~\bibnamefont
  {Frittelli}}\ and\ \bibinfo {author} {\bibfnamefont {E.~T.}\ \bibnamefont
  {Newman}},\ }\bibfield  {title} {\bibinfo {title} {Exact universal
  gravitational lensing equation},\ }\href
  {https://doi.org/10.1103/PhysRevD.59.124001} {\bibfield  {journal} {\bibinfo
  {journal} {Phys. {R}ev. {D}}\ }\textbf {\bibinfo {volume} {59}},\ \bibinfo
  {pages} {124001} (\bibinfo {year} {1999})}\BibitemShut {NoStop}%
\bibitem [{\citenamefont {Perlick}(2004)}]{Perlick2004a}%
  \BibitemOpen
  \bibfield  {author} {\bibinfo {author} {\bibfnamefont {V.}~\bibnamefont
  {Perlick}},\ }\bibfield  {title} {\bibinfo {title} {Exact gravitational lens
  equation in spherically symmetric and static spacetimes},\ }\href
  {https://doi.org/10.1103/PhysRevD.69.064017} {\bibfield  {journal} {\bibinfo
  {journal} {Phys. {R}ev. {D}}\ }\textbf {\bibinfo {volume} {69}},\ \bibinfo
  {pages} {064017} (\bibinfo {year} {2004})}\BibitemShut {NoStop}%
\bibitem [{\citenamefont {Bezanson}\ \emph {et~al.}(2017)\citenamefont
  {Bezanson}, \citenamefont {Edelman}, \citenamefont {Karpinski},\ and\
  \citenamefont {Shah}}]{Bezanson2017}%
  \BibitemOpen
  \bibfield  {author} {\bibinfo {author} {\bibfnamefont {J.}~\bibnamefont
  {Bezanson}}, \bibinfo {author} {\bibfnamefont {A.}~\bibnamefont {Edelman}},
  \bibinfo {author} {\bibfnamefont {S.}~\bibnamefont {Karpinski}},\ and\
  \bibinfo {author} {\bibfnamefont {V.~B.}\ \bibnamefont {Shah}},\ }\bibfield
  {title} {\bibinfo {title} {Julia: {A} fresh approach to numerical
  computing},\ }\href {https://doi.org/10.1137/141000671} {\bibfield  {journal}
  {\bibinfo  {journal} {SIAM {R}ev.}\ }\textbf {\bibinfo {volume} {59}},\
  \bibinfo {pages} {65} (\bibinfo {year} {2017})}\BibitemShut {NoStop}%
\bibitem [{\citenamefont {Bohn}\ \emph {et~al.}(2015)\citenamefont {Bohn},
  \citenamefont {Throwe}, \citenamefont {H{\' e}bert}, \citenamefont
  {Henriksson}, \citenamefont {Bunandar}, \citenamefont {Scheel},\ and\
  \citenamefont {Taylor}}]{Bohn2015}%
  \BibitemOpen
  \bibfield  {author} {\bibinfo {author} {\bibfnamefont {A.}~\bibnamefont
  {Bohn}}, \bibinfo {author} {\bibfnamefont {W.}~\bibnamefont {Throwe}},
  \bibinfo {author} {\bibfnamefont {F.}~\bibnamefont {H{\' e}bert}}, \bibinfo
  {author} {\bibfnamefont {K.}~\bibnamefont {Henriksson}}, \bibinfo {author}
  {\bibfnamefont {D.}~\bibnamefont {Bunandar}}, \bibinfo {author}
  {\bibfnamefont {M.~A.}\ \bibnamefont {Scheel}},\ and\ \bibinfo {author}
  {\bibfnamefont {N.~W.}\ \bibnamefont {Taylor}},\ }\bibfield  {title}
  {\bibinfo {title} {What does a binary black hole merger look like?},\ }\href
  {https://doi.org/10.1088/0264-9381/32/6/065002} {\bibfield  {journal}
  {\bibinfo  {journal} {Classical {Q}uantum {G}ravity}\ }\textbf {\bibinfo
  {volume} {32}},\ \bibinfo {pages} {065002} (\bibinfo {year}
  {2015})}\BibitemShut {NoStop}%
\bibitem [{\citenamefont {Straumann}(2013)}]{Straumann2013}%
  \BibitemOpen
  \bibfield  {author} {\bibinfo {author} {\bibfnamefont {N.}~\bibnamefont
  {Straumann}},\ }\href {https://doi.org/10.1007/978-94-007-5410-2} {\emph
  {\bibinfo {title} {General {R}elativity}}},\ \bibinfo {edition} {2nd}\ ed.,\
  Graduate {T}exts in {P}hysics\ (\bibinfo  {publisher} {Springer},\ \bibinfo
  {address} {Heidelberg},\ \bibinfo {year} {2013})\BibitemShut {NoStop}%
\bibitem [{\citenamefont {Fohlmeister}\ \emph {et~al.}(2013)\citenamefont
  {Fohlmeister}, \citenamefont {Kochanek}, \citenamefont {Falco}, \citenamefont
  {Wambsganss}, \citenamefont {Oguri},\ and\ \citenamefont
  {Dai}}]{Fohlmeister2013}%
  \BibitemOpen
  \bibfield  {author} {\bibinfo {author} {\bibfnamefont {J.}~\bibnamefont
  {Fohlmeister}}, \bibinfo {author} {\bibfnamefont {C.~S.}\ \bibnamefont
  {Kochanek}}, \bibinfo {author} {\bibfnamefont {E.~E.}\ \bibnamefont {Falco}},
  \bibinfo {author} {\bibfnamefont {J.}~\bibnamefont {Wambsganss}}, \bibinfo
  {author} {\bibfnamefont {M.}~\bibnamefont {Oguri}},\ and\ \bibinfo {author}
  {\bibfnamefont {X.}~\bibnamefont {Dai}},\ }\bibfield  {title} {\bibinfo
  {title} {A two-year time delay for the lensed quasar {SDSS} {J}1029$+$2623},\
  }\href {https://doi.org/10.1088/0004-637X/764/2/186} {\bibfield  {journal}
  {\bibinfo  {journal} {{A}strophys. {J}.}\ }\textbf {\bibinfo {volume}
  {764}},\ \bibinfo {pages} {186} (\bibinfo {year} {2013})}\BibitemShut
  {NoStop}%
\bibitem [{\citenamefont {Koptelova}\ \emph {et~al.}(2012)\citenamefont
  {Koptelova} \emph {et~al.}}]{Koptelova2012}%
  \BibitemOpen
  \bibfield  {author} {\bibinfo {author} {\bibfnamefont {E.}~\bibnamefont
  {Koptelova}} \emph {et~al.},\ }\bibfield  {title} {\bibinfo {title} {Time
  delay between images of the lensed quasar {UM}673},\ }\href
  {https://doi.org/10.1051/0004-6361/201116645} {\bibfield  {journal} {\bibinfo
   {journal} {Astron. {A}strophys.}\ }\textbf {\bibinfo {volume} {544}},\
  \bibinfo {pages} {A51} (\bibinfo {year} {2012})}\BibitemShut {NoStop}%
\bibitem [{\citenamefont {Suyu}\ \emph {et~al.}(2017)\citenamefont {Suyu} \emph
  {et~al.}}]{Suyu2017}%
  \BibitemOpen
  \bibfield  {author} {\bibinfo {author} {\bibfnamefont {S.~H.}\ \bibnamefont
  {Suyu}} \emph {et~al.},\ }\bibfield  {title} {\bibinfo {title} {H0{L}i{COW} -
  {I}. ${H}_0$ {L}enses in {COSMOGRAIL}’s {W}ellspring: Program overview},\
  }\href {https://doi.org/10.1093/mnras/stx483} {\bibfield  {journal} {\bibinfo
   {journal} {Mon. {N}ot. {R}. {A}stron. {S}oc.}\ }\textbf {\bibinfo {volume}
  {468}},\ \bibinfo {pages} {2590} (\bibinfo {year} {2017})}\BibitemShut
  {NoStop}%
\bibitem [{\citenamefont {Milne-Thomson}(1972)}]{MilneThomson1972}%
  \BibitemOpen
  \bibfield  {author} {\bibinfo {author} {\bibfnamefont {L.~M.}\ \bibnamefont
  {Milne-Thomson}},\ }\bibfield  {title} {\bibinfo {title} {Elliptic
  integrals},\ }in\ \href@noop {} {\emph {\bibinfo {booktitle} {Handbook of
  {M}athematical {F}unctions {W}ith {F}ormulas, {G}raphs, and {M}athematical
  {T}ables}}},\ \bibinfo {series and number} {Applied {M}athematics {S}eries},\
  \bibinfo {editor} {edited by\ \bibinfo {editor} {\bibfnamefont
  {M.}~\bibnamefont {Abramowitz}}\ and\ \bibinfo {editor} {\bibfnamefont
  {I.~A.}\ \bibnamefont {Stegun}}}\ (\bibinfo  {publisher} {U. {S}. {D}ept. of
  {C}ommerce, {N}ational {B}ureau of {S}tandards},\ \bibinfo {address}
  {Washington D. C.},\ \bibinfo {year} {1972})\ \bibinfo {edition} {10th}\
  ed.,\ pp.\ \bibinfo {pages} {587--607}\BibitemShut {NoStop}%
\end{thebibliography}%

\end{document}